\documentclass[aps,amsfonts,pra,twocolumn,showpacs]{revtex4-1}
\usepackage{epsfig,amsmath,amssymb,bm,epsf,graphicx,psfrag,color,dsfont,natbib,subcaption,epstopdf}

\usepackage{bbold}

\def\bra#1{\langle#1\vert}
\def\ket#1{\vert#1\rangle}
\newcommand{\bc}{\begin{center}}
\newcommand{\ec}{\end{center}}
\newcommand{\be}{\begin{equation}}
\newcommand{\ee}{\end{equation}}

\newcommand{\outerproduct}[2]{|#1\rangle\langle #2|}

\newcommand{\ztwo}{\mathbb{Z}_2}
\newcommand{\zd}{\mathbb{Z}_d}

\newcommand{\thatis}{$i.e.~$}

\begin{document}

\title{Universal quantum computing using  $(\zd)^3$ symmetry-protected topologically ordered states }
\author{Yanzhu Chen}
\affiliation{C. N. Yang Institute for Theoretical Physics and Department of Physics and Astronomy, State University of New York at Stony Brook, Stony Brook, NY 11794-3840, USA}
\author{Abhishodh Prakash}
\affiliation{C. N. Yang Institute for Theoretical Physics and Department of Physics and Astronomy, State University of New York at Stony Brook, Stony Brook, NY 11794-3840, USA}
\author{Tzu-Chieh Wei}
\affiliation{C. N. Yang Institute for Theoretical Physics and Department of Physics and Astronomy, State University of New York at Stony Brook, Stony Brook, NY 11794-3840, USA}
\date{\today}

\begin{abstract}
Measurement-based quantum computation describes a scheme where entanglement of resource states is utilized to simulate arbitrary quantum gates via local measurements. Recent works suggest that symmetry-protected topologically non-trivial, short-ranged entangled states are promising candidates for such a resource. Miller and Miyake [NPJ Quantum Information \textbf{2}, 16036 (2016)] recently constructed a particular $\mathbb{Z}_2 \times \mathbb{Z}_2 \times \mathbb{Z}_2$ symmetry-protected topological state on the Union-Jack lattice and established its quantum computational universality. However, they suggested that the same construction on the triangular lattice might not lead to a universal resource. Instead of qubits, we generalize the construction to qudits and show that the resulting $(d-1)$ qudit  nontrivial $\mathbb{Z}_d \times \mathbb{Z}_d \times \mathbb{Z}_d$ symmetry-protected topological states  are universal on the triangular lattice, for $d$ being a prime number greater than 2. The same construction also holds for other 3-colorable lattices, including the Union-Jack lattice.  
\end{abstract}

\maketitle

\section{Introduction}

Raussendorf and Briegel~\cite{Raussendorf2001} described a one-way computation scheme where the flow of quantum information is driven by local projective measurements on some suitable entangled state (also referred to as a resource state). It is one-way in the sense that the entanglement in the resource state is destroyed irreversibly as measurements are carried out. Such a quantum computation scheme is called measurement-based quantum computation (MBQC)~\cite{Raussendorf2003, Nielsen2005, Briegel2009, RaussendorfWei}. A universal resource state is one on which any gate, including the universal set, can be simulated by performing only local measurements. The first proposed and most well studied universal resource state is the cluster state, defined on a regular lattice,
	\be
		|\phi\rangle_C = \prod_{\langle a, b\rangle} CZ_{(ab)} \underset{{\rm all \, sites} \, i}{\bigotimes} |+\rangle_i	\label{cluster_state},
	\ee
where $|+\rangle \equiv \frac{1}{\sqrt{2}}(|0\rangle+|1\rangle), CZ \equiv |0\rangle\langle 0| \otimes \mathds{1} + |1\rangle\langle 1| \otimes Z$ is the controlled-Z gate, and $\langle a, b\rangle$ denotes an edge connecting vertices $a$ and $b$. This can be defined on any graph, and is in general called a graph state.

However, the complete set of universal resource states is not yet known, nor characterized. Ever since then, there has been effort to find and characterize other universal resource states. For example, the list includes cluster states on all regular lattices~\cite{VandenNest06}, certain tensor-network states~\cite{Gross2007, Gross2007_1}, the TriCluster state~\cite{TriCluster},  certain 2D AKLT and AKLT-like states \cite{Wei2011,Miyake2011,Wei2012,Wei2014,Wei2015,Cai2010}, as well as their deformation \cite{Darmawan2012, Wei2017}. 
It has been established that presence of entanglement in the resource state is necessary for it to be universal, but there is no conclusion on what the general criterion to search for MBQC resource states~\cite{VandenNest07,Gross09,Bremner09}.
Recent works have suggested that certain phases of matter may host computational capability, such as the
 symmetry protected topological (SPT) phases~\cite{GuWenSPT,ChenComplete,PollmannSPT,SchuchSPT,ChenScience}, whose ground states possess short-range entanglement.

\begin{figure}[h]
	\begin{subfigure}{0.22\textwidth}
		\includegraphics[height=35mm,width=40mm]{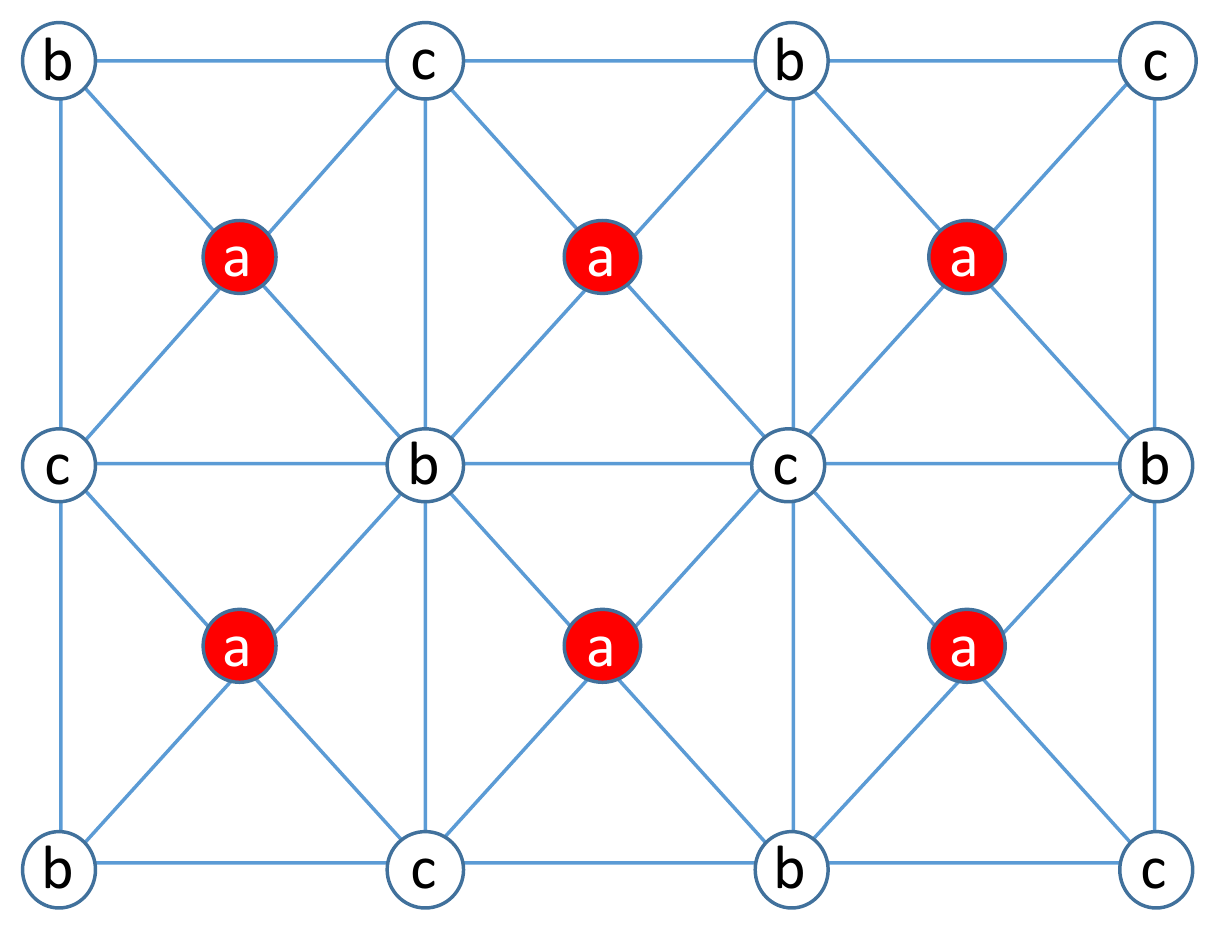}
		\caption{Each group of qubits labelled by $a, b, c$ are entangled by $CCZ$ gate.}
	\end{subfigure}
	\hspace{1mm}
	\begin{subfigure}{0.22\textwidth}
		\includegraphics[height=35mm,width=40mm]{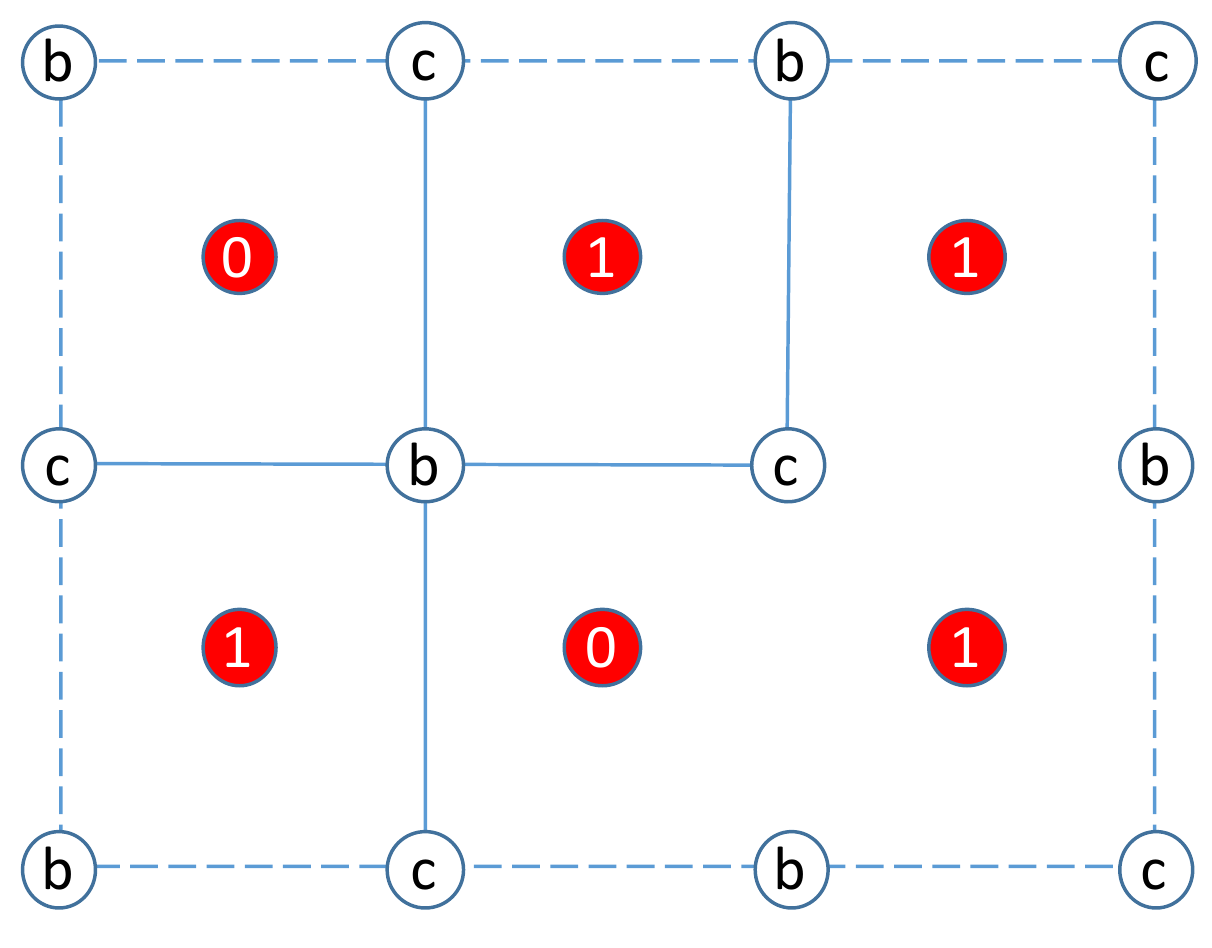}
		\caption{$CZ$ gate remains between different adjacent outcomes.}
	\end{subfigure}
	\caption{Union-Jack state proposed in Ref.~\cite{Miller2016}. Measuring the red qubits (marked by `a') at centre of each square results in a random graph state embedded in a square lattice. \label{MM}}
\end{figure}
The connection of SPT phases to MBQC was first discovered by Else et al. in 1D as a quantum wire for perfect transmission of quantum information~\cite{Else,Prakash}, and  subsequently strengthened for gate simulations in more generality~\cite{MillerMiyake2015,Stephen17,Raussendorf17}. For two dimensions, the universal quantum computation was only found to be possible on certain fixed-point SPT states~\cite{Nautrup2015,Miller2016,Miller2016_2}, as well as certain deformation around fixed points~\cite{Wei2017}.   

\begin{figure}
	\begin{subfigure}{0.22\textwidth}
		\centering
		\includegraphics[height=30mm,width=30mm]{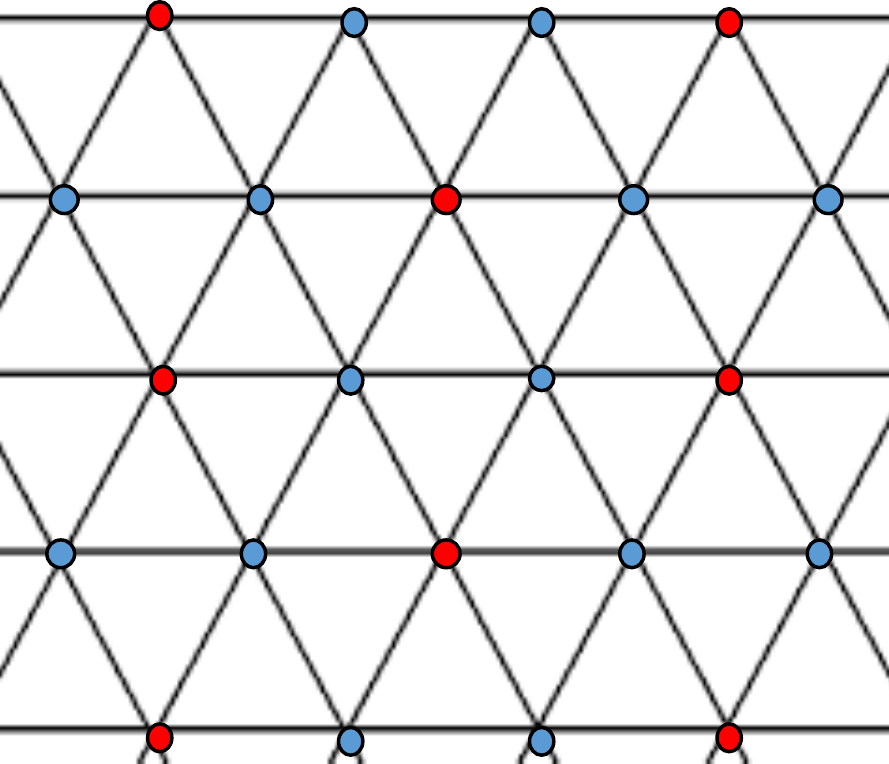}
		\caption{Qubit triangular SPT state. \label{triangular}}
	\end{subfigure}
	\hspace{5mm}
	\begin{subfigure}{0.22\textwidth}
		\centering
		\includegraphics[height=30mm,width=30mm]{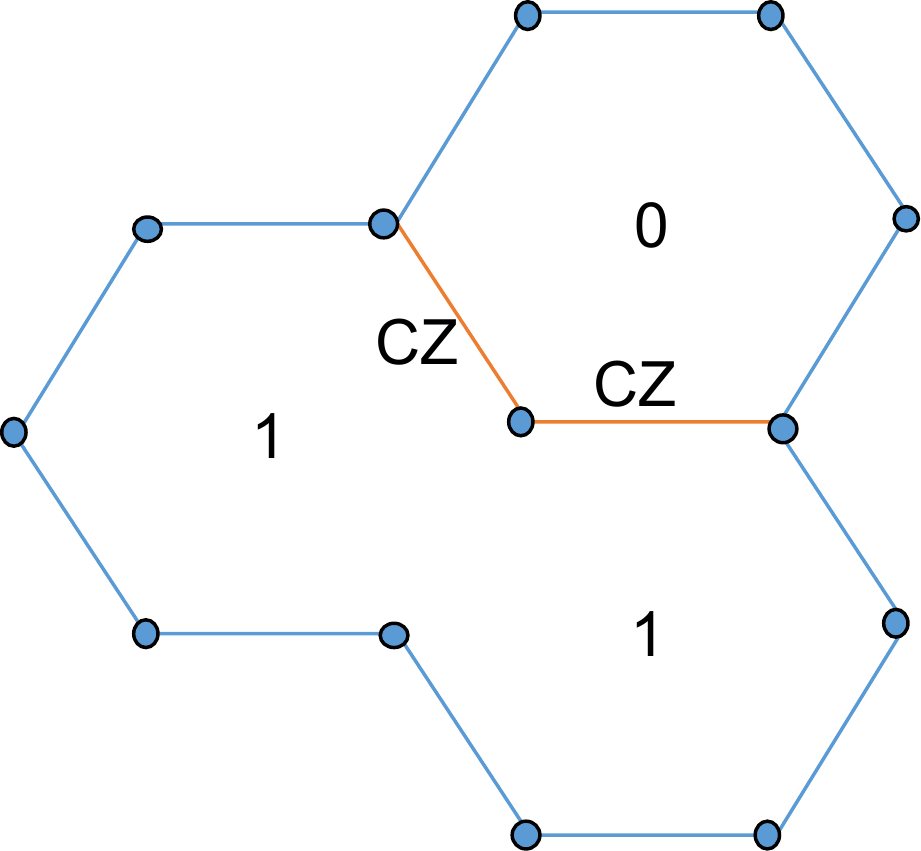}
		\caption{Three adjacent plaquettes and the edges between them for qubit case. \label{measure0}}
	\end{subfigure}
	\caption{Qubit triangular SPT state and demonstration of how network structure is lost.}
\end{figure}

Among the above examples, Miller and Miyake proposed a state that possesses 2D SPT order and showed that it is a universal resource for MBQC~ \cite{Miller2016}. It is named the Union-Jack state as it is defined on the structure shown in Fig.~\ref{MM}:
	\be
	|\phi\rangle_{UJ} = \prod_{\langle a, b, c\rangle} CCZ_{(abc)} \underset{{\rm all \, sites} \, i}{\bigotimes} |+\rangle_i, 
	\ee
where $CCZ\equiv|0\rangle\langle 0| \otimes \mathds{1} \otimes \mathds{1} + |1\rangle\langle 1| \otimes CZ$ is the control-control-Z gate and $\langle a, b, c\rangle$ denotes a triangle with vertices $a$, $b$ and $c$. The state has $\ztwo \times \ztwo \times \ztwo$ symmetry, where each $\ztwo$ factor represents symmetry action (generated by spin flip) on all sites $a$, all sites $b$, or all sites $c$, respectively. Their proof of universality involves measuring the qubits at  the center of each square (e.g. sites $a$), which leads to a (random) graph state whose graph is embedded in a square lattice, with the edges being occupied or not depending on the neighboring measurement outcomes. A similar state can be constructed with the same definition on the triangular lattice, given in Fig.~\ref{triangular}, and will be referred to as the triangular SPT state, which was also constructed by Yoshida in Ref.~\cite{Yoshida2016}. Following Miller and Miyake, if we measure those qubits marked in red, as shown in Fig.~\ref{triangular}, the resultant state will become a random graph state on a honeycomb lattice. In particular, as two of the three adjacent plaquettes at a vertex junction will necessarily have the same outcome, this implies that the edge between them experiences  no CZ action effectively. Thus, around each vertex there can be either two edges or no edge, and therefore we cannot obtain the network structure needed for universal MBQC, indicating that the qubit triangular SPT state may not be universal. In contrast, the measurements on the Union-Jack lattice result in random graphs on a square lattice and there can be zero, two or four edges occupied around a vertex, which turns out to be sufficient for the universality~\cite{Miller2016}.

In this paper, we consider qudit SPT states on triangular lattices, defined in Eq.~\ref{eq:state_defn} below (and other 3-colorable lattices, including the Union Jack), and aim to explore their quantum computational universality. 
Continuing the reasoning above, for the physical entity being a qubit, there is only one kind of domain wall, \thatis between 0 and 1. If we use qudit, whose Hilbert space has dimensionality $d \geq 3$, there are more than one kind of domain walls. In the qudit triangular state, measuring all the qudits marked in red would lead us to some random graph-like state (with each edge representing $CZ$ raised to some power; see Eq.~(\ref{eq:Gl})) embedded in the honeycomb lattice. As we shall prove in Secs.~\ref{sec:main}, the universality of qudit graph-like states depends on their graphical percolation property, in the same way as the qubit case~\cite{Browne2008,Wei2012}. From the percolation perspective, we can intuitively see why the larger $d$ will help. At each junction depending on measurement outcomes on the qudits located at the three nearest hexagons,  there are three scenarios: (i) no edge is occupied, as all three measurement outcomes are equal, and this occurs with probability ${1}/{d^2}$; (ii) two edges  are occupied with probability ${3(d-1)}/{d^2}$, when only two of the measurement outcomes are the same; (iii) three edges are occupied with probability ${(d-1)(d-2)}/{d^2}$, when all three outcomes are distinct. From this the average probability of one edge being present is $1-{1}/{d}$. For $d=2$ this is $0.5$, below the edge percolation threshold of honeycomb lattice~\cite{Sykes1964}; for $d \geq 3$ it is above the threshold, and approaches to unity as $d$ increases. This simple estimate therefore motivates us to study qudit triangular SPT state. But we remark that the use of the percolation threshold is only a crude estimate, as edge occupation is correlated and not independent of other edges. Numerical simulations will be needed to ascertain the percolation property; see Figs.~\ref{numerical} and~\ref{numericalUJ} below.

As shown in Sec.~\ref{sec:main}, our proof of universality for qudit SPT states relies on their reduction to qudit graph-like states. Here we make a distinction between graph states and graph-like states: for the former, each edge in the graph represents uniformly the action of a CZ gate, whereas for the latter, each edge can have different powers of the CZ gate, \thatis $CZ^r$, where $r$ is edge dependent. We also make similar distinction between cluster states and cluster-like states. To show qudit graph-like states are universal we need to establish graphical rules of how the graphs transform as some qudits are measured in the qudit Pauli bases and use them to reduce the graph-like states to cluster-like states (\thatis the latter graphs are regular lattices, such as the square lattice). Such rules have been studied in both the qubit case and the qudit case (see Refs.~\cite{Hein2006, Bahramgiri2007}). We will give simple derivation for those rules that are needed for the reduction in Sec.~\ref{sec:rules}. After the chain of reduction, we need to show that cluster-like states are indeed universal for quantum computation. The qudit cluster state (where the edges represent uniformly the qudit CZ gate) on the square lattice was shown by Zhou et al.~\cite{Zhou2003} to be universal, by generalizing the qubit cluster-state formalism by Raussendorf, Browne and Briegel~\cite{Raussendorf2003}. By further generalizing these results~\cite{Raussendorf2003,Zhou2003}, we  can show that  qudit cluster-like states are indeed universal. Although our proof only works for $d$ being prime, we do expect that they are universal for all $d$.  

The remaining of the paper is organized as follows. In Sec.~\ref{sec:triangularSPT} we describe the construction of qudit triangular SPT states. The discussion of their symmetry properties is relegated to Appendices~\ref{app:SPT} and \ref{app:DDW}. It applies to lattices or graphs that are 3-colorable. 
 In Sec.~\ref{sec:main} we discuss the procedure we use to establish quantum computational universality of qudit SPT states. This section contains important ingredients, discussed above, including (i) the reduction, via local measurements, of qudit SPT states to qudit graph-like states; (ii) numerical results that confirm the required percolation properties of the random graphs; (iii) the reduction of qudit graph-like states to qudit cluster-like states; (iv) proof that cluster-like states are universal.  The details for (iii) are elaborated in Appendix~\ref{app:convert_procedure} and the those for (iv) are described in Appendix~\ref{app:cluster-like}. In Sec.~\ref{sec:discussion} we summarize our results and discuss a few unresolved issues and possible future work.

\begin{table*}[ht]
\begin{center}
	\begin{tabular}{ l | r }
	\hline
	\textbf{Object} & \textbf{Definition} \\
	\hline
	Fourier transform basis & $|+_j\rangle = \frac{1}{\sqrt{d}} \sum_{k=0}^{d-1} \varpi^{jk} |k\rangle$ \\
	& $|+\rangle=|+_0\rangle$ \\
	\hline
	Generalized Pauli operators & $Z = \sum_{k=0}^{d-1} \varpi^k |k\rangle \langle k|$ \\
	\cline{2-2}
	& $X = \sum_{k=0}^{d-1} |k-1\rangle \langle k|$ \\
	\hline
	Fourier gate & $F = \frac{1}{\sqrt{d}} \sum_{j,k=0}^{d-1} \varpi^{jk} |j\rangle \langle k| = \sum_{k=0}^{d-1} |+_k\rangle \langle k|$ \\
	& (in $d=2$ $F$ is the Hadamard gate) \\
	\hline
	Control-Z & $CZ = \sum_{k=0}^{d-1} |k\rangle \langle k| \otimes Z^k$ \\
	& $= \sum_{k,l=0}^{d-1} \varpi^{kl} |k\rangle \langle k| \otimes |l\rangle \langle l|$ \\
	\hline
	Generalized Control-Z & $CZ^q = \sum_{k=0}^{d-1} |k\rangle \langle k| \otimes Z^{qk}$ \\
	& $= \sum_{k,l=0}^{d-1} \varpi^{qkl} |k\rangle \langle k| \otimes |l\rangle \langle l|$ \\
	\hline
	Control-Control-Z & $CCZ = \sum_{k=0}^{d-1} |k\rangle \langle k| \otimes CZ^k$ \\
	& $= \sum_{k,l,m=0}^{d-1} \varpi^{klm} |k\rangle \langle k| \otimes |l\rangle \langle l| \otimes |m\rangle \langle m|$ \\
	\hline
	Generalized Pauli group & $\mathcal{P}_d = \{\varpi^aX^bZ^c | a, b, c \in \zd\}$ \\
	\hline
	Mutually unbiased bases $\{|a_k\rangle\}^{d-1}_{k=0}$ and $\{|b_k\rangle\}^{d-1}_{k=0}$ & if $|\langle a_i | b_j \rangle| = 1/\sqrt{d}$ for all $i, j$ \\
	\hline
	\end{tabular}
	\caption{Definitions for states and operators in the qudit case; see e.g.  Ref.~\cite{Hall2006}. \label{defn}}
\end{center}
\end{table*}

\section{Qudit triangular SPT state}
\label{sec:triangularSPT}
From now on we will be dealing with qudits only, and modulo $d$ arithmetic is used in most cases where $d$ is the dimension of Hilbert space for a single qudit. For convenience we will define $\varpi \equiv \exp{(2\pi i/d)}$, and denote the computational basis by $\{ \ket k \}$ with $k=0, ..., d-1$. We summarize some definitions for qudits in Table \ref{defn}. Unless stated otherwise, the summation over states is always from 0 to $d-1$. Measuring in the basis defined by unitary $U \in U(d)$ means measuring in the basis $\{ U \ket k \}, \, k=0, ..., d-1$.
The state of interest is defined on a triangular lattice (see Fig.~\ref{state}). For it to possess symmetry similar to the qubit Union-Jack state, we need to modify the entangling operation in the definition for the qudit case. With one qudit at each site in the state $|+\rangle\equiv \sum_{j=0}^{d-1}|j\rangle/\sqrt{d}$, we entangle the three qudits on the vertices of each upward triangle by $CCZ^k$ gate, and those on each downward triangle by $CCZ^{\dagger k}$ gate. We thus arrive at states~\cite{Yoshida2016}
	\be
	|\phi_k\rangle_T = \prod_{\vartriangle(a, b, c)} CCZ_{(abc)}^k \prod_{\triangledown(d, e, f)} CCZ^{\dagger k}_{(def)} \underset{\mathrm{site} \, i}{\bigotimes} |+\rangle_i,
	\label{eq:state_defn}
	\ee
where $\vartriangle(a, b, c)$ and $\triangledown(d, e, f)$ denote the upward and downward triangles, respectively. The qudit CCZ gate is defined as 
	\be
	CCZ \equiv \sum_{m=0}^{d-1} |m\rangle \langle m| \otimes CZ^m,
	\ee
	and the qudit CZ gate is defined as
	\be
	CZ \equiv \sum_{m=0}^{d-1} |m\rangle \langle m| \otimes Z^{m}=\sum_{m,q} \varpi^{mq}|m,q\rangle\langle m,q|.
	\ee	
 The index $k$ can take value from 1 to $d-1$ and denotes certain nontrivial SPT classes and   our treatment for universality holds for all $k \neq 0 \, \mathrm{mod} \, d$ cases. To see the symmetry group $\zd \times \zd \times \zd$, we refer to the example in Fig.~\ref{symmetry}. Two adjacent triangles are necessarily one upward and one downward. Without loss of generality we examine sites $a$. Neglecting all the rest of the state, this part shown in the figure has wavefunction:
	\be
	|\psi\rangle = CCZ^k_{(a_1bc)} CCZ^{\dagger k}_{(a_2bc)} \dots |+\rangle_{(a_1bca_2)} \dots,
	\label{example_state}
	\ee
where $\dots$ indicates other part of the state which can have entanglement with qudits $a_1$, $b$, $c$ and $a_2$. We make use of the following equation:
	\begin{align}
	CCZ^{\dagger k}_{(abc)} X_a CCZ^k_{(abc)} = CZ_{(bc)}^k X_a \\
	CCZ^k_{(abc)} X_a CCZ^{\dagger k}_{(abc)} = CZ_{(bc)}^{\dagger k} X_a.
	\end{align}
Operating $X_{a_1} \otimes X_{a_2}$ on the state therefore gives:
	\begin{align}
	X_{a_1} X_{a_2} |\psi\rangle & = [CCZ^k_{(a_1bc)} CCZ^{\dagger k}_{(a_2bc)}] \times \nonumber \\
		& [CZ_{(bc)}^k X_{a_1} CZ^{\dagger k}_{(bc)} X_{a_2}] \dots |+\rangle_{(a_1bca_2)} \dots \nonumber \\
		& = [CCZ^k_{(a_1bc)} CCZ^{\dagger k}_{(a_2bc)}] \dots |+\rangle_{(a_1bca_2)} \dots \nonumber \\
		& = |\psi\rangle,
	\end{align}
where we have also used the fact $X\equiv\sum_{m=0}^{d-1} |m-1\rangle \langle m|$ and thus $X \ket+=\ket+$. If we extend the consideration to all sites, we can label each site by one of the labels $a, b, c$ such that  the three vertices of each triangle are labelled by $(a, b, c)$. From the discussion above we thus see that acting $X^m$ on all sites with the same label is a symmetry, because each edge lies between an upward triangle and a downward triangle.  $\{ X^m | m=0, 1, \dots, d-1 \}$ forms a $\zd$ group. As the $CCZ$ gate is symmetric in the three qudits involved, so the same symmetry holds for sublattice of $b$ or $c$ as well. The construction of our states follows the standard way of constructing SPT states in Ref.~\cite{Chen2013, Miller2016_2}, and is a symmetry-protected topologically non-trivial state. It can be defined on any 3-colorable graph composed of triangles, including Union-Jack lattice; the detail is described explicitly in Appendix~\ref{app:SPT}. The nontrivial SPT order for these states can be understood from the so-called decorated domain-wall  (DDW) picture~\cite{Chen2014}, which we review in Appendix~\ref{app:DDW}. In fact, the measurement of qudits marked red acts to freeze the states of the domain spins. The qudits on the domain walls therefore form $\zd \times \zd$ SPT states that decorate the domains.

\section{Showing universality} \label{sec:main}

\begin{figure}[h]
	\begin{subfigure}{0.22\textwidth}
		\centering
		\includegraphics[height=25mm,width=25mm]{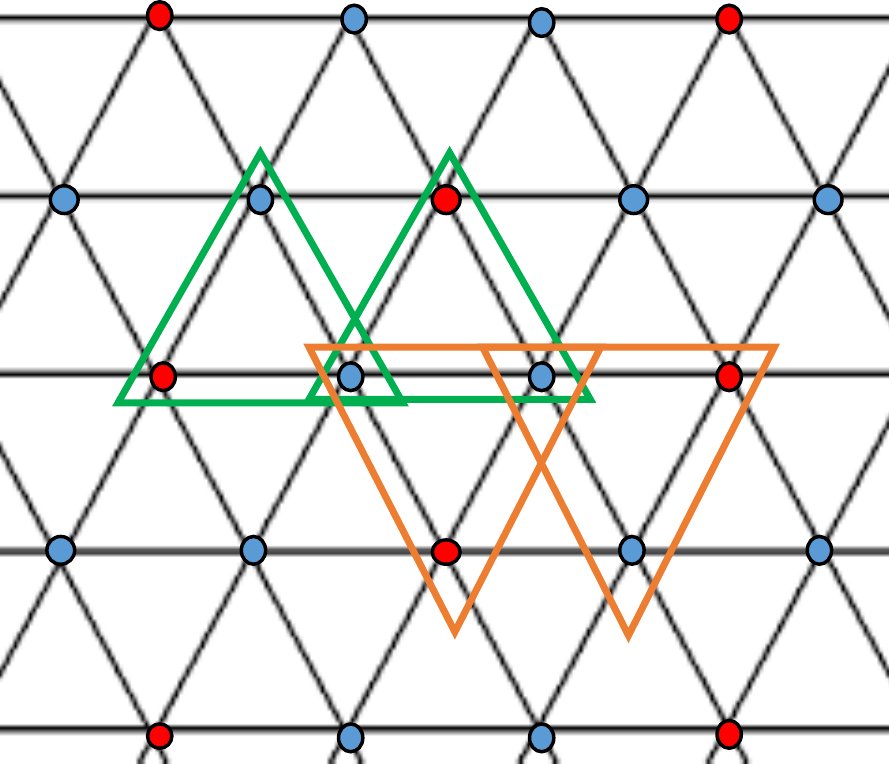}
		\caption{Green (upward)/orange (downward) triangles encircle three qudits entangled by $CCZ^k/CCZ^{\dagger k}$ gate. \label{state}}
	\end{subfigure}
	\hspace{1mm}
	\begin{subfigure}{0.22\textwidth}
		\centering
		\includegraphics[height=25mm,width=20mm]{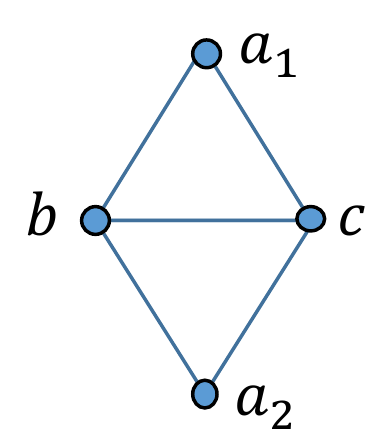}
		\caption{The state used to demonstrate the symmetry action. \label{symmetry}}
	\end{subfigure} \\
	\vspace{10mm}
	\begin{subfigure}{0.22\textwidth}
		\centering
		\includegraphics[height=25mm,width=25mm]{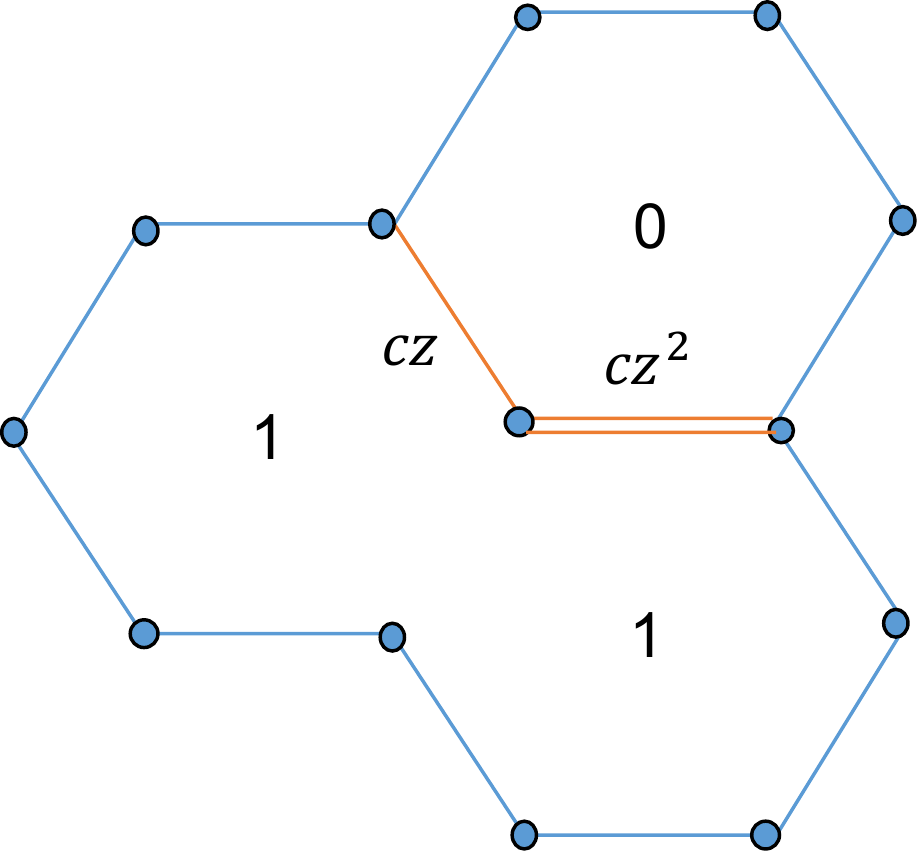}
		\caption{Three adjacent plaquettes for $d=3$, where two edges are present. \label{measure}}
	\end{subfigure}
	\hspace{1mm}
	\begin{subfigure}{0.22\textwidth}
		\centering
		\includegraphics[height=25mm,width=25mm]{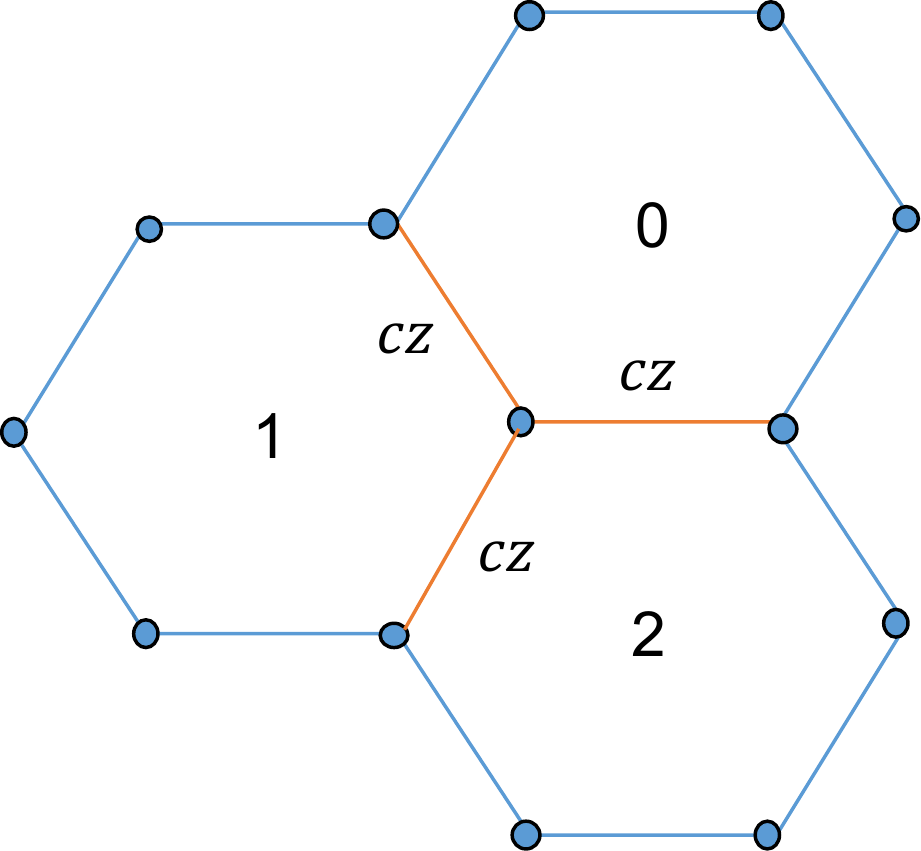}
		\caption{Three adjacent plaquettes for $d=3$, where three edges are present. \label{measure1}}
	\end{subfigure}
	\caption{Qudit triangular SPT state and example result after measurements. Note that the edges at the boundary of (c) and (d) are just guide to the eye.}
\end{figure}

Our goal is to test whether the qudit triangular SPT state can serve as a universal resource for MBQC. Following Ref.~\cite{Miller2016} we measure the red qudits in Fig.~\ref{state}, and examples of the outcomes on three adjacent plaquettes are shown in Figs.~\ref{measure} \&~\ref{measure1}. It is worth mentioning what we mean when we say a measurement in a certain basis produces an outcome $a$: after this measurement the measured qudit is projected onto the one basis state labelled by $a$, where $0 \leq a \leq d-1$. The measurement on all domain spins results in a random graph-like state embedded on the honeycomb lattice (for the triangular SPT state), which holds regardless of the index $k$ in Eq.~\ref{eq:state_defn} (as long as $k\neq0$). In order to show that the graph-like states are universal for MBQC, we need to examine their graphical properties. There have been works \cite{Browne2008, Wei2011} that identified certain MBQC resource states by studying their percolation properties. Here we follow the same idea, and list the four steps in our approach.

\begin{enumerate}
	\item Measure certain qudits on the triangular state in the computational basis, such that the remaining qudits form a honeycomb. Depending on the outcome, the remaining qudits can be either connected to their neighbours by edges representing $CZ^q$ gates (where $q$ can take value from 1 to $d-1$ and can be different for different edges), or not connected.
	\item Percolation property of the random subgraphs of the resulting honeycomb can be studied by numerical simulation.
	\item Show that the random graph-like state formed after measurements can be converted by local measurements to a cluster-like state (where $CZ$ edges are replaced by general $CZ^q$). If the idea in Ref.~\cite{Wei2012} can be generalized to the qudit case, we will be able to make the connection between universality and percolation property of the graph.
	\item The qudit cluster state where qudits are connected by $CZ$ gate is proved to be universal by Zhou et al~\cite{Zhou2003}; we need to show that the cluster-like states with general $CZ^q$ gates work as well. That is, all possible 1-qudit gates and one imprimitive 2-qudit gate can be realized on the cluster-like state.
\end{enumerate}

	\subsection{Probability of measurement outcomes} \label{sec:1st_step}
	
	From the definition of our triangular SPT state Eq.~(\ref{eq:state_defn}) it is easy to see that all $d$ possible outcomes appear with the same probability $\frac{1}{d}$. Depending on the outcomes $m_1$ and $m_2$ for two adjacent measurements, the edge between them results in a $CZ^{m_1-m_2}$ gate entangling the two qudits on the ends of this edge, where $m_1/m_2$ together with the two remaining qudits forms an upward/downward triangle. It is clear that when $m_1-m_2=0$ qudits on the two ends will be disentangled. We can again use Fig.~\ref{symmetry} to demonstrate this. Using Eq.~(\ref{example_state}) we see that:
		\be
		(_{a_1}\bra {m_1} _{a_2}\bra {m_2}) \ket \psi_{a_1bca_2} = (\frac{1}{\sqrt{d}})^4 \sum_{j, k} \varpi^{(m_1-m_2)jk} \ket j_b \ket k_c.
		\ee
	The measurement hence projects the state to $CZ^{m_1-m_2} \ket +_b \ket +_c$.

	\subsection{Simulation results} \label{sec:simulation}
	
\begin{figure}[h]
		\centering
		\includegraphics[height=48mm,width=80mm]{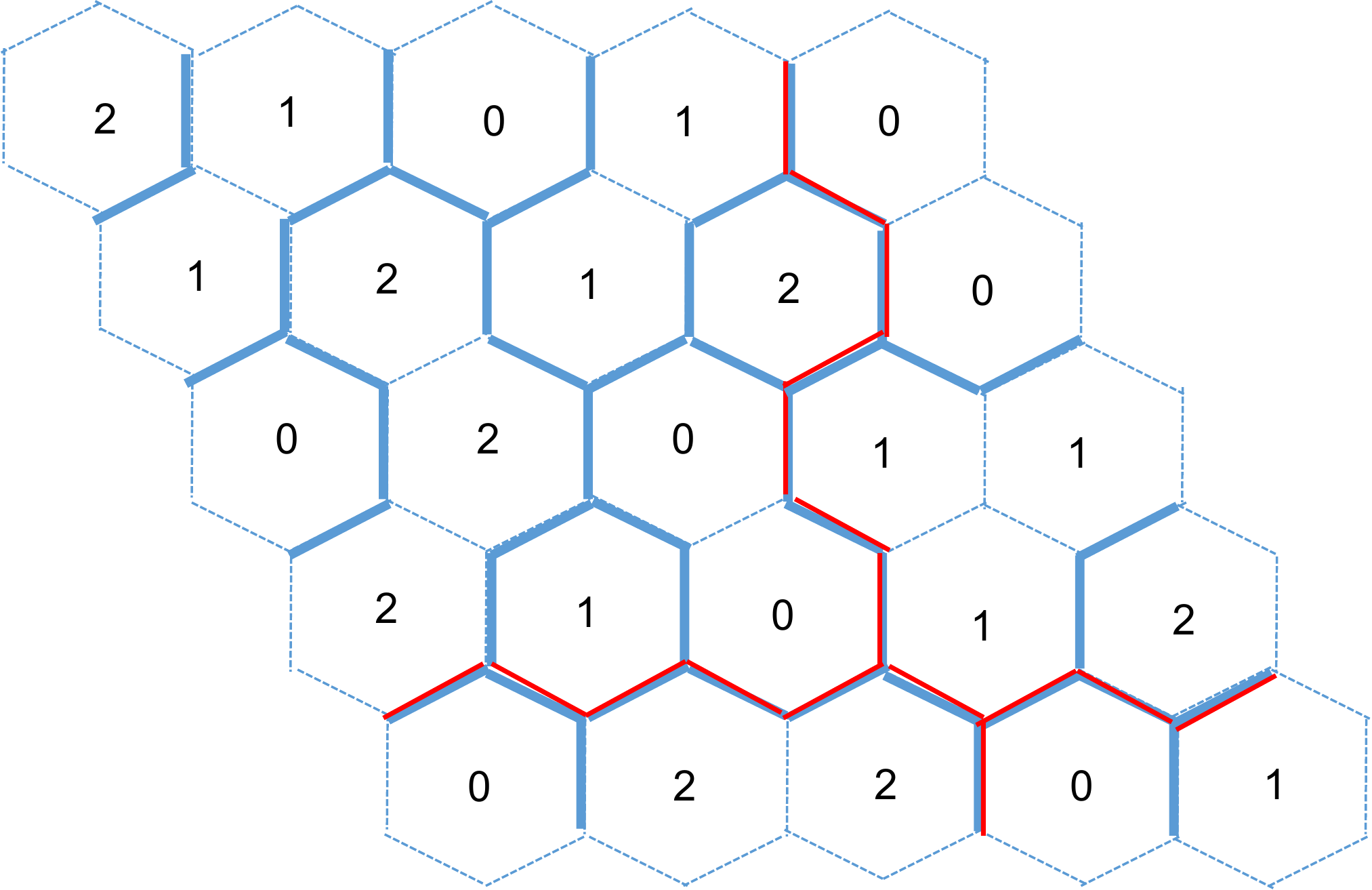}
		\caption{A `percolated' structure formed on a $5\times5$ honeycomb lattice, after obtaining random measured outcomes for $d=3$ on the plaquettes. Each filled edge is a $CZ$ or $CZ^2$ gate.}
		\label{simulation}
	\end{figure}

	We will first describe our simulation procedure, and then present the numerical results on the percolation properties of the random graphs resulting from local measurements on all domain spins. We demonstrate that the resulting graphs are in the percolated phase (\thatis supercritical phase of percolation). This is further confirmed by a stability analysis. Our simulation procedure is given as follows, and illustrated in Fig.~\ref{simulation}.
	\begin{enumerate}
		\item Generate random outcomes from $\{0, 1, ... , d-1\}$ for each plaquette.
		\item Determine whether the edge is filled from criterion $a-b\neq0$ mod $d$.
		\item Start filling the present edges; put them into existing trees if they are connected to previously filled edges. \cite{Newman2001}.
		\item Search for a path connecting left boundary  to right boundary, as well as one connecting top boundary to bottom boundary.
		\item Repeat the steps above and record the probability for existence of both paths in Step 4, which we refer to as the percolation probability.
	\end{enumerate}
	
\begin{figure}[t]
		\centering
		\includegraphics[width=0.45\textwidth]{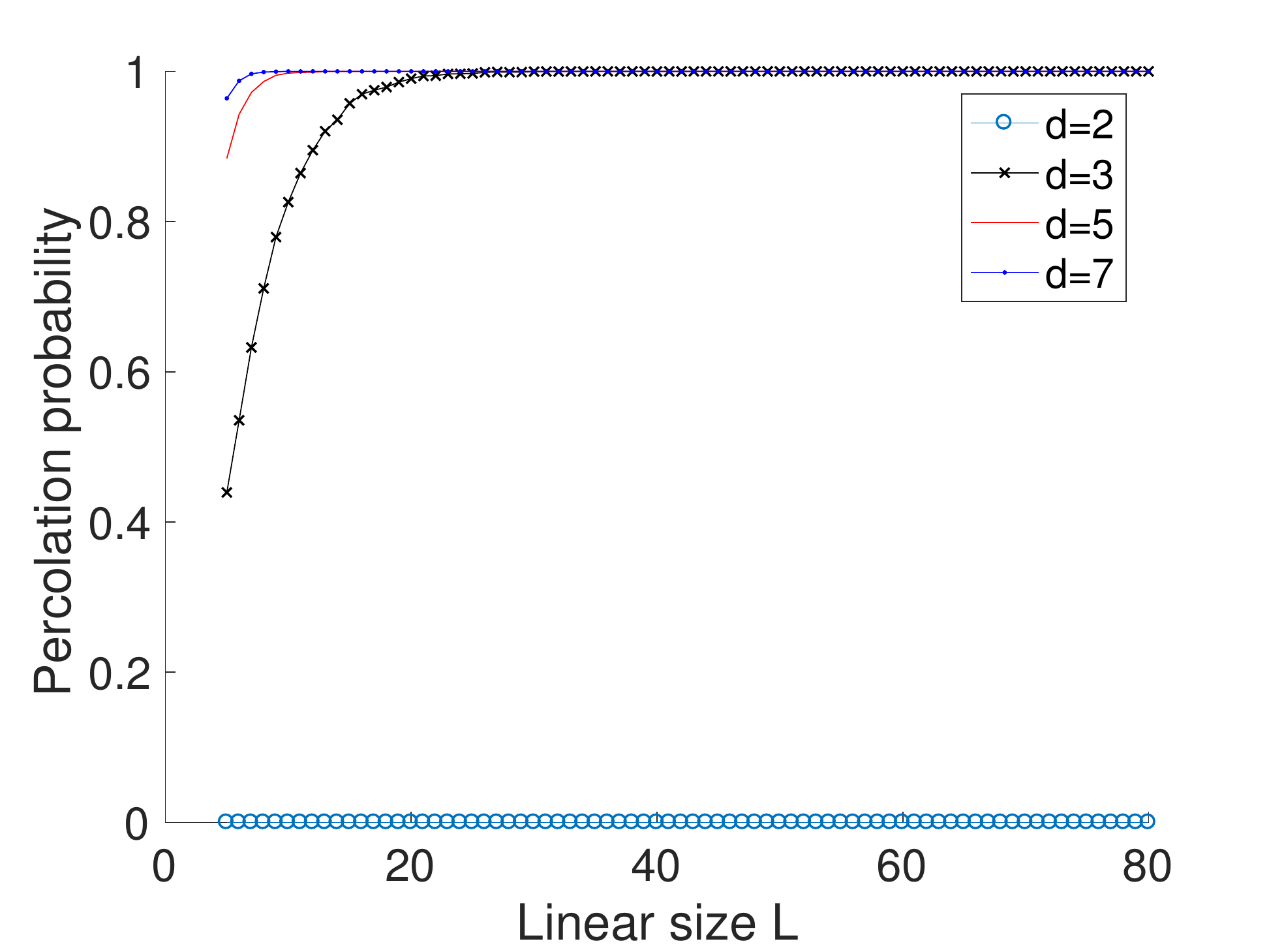}
		\caption{Percolation probability as a function of linear system size $L$ for the honeycomb lattice (as a result of measuring certain sites in the triangular lattice). The turquoise (circled), black (crossed), red (plain) and blue (dotted) lines represent different prime $d$. For each $L$ the probability is calculated from 10000 randomly generated measurement outcome patterns. The percolation probability seems to approach unity exponentially fast in $L$.}
		\label{numerical}
	\end{figure}

\begin{figure}[h]
		\begin{subfigure}{0.48\textwidth}
			\centering
			\includegraphics[height=50mm,width=75mm]{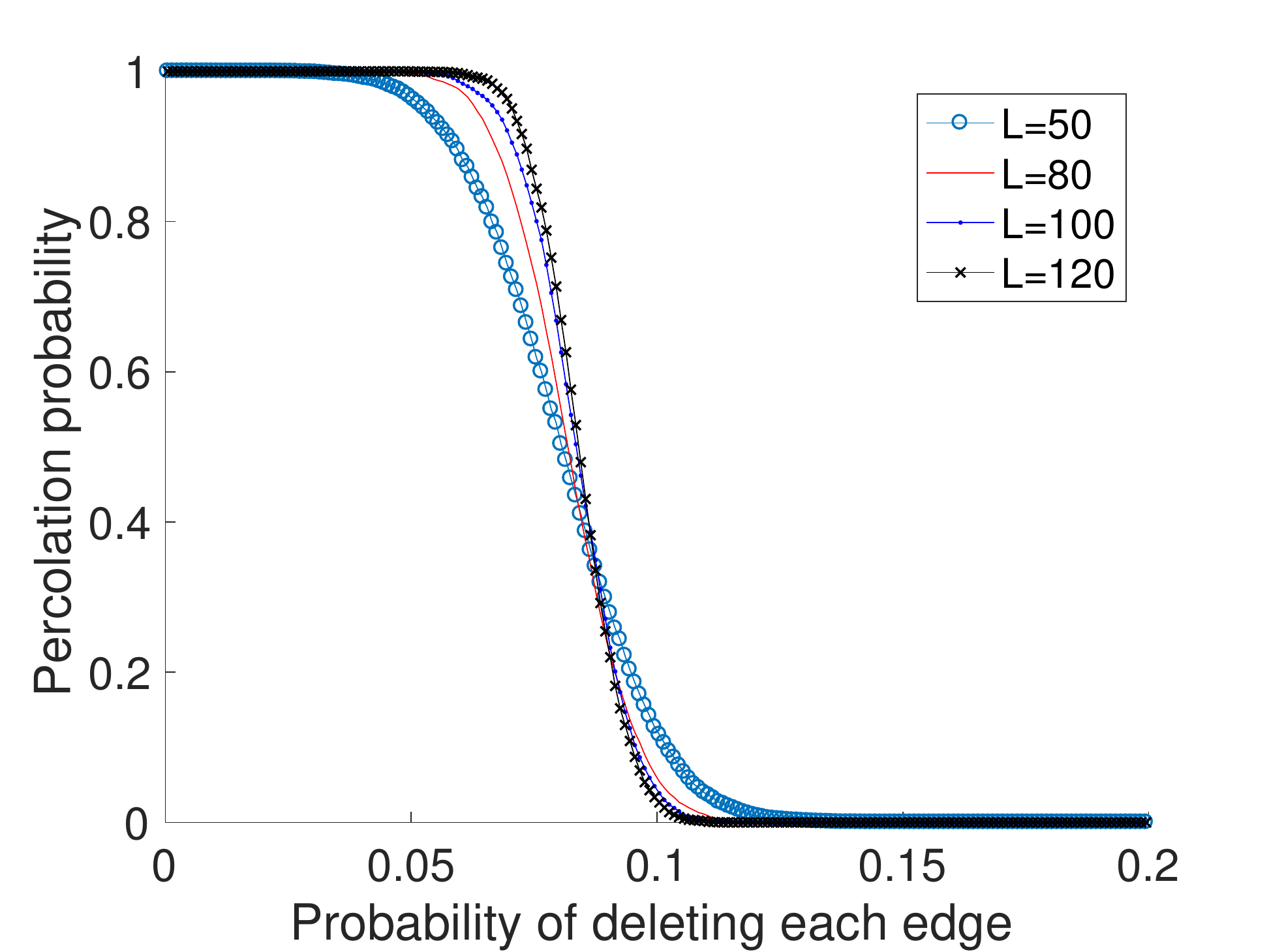}
		\end{subfigure}
		\vspace{1mm}
		\begin{subfigure}{0.48\textwidth}
			\centering
			\includegraphics[height=50mm,width=75mm]{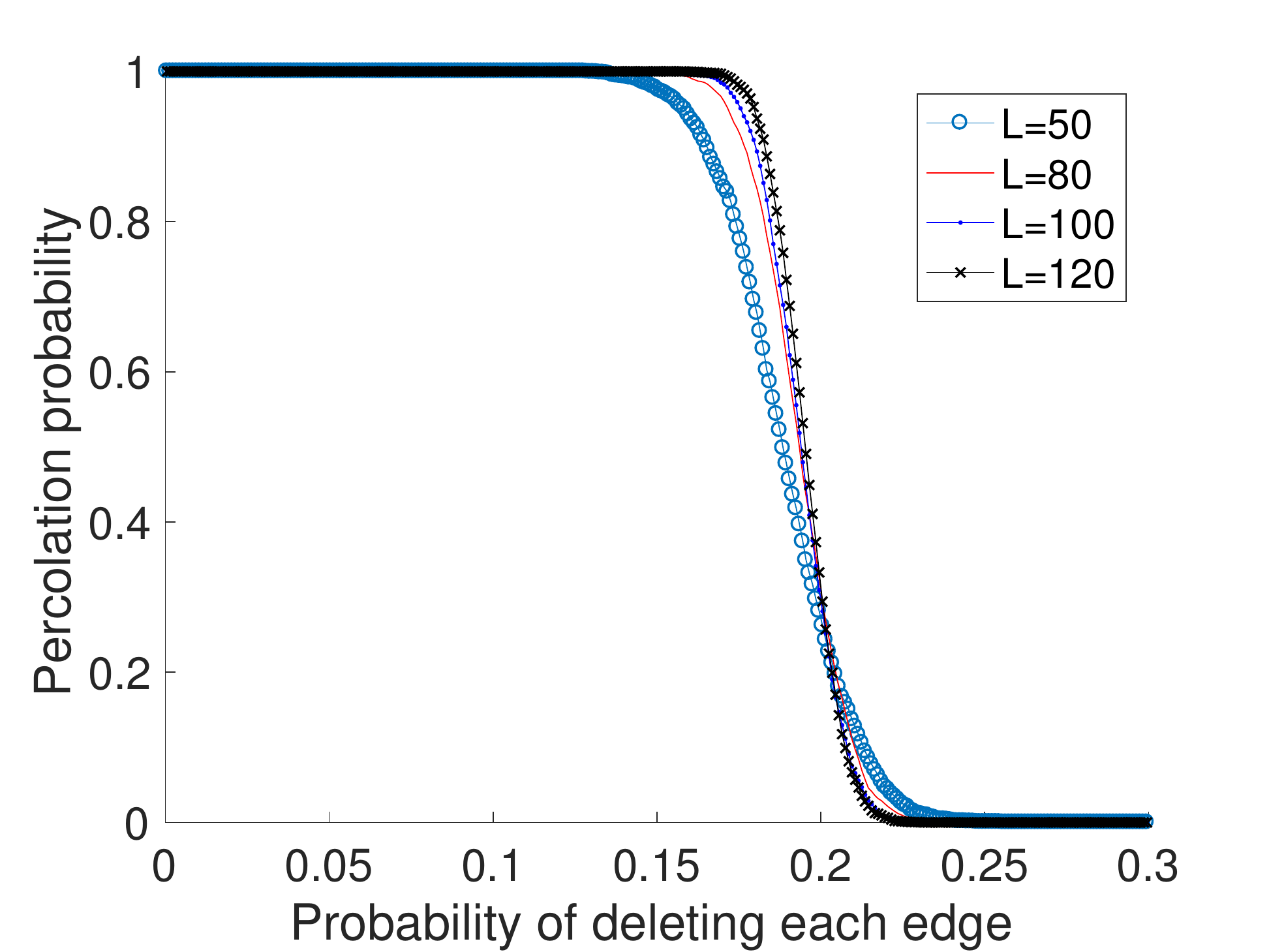}
		\end{subfigure}
		\caption{Percolation probability as a function of probability of deleting each edge. The upper and lower figures are the cases $d=3$ and $d=5$ respectively. On each figure different lines represent different linear system sizes $L$. We used the histogram method mentioned in Ref.~\cite{Newman2001}. For each data point we generated 50 measurement outcome patterns (random graphs) and for each pattern 50 instances of deleting occupied edges were averaged.}
		\label{fig:stab}
	\end{figure}

	In Fig.~\ref{numerical} we present our simulated percolation probability for four different values of $d$. It is clear that probability grows as system size increases, and approaches 1 faster for larger values of $d$. We restrict to prime $d$; the reason will become clear later when we study qudit cluster-like state. To see if this percolation is stable, we delete all the edges independently with a fixed probability $p$, and test the connectivity again. We found that the percolation probability drops from close to 1, to close to 0, at some finite deleting probability $p$, as shown in Fig.~\ref{fig:stab}. The transition becomes sharper as the system size grows. The existence of a phase transition~\cite{Browne2008} shows that our graphs are in the percolated phase, \thatis the supercritical phase of percolation.

	\subsection{Reduction of random graph-like states to cluster-like states} \label{sec:rules}

	\begin{figure}[h]
		\centering
		\begin{subfigure}{0.45\textwidth}
			\centering
			\includegraphics[height=15mm,width=85mm]{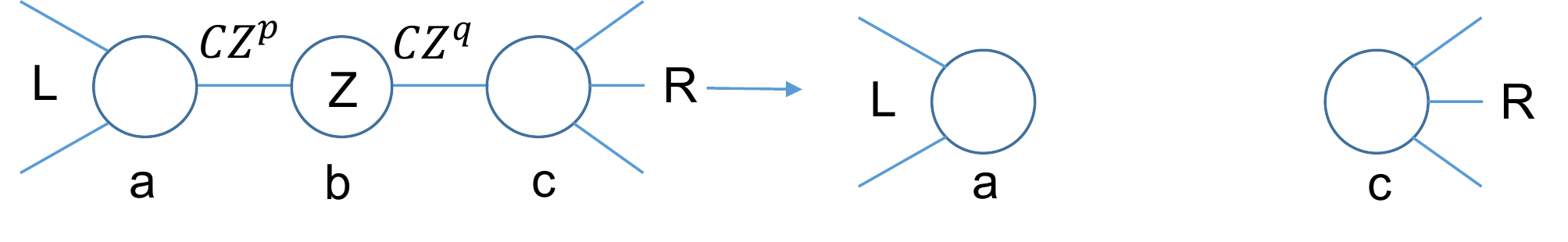}
			\caption{} \label{fig:rule1}
		\end{subfigure} \\
		\vspace{10mm}
		\begin{subfigure}{0.45\textwidth}
			\centering
			\includegraphics[height=15mm,width=70mm]{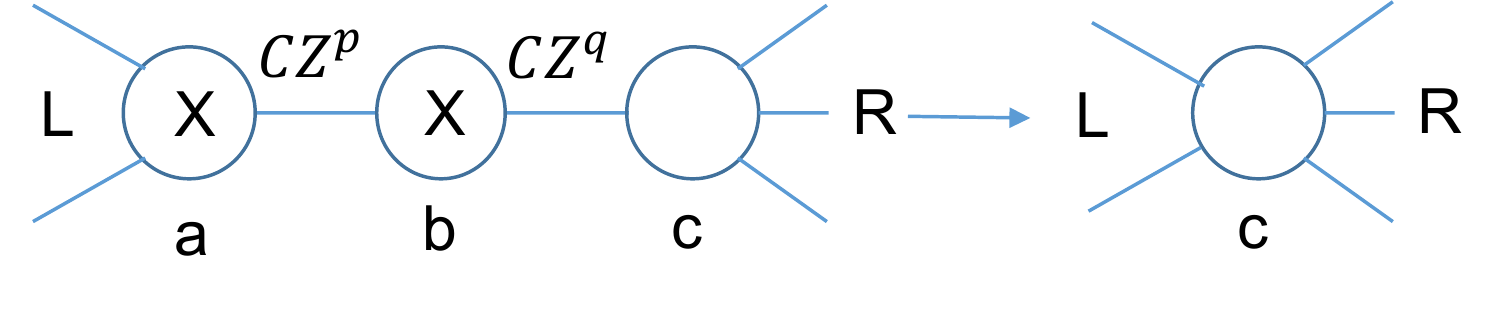}
			\caption{} \label{fig:rule2}
		\end{subfigure} \\
		\vspace{10mm}
		\begin{subfigure}{0.45\textwidth}
			\centering
			\includegraphics[height=18mm,width=80mm]{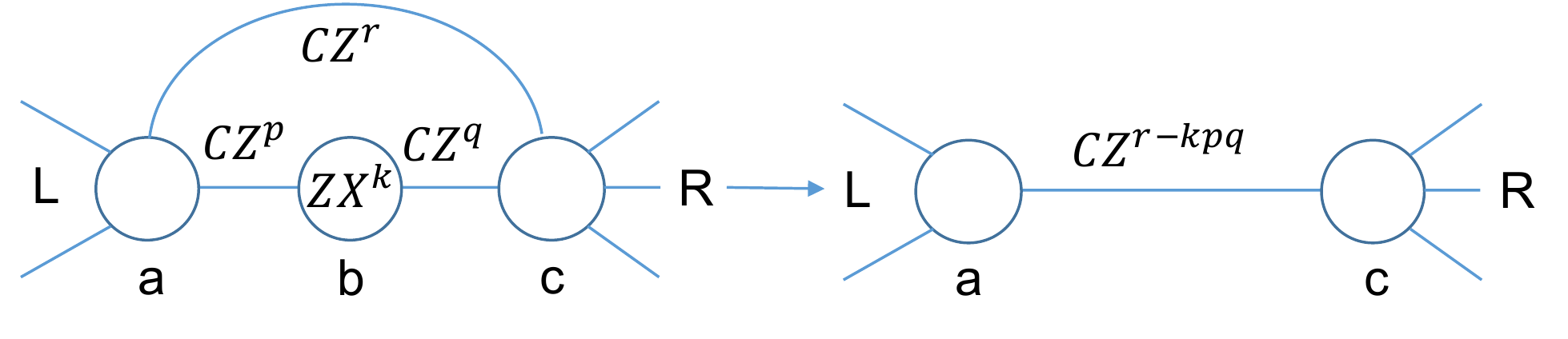}
			\caption{} \label{fig:rule3}
		\end{subfigure}
		\caption{Rules of manipulating the graph state by local measurements. The diagrams to the left of the arrow represent the states prior to measurements and the measuring pattern, and the corresponding outcomes are on the right.} \label{graph}
	\end{figure}
	In Refs.~\cite{Wei2011, Wei2012}, to show universality of random qubit graph states resulting from local generalized measurements on AKLT state, the crucial criterion hinges on the percolation property, and whether a graph is in the supercritical phase determines whether it can be converted to a coarse-grained cluster state (see also Ref.~\cite{Browne2008}). Although we have different random graph-like states, the same idea should apply provided we can establish similar rules for the qudit case. Let us first review the definitions for qudit graph state and its variation graph-like state. The qudit cluster state can be defined in a similar way as the qubit cluster state in Eq.~(\ref{cluster_state}), with definitions for $CZ$ gate and $\ket +$ given in Table \ref{defn}. If instead of on a regular lattice, the state is defined on a general graph with:
		\be
		|\phi\rangle_G = \prod_{\langle a, b\rangle} CZ_{(ab)} \underset{\mathrm{all \, sites} \, i}{\bigotimes} |+\rangle_i,
		\ee
	where $\langle a, b\rangle$ represents an edge in $G$ connecting vertices $a, b$, then it will be called a qudit graph state. Alternatively, it can also be defined through the stabilizer operator for all sites:
		\be
		X^\dagger_a \underset{b\in Nb(a)}{\bigotimes} Z_b |\phi\rangle_G = |\phi\rangle_G,
		\ee
	where $Nb(a)$ is the collection of neighbours of any site $a$. Note that in this expression $X^\dagger=X$ for qubit. The main difference between the qubit and qudit cases is that the operators $X$ and $Z$ are not Hermitian for qudit. Now we introduce the graph-like state given by:
		\be \label{eq:Gl}
		|\phi\rangle_{Gl} = \prod_{\langle a, b\rangle} CZ_{(ab)}^{r_{(a, b)}} \underset{\mathrm{all \, sites} \, i}{\bigotimes} |+\rangle_i,
		\ee
	where for each edge $\langle a, b\rangle$ in graph $G$ we assign an integer $r_{(a, b)} \in \{1, ..., d-1\}$. The corresponding stabilizer definition is, for any site $a$,
		\be
		X_a^\dagger \underset{b \in Nb(a)}{\bigotimes} Z_b^{r_{(a, b)}} |\phi\rangle_{Gl} = |\phi\rangle_{Gl},
		\label{stabilizer}
		\ee
	where sites $a$ and $b$ are connected with a $CZ^{r_{(a, b)}}$ gate. In some previous studies the states we call graph-like states were termed graph states~\cite{Bahramgiri2007, Looi2008, Keet2010, Helwig2013} but here we make a distinction as remarked earlier. If the underlying graph is a regular lattice, we call the state a qudit cluster-like state.
	
	\begin{figure}[h]
		\centering
		\begin{subfigure}{0.45\textwidth}
			\centering
			\includegraphics[height=50mm,width=90mm]{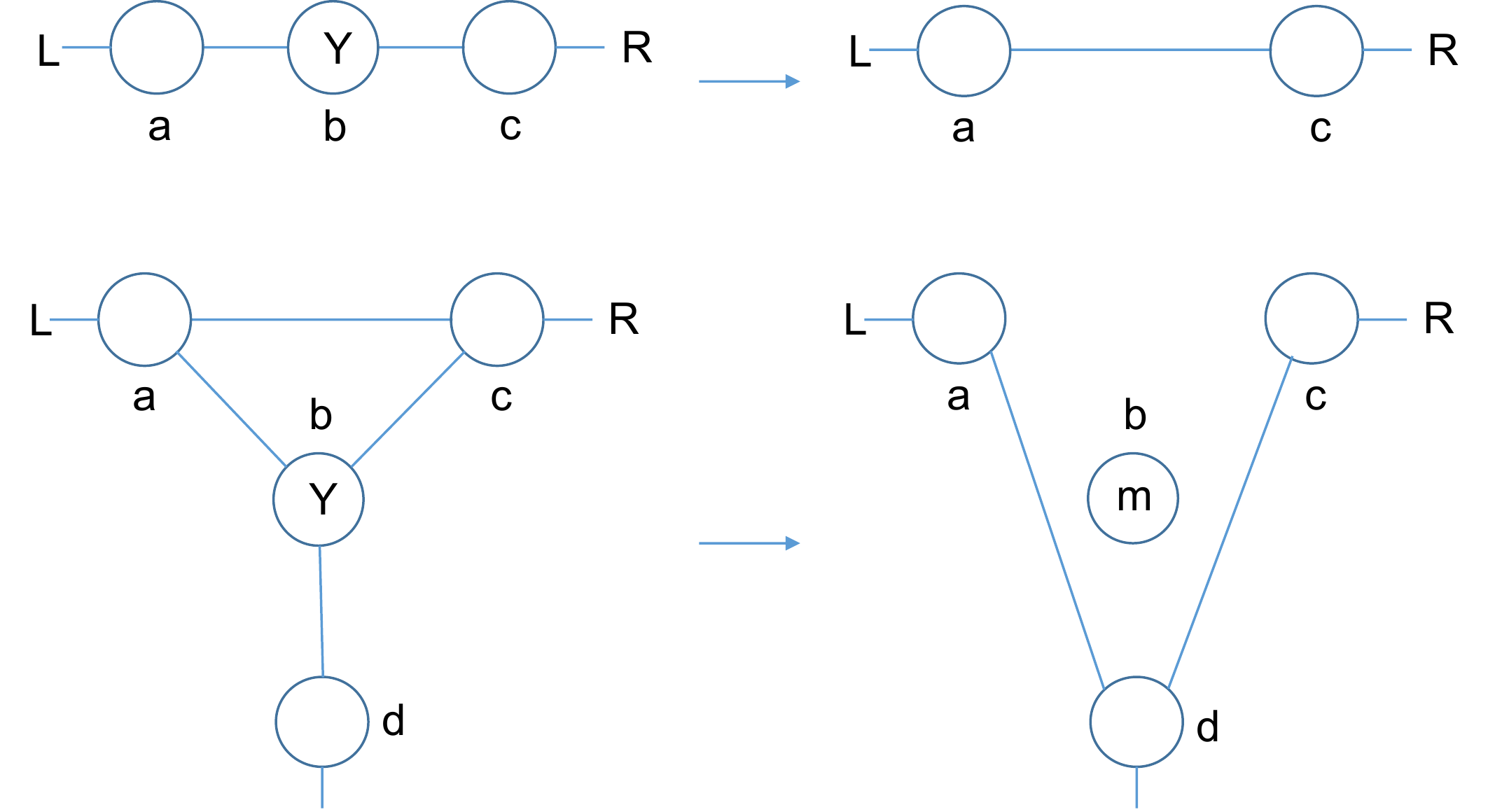}
			\caption{Qubit case.}
		\end{subfigure} \\
		\vspace{2mm}
		\begin{subfigure}{0.45\textwidth}
			\centering
			\includegraphics[height=50mm,width=90mm]{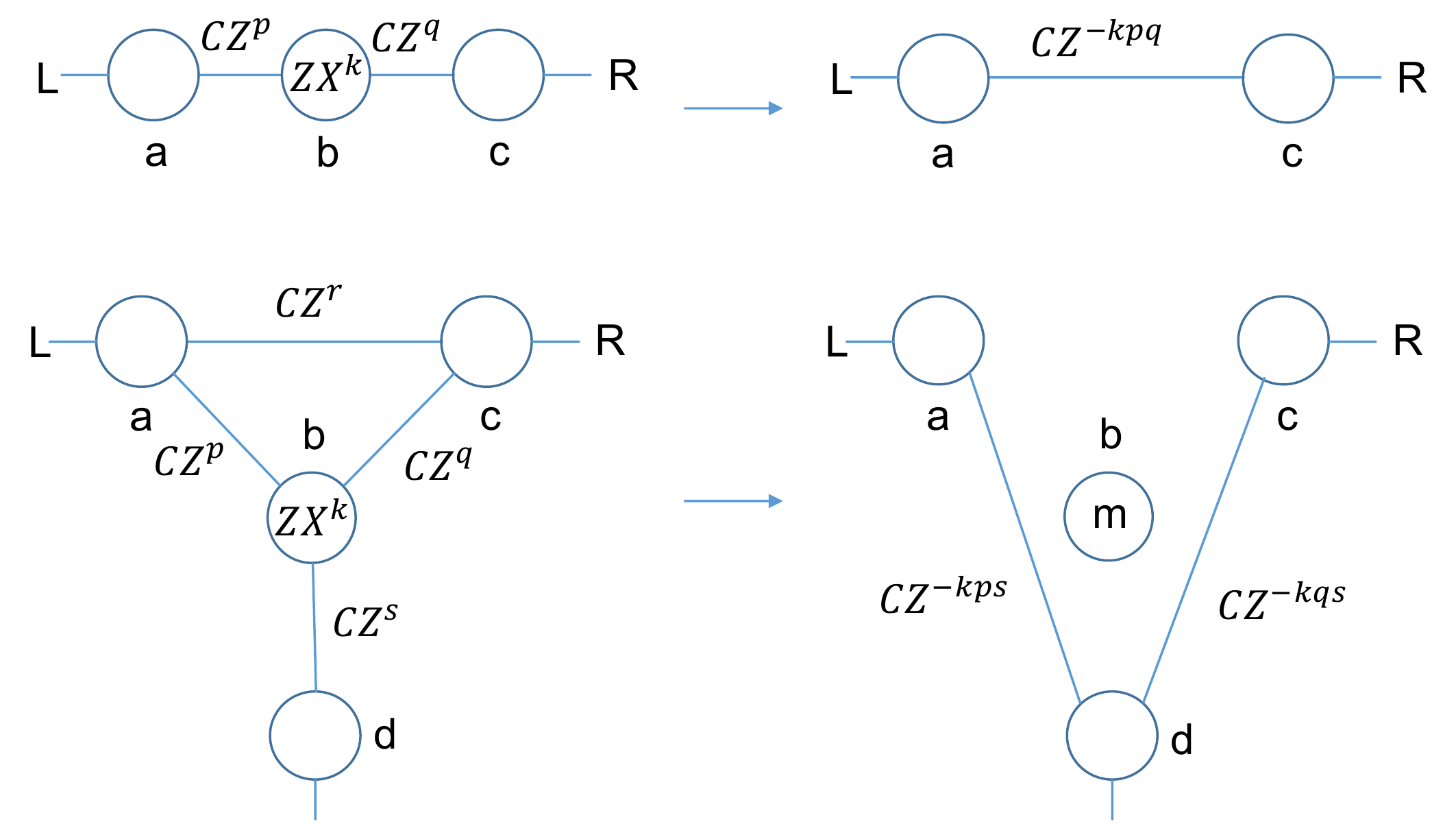}
			\caption{Qudit case.}
		\end{subfigure}
		\caption{Situations where $Y$ (in qubit case) and $ZX^k$ (in qudit case) measurements are needed.} \label{convert}
	\end{figure}
	
	For the reduction to work, we first present three rules of manipulating the graph-like state which are the qudit counterparts of the rules in Ref.~\cite{Wei2012}. By making measurements in some specific basis on qudits, we alter the form of the graph-like state we obtain from the first step. Note that the measurements will possibly introduce local unitary transformation on the qudits surrounding the measured ones. We are going to prove them explicitly for the cases shown in Fig.~\ref{graph}. General consideration can be found in Ref.~\cite{Bahramgiri2007} using the qudit stabilizer formalism. The state prior to any measurement is:
		\be
		|\phi\rangle_C = \sum_{j_a, j_b, j_c} |j_a\rangle_a |j_b\rangle_b |j_c\rangle_c |\phi_L(j_a)\rangle |\phi_R(j_c)\rangle \varpi^{pj_aj_b+qj_bj_c}, \label{prior_measure}
		\ee
	where we use $|\phi_L(j_a)\rangle$ and $|\phi_R(j_c)\rangle$ to represent the states to the left and right of the three qudits, and ignore the overall normalization of $|\phi\rangle_C$. If we measure qudit $b$ in $Z$ basis with outcome $m$, then we obtain, up to overall normalization,
		\be
		|\phi\rangle_C^{(1)} = \left( \sum_{j_a}\varpi^{mpj_a}|j_a\rangle_a|\phi_L(j_a)\rangle \right) \left( \sum_{j_c}\varpi^{mqj_c}|j_c\rangle_c|\phi_R(j_c)\rangle \right).
		\ee
	This is essentially removing qudit $b$ from the state. If we measure qudits $a$ and $b$ in $X$ basis instead, obtaining $m$ and $n$ respectively, then the result will be effectively merging the left and right:
		\begin{align}
		|\phi\rangle_C^{(2)} & = \frac{1}{d} \sum_{j_a, j_b, j_c, k, l} \langle k|j_a\rangle_a \langle l|j_b\rangle_b |j_c\rangle_c |\phi_L(j_a)\rangle |\phi_R(j_c)\rangle \times \nonumber \\
			& \varpi^{-mk-nl} \varpi^{pj_aj_b+qj_bj_c} \nonumber \\
			& = \sum_{j_a, j_c} \left( \sum_{j_b} \frac{1}{d} \varpi^{j_b(pj_a+qj_c-n)} \right) \times \nonumber \\
			& \varpi^{-mj_a} |j_c\rangle_c |\phi_L(j_a)\rangle |\phi_R(j_c)\rangle \nonumber \\
			& = \sum_{j_c} \varpi^{-mh(j_c)} |j_c\rangle_c |\phi_L(h(j_c))\rangle |\phi_R(j_c)\rangle,
		\end{align}
where $h(j_c) = p^{-1}(n-qj_c)$. Thus we prove the second rule (rule b).
	
	To see the third rule, where $a$ and $c$ are possibly entangled, instead of Eq.~(\ref{prior_measure}), the initial state is given by:
		\be
		|\phi\rangle_C = \sum_{j_a, j_b, j_c} |j_a\rangle_a |j_b\rangle_b |j_c\rangle_c |\phi_L(j_a)\rangle |\phi_R(j_c)\rangle \varpi^{pj_aj_b+qj_bj_c+rj_aj_c},
		\ee
	and we need to  examine the result of measuring qudit $b$ in $ZX^k$ basis for $k\neq0 \, \mathrm{mod} \, d$. We will not derive explicitly what the eigenstates of $ZX^k$ are, but use the result in Ref.~\cite{Hall2006}, which states that the eigenstate of $ZX^k$ with eigenvalue $\varpi^m$ is given by:
		\be
		|\psi_k^m\rangle = \frac{1}{\sqrt{d}} \sum_l \varpi^{\alpha_l} |l+m\rangle,
		\ee
where $\alpha$'s are integers defined by:
		\be
		\alpha_{k+l}+l=\alpha_l.
		\label{alpha}
		\ee
	Now suppose there is a gate $CZ^r$ between qudits $a$ and $c$, and $r$ can be 0 if there is actually no gate present. The state after measurement on qudit $b$ with outcome $m$ in $ZX^k$ basis is:
		\begin{align}
		|\phi\rangle_C^{(3)} & = \frac{1}{\sqrt{d}} \sum_{j_a, j_c, l} |j_a\rangle_a| \phi_L(j_a)\rangle |j_c\rangle_c|\phi_R(j_c)\rangle \times \nonumber \\
			& \varpi^{-\alpha_l+pj_a(l+m)+q(l+m)j_c+rj_aj_c}.
		\end{align}
	Let us define $R=pj_a+qj_c$, and $f(R)=\frac{1}{\sqrt{d}} \sum_{l} \varpi^{-\alpha_l+lR}$ then we have:
		\be
		\frac{1}{\sqrt{d}} \sum_l \varpi^{-\alpha_l+pj_a(l+m)+q(l+m)j_c} = \varpi^{mR} f(R).
		\ee
	We can see from Eq.~(\ref{alpha}) that $f(R)$ obeys the recursive relation:
		\begin{align}
		f(R) & = \sum_{l} \varpi^{-\alpha_{l+k}-l+lR} \nonumber \nonumber \\
			& = \varpi^{-k(R-1)} \sum_{l+k} \varpi^{-\alpha_{l+k}+(l+k)(R-1)} \nonumber \\
			& = \varpi^{-k(R-1)} f(R-1).
		\end{align}
	Therefore $f(R)=\varpi^{-\frac{1}{2}kR(R-1)}f(0)$ where $f(0)$ is a constant dependent on $k$. Rearranging the above expression, we arrive at:
		\begin{align}
		|\phi\rangle_C^{(3)} & = \sum_{j_a, j_c} |j_a\rangle_a| \phi_L(j_a)\rangle |j_c\rangle_c|\phi_R(j_c)\rangle \times \nonumber \\
			& \varpi^{rj_aj_c} \varpi^{mR-\frac{1}{2}kR(R-1)}f(0) \nonumber \\
			& = \sum_{j_a, j_c} |j_a\rangle_a| \phi_L(j_a)\rangle |j_c\rangle_c|\phi_R(j_c)\rangle \varpi^{(r-kpq)j_aj_c} \times \nonumber \\
			& \varpi^{(m+\frac{k}{2})pj_a-\frac{1}{2}kp^2j_a^2} \varpi^{(m+\frac{k}{2})qj_c-\frac{1}{2}kq^2j_c^2}.
		\end{align}
	This is effectively entangling qudits $a$ and $c$ by gate $CZ^{r-kpq}$, additionally with local unitary transformations $U_a$ on $a$ and $U_c$ on $c$ respectively, where
		\begin{align}
		U_a = \sum_{j_a} \varpi^{(m+\frac{k}{2})pj_a-\frac{1}{2}kp^2j_a^2} \ket{j_a} \bra{j_a} \\
		U_c = \sum_{j_c} \varpi^{(m+\frac{k}{2})qj_c-\frac{1}{2}kq^2j_c^2} \ket{j_c} \bra{j_c}.
		\end{align}
	If $r=0$ we can choose $k=-p^{-1}q^{-1}$ such that $a$ and $c$ are entangled by $CZ$, whereas if $r\neq0$ we choose $k=rp^{-1}q^{-1}$ such that $a$ and $c$ are disentangled. We summarize the three rules in Table \ref{tab:rules}.

	We will now examine if this type of measurement can be utilized to convert the random graph state to cluster-like state. First let us review the qubit situation where $Y$ measurement is needed. There are two such situations, which we illustrate in Fig.~\ref{convert}: (i) when we need to reduce number of qubits on a segment by one; (ii) when there is a junction as shown. In the qudit equivalent of the former case, we have $r=0$, so we can choose $k=-p^{-1}q^{-1}$ such that the initially unentangled neighbours of the measured qudit will become entangled. For the case of the qudit junction, we choose $k=rp^{-1}q^{-1}$, and disentangle $a$ and $c$. Since initially $a-d$ and $c-d$ are unentangled, the result is that they are now entangled by $CZ^{-kps}$ and $CZ^{-kqs}$, respectively. Therefore, with an appropriate choice of basis $ZX^k$ all the necessary converting rules in the qubit case also work in the qudit case. The reduction of qudit graph-like states (whose graphs are percolated) therefore follows the same procedure as that in Ref.~\cite{Wei2012}, which we summarize in Appendix~\ref{app:convert_procedure}.
	
	We would like to mention that more general graphic rules for qubit measurement in $X$ and $Y$ bases, are proved using a local unitary that gives local complementation~\cite{Hein2006}. This has been generalized to graphic rules for qudit Clifford group operations in Ref.~\cite{Bahramgiri2007}. The rules we proved above explicitly are special cases of such operations, but those are all we need for the reduction.

\begin{table*}[ht]
\begin{center}
	\begin{tabular}{ l | c | c | c }
	\hline
	\textbf{State prior to} & \multicolumn{2}{|c|}{$\sum_{j_a, j_b, j_c} |j_a\rangle_a |j_b\rangle_b |j_c\rangle_c |\phi_L(j_a)\rangle |\phi_R(j_c)\rangle$} & $|\phi\rangle_C = \sum_{j_a, j_b, j_c} |j_a\rangle_a |j_b\rangle_b |j_c\rangle_c |\phi_L(j_a)\rangle$ \\
	\textbf{measurement} & \multicolumn{2}{|c|}{$\times \varpi^{pj_aj_b+qj_bj_c}$} & $\times |\phi_R(j_c)\rangle \varpi^{pj_aj_b+qj_bj_c+rj_aj_c}$ \\
	\hline
	\textbf{Measured qudit} & $b$ & $a$ and $b$ & $b$ \\
	\hline
	\textbf{Measurement basis} & $Z$ & $X$ & $ZX^k$ \\
	\hline
	\textbf{State after} & $\left( \sum_{j_a}\varpi^{mpj_a}|j_a\rangle_a|\phi_L(j_a)\rangle \right)$ & $\sum_{j_c} |j_c\rangle_c |\phi_L(h(j_c))\rangle |\phi_R(j_c)\rangle$ & $\sum_{j_a, j_c} |j_a\rangle_a| \phi_L(j_a)\rangle |j_c\rangle_c|\phi_R(j_c)\rangle \varpi^{(r-kpq)j_aj_c}$ \\
	\textbf{measurement} & $\times \left( \sum_{j_c}\varpi^{mqj_c}|j_c\rangle_c|\phi_R(j_c)\rangle \right)$ & $\times \varpi^{-mh(j_c)}$ & $\times \varpi^{(m+\frac{k}{2})pj_a-\frac{1}{2}kp^2j_a^2} \varpi^{(m+\frac{k}{2})qj_c-\frac{1}{2}kq^2j_c^2}$ \\
	\hline
	\textbf{Graphical rule} & Fig.~\ref{fig:rule1} & Fig.~\ref{fig:rule2} & Fig.~\ref{fig:rule3} \\
	\hline
	\end{tabular}
	\caption{Rules of different qudit measurements needed for coarse-graining. \label{tab:rules}}
\end{center}
\end{table*}

	\subsection{Qudit cluster-like states are universal} \label{sec:cluster-like}

	The qudit triangular SPT state after local measurements on one third of qudits is randomly reduced to a qudit graph-like state. We have done simulations showing that for sufficiently large system size, any such random graph is, with probability 1, in the supercritical phase of percolation. Combining this and the qudit graphical rules of qudit Pauli measurement allows us to reduce the graph-like state to a qudit cluster-like state on a square lattice. This section focuses on the last step in our approach, that is to show the qudit cluster-like state is universal for MBQC. Brylinski and Brylinski~\cite{Brylinski2001} proved the theorem that to simulate all possible gates, one only needs to simulate all possible one-qudit gates and one imprimitive two-qudit gate. A two-qudit gate is defined as primitive if it maps any decomposable state to another decomposable state. It is equivalent to saying a primitive gate $V=S\otimes T$ or $V=(S\otimes T)P$ where $S$ and $T$ are one-qudit gates, and $P|xy\rangle=|yx\rangle$. Based on the aforementioned theorem, both qubit cluster state and qudit cluster state on the square lattice have been shown to be universal resource states~\cite{Raussendorf2003, Hall2006, Zhou2003}.

	However, the final state that we reduce to by local measurements from the qudit SPT state is not the qudit cluster state, but the qudit cluster-like state, where the $CZ$ edges in the cluster state are replaced by general $CZ^q$ edges, where $q$ can take different values at different edges. Is the modified state still a universal resource? Some special cases in one dimension have been studied by Wang et al.~\cite{Wang2017}. First let us see how teleportation can be realized on this state. Then we proceed with the stabilizer approach in Ref.~\cite{Raussendorf2003, Zhou2003}, and show that their gates can also be realized with modifications on this state, provided that $d$ is prime.

	We follow the procedure described by Hall~\cite{Hall2006} with slight modifications. Suppose we have two qudits entangled with each other by $CZ^q$ ($1\leq q \leq d-1$) (see Fig.~\ref{teleportation1}) and the first one is prepared as some input $|in\rangle = \sum_k a_k |k\rangle$.
		\be
		|\phi\rangle =  CZ^q |in\rangle_1 |+\rangle_2 = \sum_k a_k |k\rangle_1 |+_{qk}\rangle_2
		\ee
	Now it is useful to look at two other types of gates. One is modified Fourier gate:
		\be
		F_q = \sum_{k=0}^{d-1} |+_{qk}\rangle \langle k|, \, 1 \leq q \leq d-1,
		\ee
and the other is:
		\be
		S_q = \sum_k |qk\rangle \langle k|, \,  1 \leq q \leq d-1.
		\ee
	We see that for prime $d$, $S_q$ is a permutation of the computational basis for any $q$. When $d$ is not prime, however, $S_q$ projects to a smaller Hilbert space for some values of $q$. We also have $F_q = F S_q$. We measure the first qudit in the basis defined by $F_q^\dagger$, which is equivalent to applying $F_q$ gate and then measuring in computational basis.
		\begin{align}
		(F_q \otimes \mathds{1}) |\phi\rangle &= \sum_k a_k |+_{qk}\rangle_1 |+_{qk}\rangle_2 \nonumber \\
			&= \frac{1}{\sqrt{d}} \sum_{j, k} a_k \varpi^{qjk} |j\rangle_1 |+_{qk}\rangle_2 \nonumber \\
			&= \frac{1}{\sqrt{d}} \sum_j (\mathds{1} \otimes X^jF_q) |j\rangle_1 \sum_k a_k|k\rangle_2.
		\end{align}
	We thus observe that when $j$ is measured on the first qudit, the second qudit will be left in the state $X^jF_q |in\rangle$. If we wish to recover the input information, we need $d$ to be prime. In our discussion we would henceforth focus on prime $d$ only. Similar to the realization of identity gate in Ref.~\cite{Raussendorf2003}, an identity up to some byproduct operators can be constructed on a three-qudit cluster-like state. Assuming the two edges are $CZ^p$ and $CZ^q$ respectively,  as shown in Fig.~\ref{identity1}, applying the procedure first with basis defined by $F_p^\dagger$ (outcome $m$) and then with basis defined by $F_q^\dagger$ (outcome $n$) gives us:
		\begin{align}
		X^nF_qX^mF_p |in\rangle &= X^n Z^{\dagger qm} F_q F_p |in\rangle \nonumber \\
			&= X^n Z^{\dagger qm} S_c |in\rangle,
		\end{align}
	where $F_qF_p = \sum_k |j_k\rangle \langle k|$, and $j_k$ is uniquely (in the case of prime $d$) determined by $qj_k+pk=0$. In this case we can find an integer $c$ between 1 and $d-1$ such that $qc+p=0$, which means that $F_qF_p$ is a permutation of computational basis $S_c$. If we include the permutations as another type of byproduct operator, we have realized identity. Similar gate constructions were also discussed in Ref.~\cite{Wang2017}.

	Now we will show that the qudit cluster-like state is universal from stabilizer approach. Our state satisfies the eigenvalue equations Eq.~(\ref{stabilizer}). Our proof proceeds in a way similar to that for the qudit cluster state~\cite{Zhou2003}. First notice that our qudit cluster-like state state is:
		\be
		|\phi\rangle_C = S \underset{{\rm all \, sites} \, s}{\bigotimes}|+\rangle_s,
		\ee
	and $S$ is the product of all the entangling operations and is itself unitary:
		\be
		S = \prod_{(a, b) \in Edges} CZ_{ab}^{q_{(a, b)}}.
		\ee
	On any site $a$ we have $X^\dagger_a |+\rangle_a = |+\rangle_a$, and therefore:
		\be
		S X^\dagger_a S^\dagger |\phi\rangle_C = |\phi\rangle_C.
		\ee
	Making use of the following identities,
		\begin{align}
		& CZ_{ab}^q X_a^\dagger (CZ_{ab}^q)^\dagger = X_a^\dagger \otimes Z_b^q \\
		& CZ_{ab}^q X_b^\dagger (CZ_{ab}^q)^\dagger = Z_a^q \otimes X_b^\dagger \\
		& CZ_{ab}^q X_c^\dagger (CZ_{ab}^q)^\dagger = X_c^\dagger, \, c \neq a, b,
		\end{align}
	we obtain the result $S X^\dagger_a S^\dagger = X_a^\dagger \underset{b \in Nb(a)}{\bigotimes} Z_b^{q_{(a, b)}}$. Notice that in general an eigenvalue equation $O|\psi\rangle=\lambda|\psi\rangle$ does not imply $O^\dagger|\psi\rangle=\lambda^*|\psi\rangle$. However Eq.~(\ref{stabilizer}) still holds when the stabilizer is changed to its Hermitian conjugate, because we have $X_a |+\rangle_a = |+\rangle_a$ too.
	
	Now we describe the computation scheme adopted in Ref.~\cite{Raussendorf2003, Zhou2003}: to simulate a gate $g$, we first remove the unwanted qudits by measuring them in computational basis, and obtain the graph containing input section $C_I$, body section $C_M$, and output section $C_O$; input state is prepared on $C_I$; we then measure the $n$ qudits in $C_I$ in $X$ basis, and measure all the qudits in $C_M$ according to a measurement pattern $M_{C_M}$, denoting all the outcomes by $\{s\}$; the $n$ qudits in $C_O$ will be in the input state after gate operation $U(g)$, subject to some byproduct operators. If we define projector $P^{\{s\}}_M$ as the operator that projects the qudits onto the states denoted by $\{s\}$ in the basis specified by $M$, the previous measuring process is then equivalent to acting on all qudits in $C_I \cup C_M \cup C_O$ with projectors $P^{\{s\}}_{M_{C_I}} P^{\{s\}}_{M_{C_M}}$.
	
	Since we have replaced the $CZ$ gates with general $CZ^q$ gates, we need an altered version of Theorem 1 in Ref.~\cite{Raussendorf2003} and Theorem 2 in Ref.~\cite{Zhou2003}. Let us change the $X$ and $Z$ operators to some powers of themselves, add the new type of byproduct operator and prove the altered theorem.
	
	\textit{Theorem}. If the state $|\psi\rangle_{C(g)} = P^{\{s\}}_{M_{C_M}} |\phi\rangle_{C(g)}$ satisfies the following $2n$ eigenvalue equations:
		\begin{align}
		X^{p_i}_{C_I, i}(UX_i^{q_i}U^\dagger)_{C_O} |\psi\rangle_{C(g)} = \varpi^{-\lambda_{x, i}} |\psi\rangle_{C(g)}
		\label{thm1} \\
		Z^{\dagger q_ir_i}_{C_I, i}(UZ_i^{p_ir_i}U^\dagger)_{C_O} |\psi\rangle_{C(g)} = \varpi^{-\lambda_{z, i}} |\psi\rangle_{C(g)},
		\label{thm2}
		\end{align}
	where $\lambda_{x, i}, \lambda_{z, i} \in \zd$, $1 \leq p_i, q_i, r_i \leq d-1$ and $1 \leq i \leq n$, then the output state on $C_O$ will be:
		\be
		|\psi(\mathrm{out})\rangle = U U_\Sigma |\psi(\mathrm{in})\rangle,
		\ee
	where $U_\Sigma$ is given by:
		\be
		U_\Sigma = \overset{n}{\underset{i=1}{\bigotimes}} Z_i^{-(s_ip_i+\lambda_{x, i})q_i^{-1}}X_i^{\lambda_{z, i}p_i^{-1}r_i^{-1}}S_{q_ip_i^{-1}, i}.
		\label{byproduct}
		\ee
	Technically $U_\Sigma$ is not the byproduct operator because it is not in front of $U|\psi(\mathrm{in})\rangle$, but it is closely related. The proof closely follows that in Ref.~\cite{Raussendorf2003, Zhou2003} with slight changes, so we present it in Appendix \ref{proof}. This theorem can then be utilized in proving realization of specific gates, which include a universal set of gates. In Ref.~\cite{Zhou2003} the authors give a collection of gates from which all one-qudit gates can be realized. We claim that all these gates, as well as an imprimitive two-qudit gate, can be simulated on our cluster-like state in a similar way and prove this in Appendix \ref{gates_realization}. With this we have completed the proof that our qudit triangular SPT states are universal for MBQC if $d$ is a prime that is greater than 2.

	\begin{figure}[t!]
		\centering
		\includegraphics[width=0.45\textwidth]{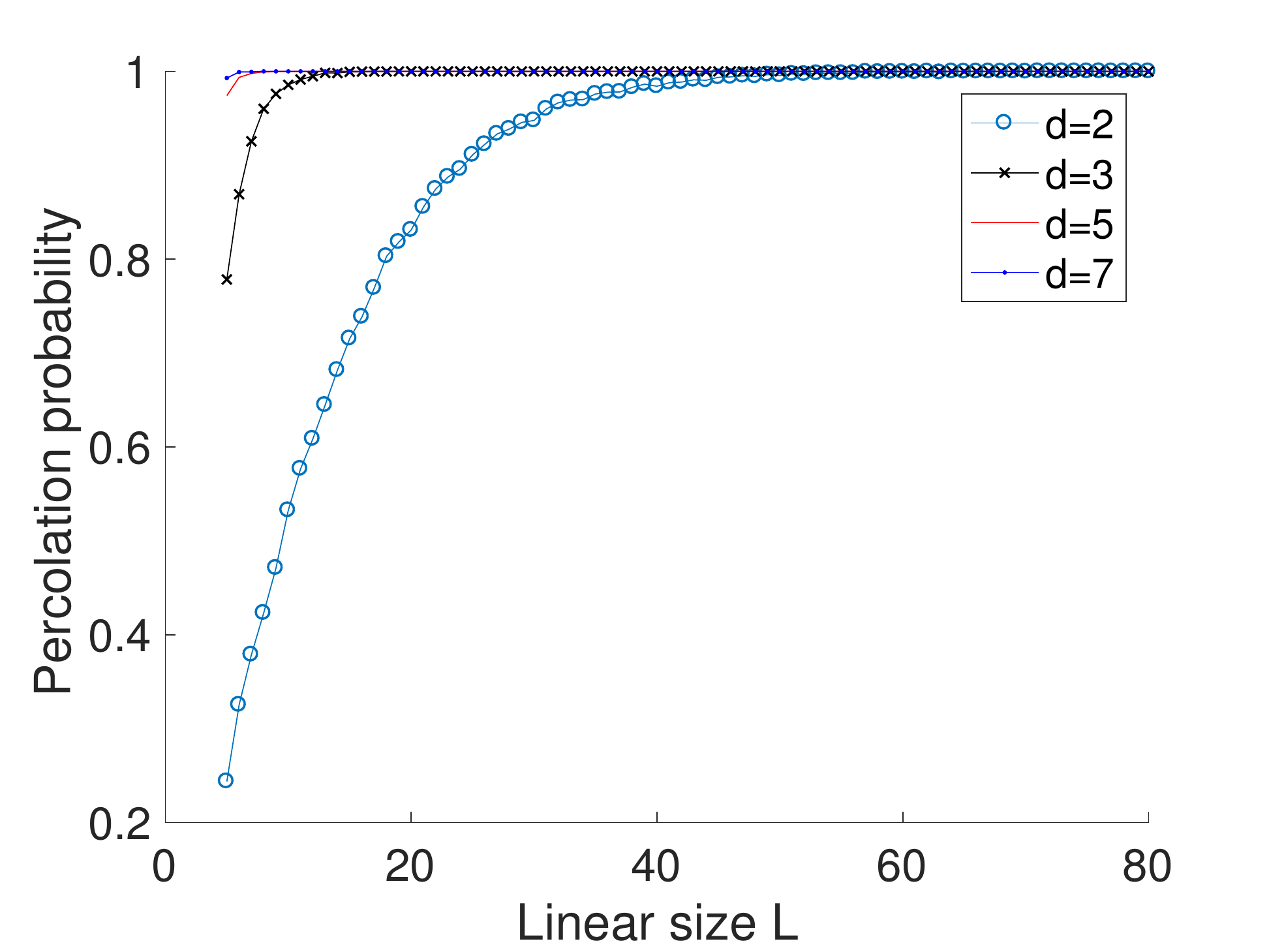}
		\caption{Percolation probability as a function of linear system size for the square lattice (as a result of measuring certain sites in the Union-Jack lattice). The turquoise (circled), black (crossed), red (plain) and blue (dotted) lines represent different prime $d$.}
		\label{numericalUJ}
	\end{figure}
	
 \section{Discussion} \label{sec:discussion}
We have considered qudit SPT states that are protected by $\zd\times \zd \times \zd$ symmetry and shown that they are universal for prime $d>3$ on the triangular lattice. Our approach extends to other 3-colorable lattices, including the Union-Jack lattice, on which the qubit SPT state by Miller and Miyake is universal. We have thus performed analysis and simulations on the qudit Union-Jack SPT states and find that they are also universal, as expected. The simulation results for the percolation study are shown in Fig.~\ref{numericalUJ}. 

We summarize that our approach in establishing the quantum computational universality for these qudit SPT states consists of a few steps: (i) reduction, via local measurements, of qudit SPT states to qudit graph-like states; (ii) showing the resultant graphs are in supercritical phase of percolation; (iii) reduction of these random graph-like states to qudit cluster-like states; and (iv) showing that qudit cluster-like states are universal. Step (i) is consistent with the so-called decorated domain-wall picture, as the graph-like states are characterized by  nontrivial $\zd \times \zd$ order (in terms of its second cohomology group). Step (ii) turns the problem of deciding the universality to the problem of percolation, but it relies on steps (iii) and (iv) to hold. Step (iii) has been previously studied in the context of qubit graph states, and we have generalized and shown it to hold in the qudit case as well. Step  (iv) was studied previously in qubit and qudit graph states, but not for qudit graph-like states, where each edge of the graph represents action of CZ gate to some power (which is edge dependent). We show that the universality also holds for qudit cluster-like states. 

We note that the results regarding qudit graph-like and cluster-like states are only proved for prime $d$, which may only be  an issue in technicality and in particular, we have required the operator $S_c$ be full-rank for $c \in \zd$. In Ref.~\cite{Zhou2003} Zhou et al. showed that qudit cluster state is a MBQC universal resource for general $d$. It is thus interesting to extend our results to non-prime $d$, by establishing that these qudit-like cluster states are universal.

In Ref.~\cite{Miller2016}  Miller and Miyake showed that the qubit Union-Jack state is universal even if one restricts the measurements to be in Pauli bases. It would be interesting to construct gates explicitly on the qudit triangular SPT states to see if whether they are also qudit Pauli universal. This may also shed light on the universality of the non-prime $d$ case. A more challenging question is to find some way to  extend these particular states to some family of states~\cite{Miller2016_2},  or even some entire  2D phases, as  this has been achieved in one-dimensional SPT phases~\cite{Stephen17,Raussendorf17}.

\section*{Acknowledgement}
 The authors acknowledge useful discussions with Akimasa Miyake. This work was supported by the National Science Foundation under Grant No. PHY 1620252.

\bibliographystyle{unsrt}	
\bibliography{mybib}

\appendix

\section{SPT properties} \label{app:SPT}

\begin{figure}[h]
	\centering
	\includegraphics[height=20mm,width=50mm]{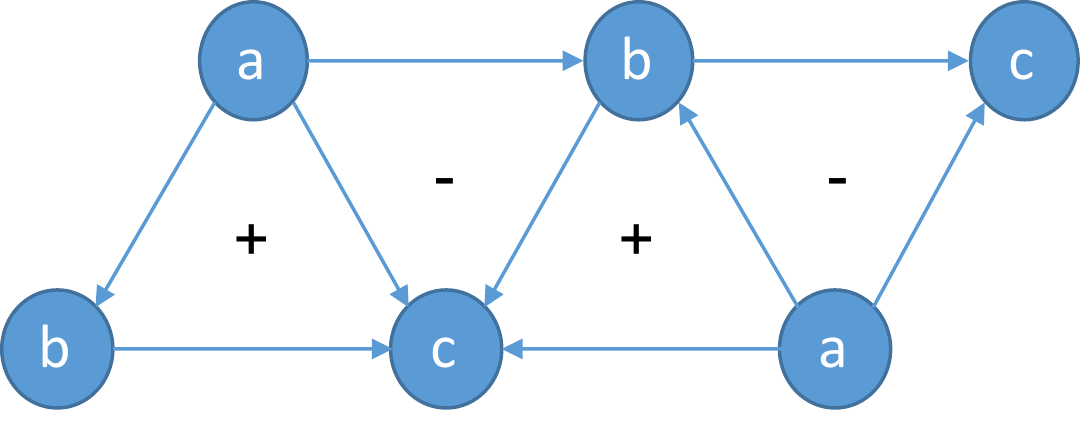}
	\caption{Branching structure. Upward triangles have a counter-clockwise orientation and hence have $s=1$. Any triangle is next to three triangles of the opposite orientation.} \label{branching}
\end{figure}

In this section we will see that the triangular state is a non-trivial SPT state. A standard procedure of constructing SPT states given by Ref.~\cite{Chen2013, Yoshida2016, Miller2016_2} is: 
	\be
	|\phi\rangle_{SPT} = \left( \prod_{\Delta_i} U(\nu_D)^{s({\Delta_i})} \right) \underset{\mathrm{site} \, j}{\bigotimes}\sum_{g \in G} |g\rangle_j,
	\ee
where we ignore some overall normalization. Here $\Delta_i$ is the $i$-th $(D-1)$-simplex, and $s({\Delta_i})$ is defined such that $s({\Delta})=-s({\Delta^\prime})$ for $\Delta$ and $\Delta^\prime$ sharing one $(D-2)$-simplex. The symmetry group is $G$, and states on each site are labelled by group elements $g \in G$. This gives a representation of $G$: action of a group element $h$ on state $|g\rangle$ is $U_h |g\rangle = |hg\rangle$. $U(\nu_D)$ is a gate formed by $D$-cocycle acting on $D$ qudits:
	\be
	U(\nu_D) = \sum_{(g_1, ... , g_D) \in G^D} \nu_D(1, g_1, ... g_D) |g_1, ... , g_D\rangle \langle g_1, ... , g_D|, 
	\ee
where $\nu_D(1, g_1, ... g_D)$ is a $D$-cocycle of $G$: $\nu_D(1, g_1, ... g_D) \in \mathcal{Z}^D(G, U(1))$. Starting from symmetry $G=\zd \times \zd \times \zd$, we will construct a non-trivial SPT state $|\phi\rangle$. First we have the triangular lattice, and define $s$ to be +1 for upward triangles and -1 for downward triangles. More generally $s$ is determined by choosing the branching rules on the triangles \cite{Chen2013}. If we assign an arrow to each edge of the triangles such that no complete loop is formed, then we have a branching structure. Next we choose values of $s$ for different triangle orientations given by the arrows. We label three sublattices $a$, $b$ and $c$ such that on each triangle we have $(a, b, c)$. We fix the rule that in any triangle, the vertex labelled $a$ will be attached to two outgoing edges; the vertex labelled $b$ will have one incoming edge and one outgoing edge; the vertex labelled by $c$ will have both edges incoming. As in Fig.~\ref{branching}, we have two orientations, either one surrounded by the other type. We can define the counter-clockwise/clockwise orientation to have phase $+1/-1$, which is the same rule as $+1/-1$ for upward/downward triangles.
	
One group element in $G$ can be labelled by $(g^{(1)}, g^{(2)}, g^{(3)})$ where $g^{(n)} \in \zd$, which can then label the state on a site. Note that the Hilbert space on each site is enlarged from the physical $d$-dimensional to $d^3$-dimensional. Later we will show that our state with product ancillas and this state in the enlarged Hilbert space are in the same phase. Now we use a specific 3-cocycle:
	\begin{align}
	\nu_3 (1, g_a, g_b, g_c) = \omega_3 (g_a, g_a^{-1}g_b, g_b^{-1}g_c) \\
	\omega_3 (g_1, g_2, g_3) = \varpi^{k g_1^{(1)} g_2^{(2)} g_3^{(3)}},
	\end{align}
where $k$ takes value from 0 to $d-1$, the case $k=0$ being trivial.

\begin{figure}[h]
	\centering
	\includegraphics[height=30mm,width=50mm]{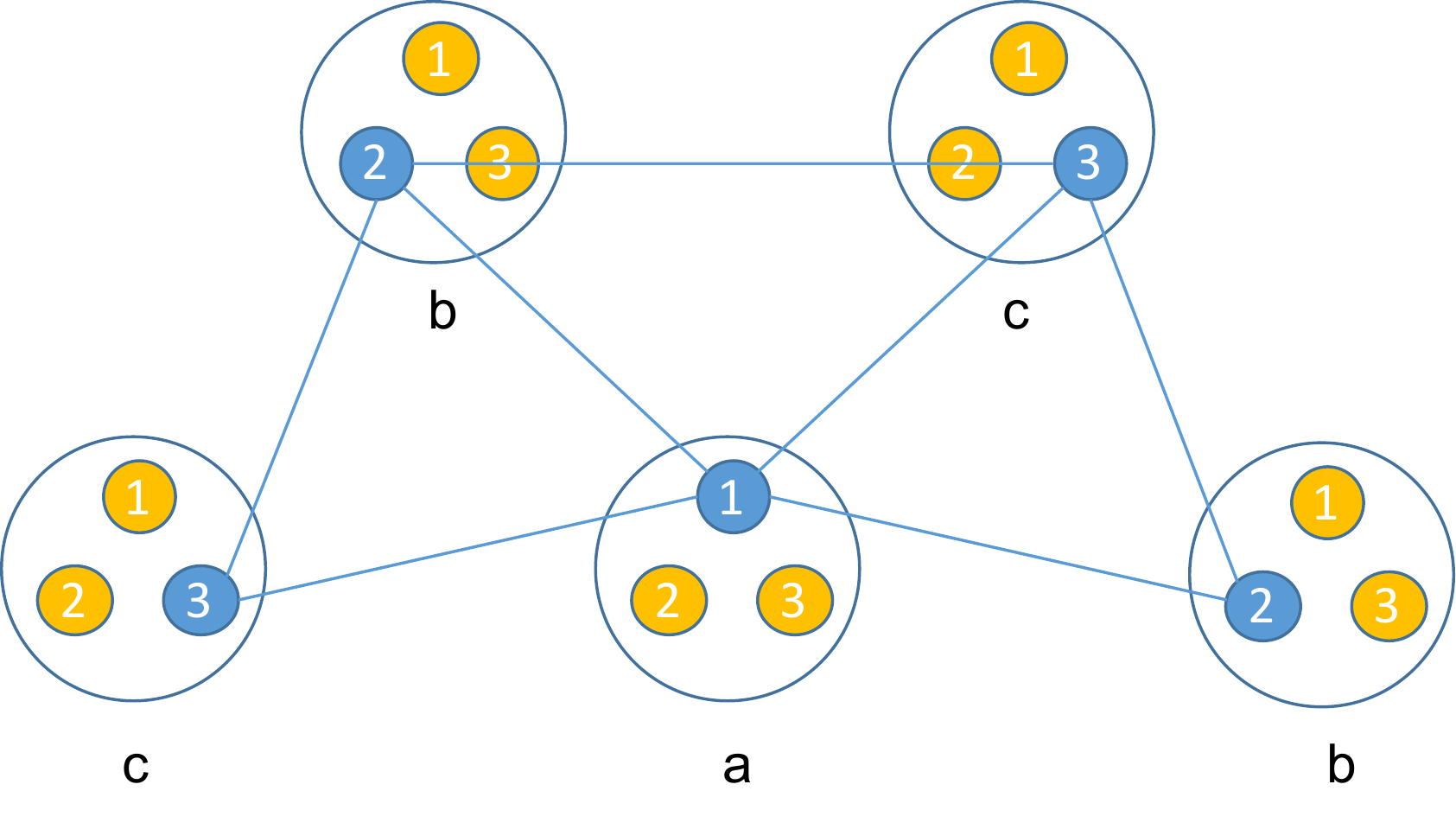}
	\caption{Hilbert space of one qudit enlarged to contain three qudits. The triangular SPT state is defined on the blue ones (connected by lines), whereas the yellow qudits are in the $\ket+$ state.} \label{fig:layer}
\end{figure}

The constructed state is:
	\be
	|\phi_k\rangle = \left( \prod_{\mathrm{triangle} \, i} \hat{\varpi}^{s_i k g_{a_i}^{(1)} (g_{b_i}-g_{a_i})^{(2)} (g_{c_i}-g_{b_i})^{(3)}} \right) \underset{\mathrm{site} \, j}{\bigotimes} \sum_{g_j \in G} \ket{g_j}_j,
	\ee
where $s_i=\pm1$ for upward/downward triangles, and $a_i, b_i, c_i$ are vertices of triangle $i$. Let us examine the phase given by an upward triangle: 
	\begin{align}
	\vartriangle & = \hat{\varpi}^{k g_a^{(1)} (g_b-g_a)^{(2)} (g_c-g_b)^{(3)}} \nonumber \\
		& = \hat{\varpi}^{k g_a^{(1)} g_b^{(2)} g_c^{(3)}} \hat{\varpi}^{k [-g_a^{(1)} g_b^{(2)} g_b^{(3)} -g_a^{(1)} g_a^{(2)} g_c^{(3)} +g_a^{(1)} g_a^{(2)} g_b^{(3)}]}.
	\end{align}
By placing $\hat{\,}$ on $\varpi$ we emphasize that it is an operator, for which the group element $g$'s act diagonally on $\ket{g_j}$, inducing a phase factor. Similarly, on an adjacent downward triangle that shares vertices $a, b$, the term is:
	\begin{align}
	\triangledown & = \hat{\varpi}^{-k g_a^{(1)} (g_b-g_a)^{(2)} (g_{c^\prime}-g_b)^{(3)}} \nonumber \\
		& = \hat{\varpi}^{-k g_a^{(1)} g_b^{(2)} g_{c^\prime}^{(3)}} \hat{\varpi}^{-k [-g_a^{(1)} g_b^{(2)} g_b^{(3)} -g_a^{(1)} g_a^{(2)} g_{c^\prime}^{(3)} +g_a^{(1)} g_a^{(2)} g_b^{(3)}]}.
	\end{align}
When multiplying together, the terms $\pm k (-g_a^{(1)} g_b^{(2)} g_b^{(3)}+g_a^{(1)} g_a^{(2)} g_b^{(3)})$ cancel. In the same way $-kg_a^{(1)} g_a^{(2)} g_c^{(3)}$ cancels with the term from a downward triangle sharing vertices $a, c$. We can do the same for the phase given by a downward triangle. Therefore, the state we get is:
	\be
	|\phi_k\rangle = \left( \prod_{\mathrm{triangle} \, i} \hat{\varpi}^{s_i k g_{a_i}^{(1)} g_{b_i}^{(2)} g_{c_i}^{(3)}} \right) \underset{\mathrm{site} \, j}{\bigotimes} \sum_{g_j \in G} \ket{g_j}_j.
	\ee
If we interpret the state $|g_j\rangle$ on each site as formed by three qudits, as in Fig.~\ref{fig:layer}):
	\be
	|g_j\rangle_j \equiv |g_j^{(1)}, g_j^{(2)}, g_j^{(3)}\rangle_j = |g^{(1)}\rangle_{j, 1} |g^{(2)}\rangle_{j, 2} |g^{(3)}\rangle_{j, 3},
	\ee
then we see that the constructed state can be written as:
	\begin{align}
	\ket{\phi_k} & = \left( \prod_{\mathrm{triangle} \, i}  \hat{\varpi}^{s_i k g_{a_i}^{(1)} g_{b_i}^{(2)} g_{c_i}^{(3)}} \right) \left( \underset{\mathrm{site} \, j \in a}{\bigotimes} \sum_{g_j^{(1)} \in \zd} \ket{g_j^{(1)}}_{j, 1} \right) \nonumber \\
		& \times \left( \underset{\mathrm{site} \, k \in b}{\bigotimes} \sum_{g_k^{(2)} \in \zd} \ket{g_k^{(2)}}_{k, 2} \right) \left( \underset{\mathrm{site} \, l \in c}{\bigotimes} \sum_{g_l^{(3)} \in \zd} \ket{g_l^{(3)}}_{l, 3} \right) \nonumber \\
		& \times \ket{+...+} \nonumber \\
		& = \ket{\phi_k}_T \ket{+...+},
	\end{align}
where we use $j \in a$ to denote site $j$ labelled by $a$. This state has been considered and generalized to higher dimensions in Ref.~\cite{Yoshida2016}. Here the triangular SPT state $|\phi\rangle_T$ (as in Eq.~(\ref{eq:state_defn})) is defined on the the collection of qudits marked `$(1)$' on sites labelled $a$, marked `$(2)$' on sites labelled $b$ and marked `$(3)$' on sites labelled $c$. All the other qudits are in the state:
	\begin{align}
	\ket{+...+} & \equiv \left( \underset{\mathrm{site} \, j \in a}{\bigotimes} \ket+_{j, 2} \ket+_{j, 3} \right) \left( \underset{\mathrm{site} \, k \in b}{\bigotimes} \ket+_{k, 1} \ket+_{k, 3} \right) \nonumber \\
		& \times \left( \underset{\mathrm{site} \, l \in c}{\bigotimes} \ket+_{l, 1} \ket+_{l, 2} \right).
	\end{align}
$\ket\phi$ is in a non-trivial topological phase protected by symmetry $\zd \times \zd \times \zd$. Here elements in each factor $\zd$ act on all sites simultaneously. Since $\ket{+...+}$ is a trivial product state, our state $|\phi\rangle_T$ and the constructed state $|\phi\rangle$ fall in the same phase. To give a physical understanding of this construction, we discuss the decorated domain-wall picture in the next section.

\section{Decorated domain-wall SPTs} \label{app:DDW}
In this section we review  the decorated domain-wall (DDW) construction of SPT phases~\cite{Chen2014} and then use this to demonstrate that our states have nontrivial SPT order.

\subsection{Overview}
Consider a $d$ dimensional bosonic system with a global symmetry given by a unitary on-site representation of a group $G$. SPT phases classified by group cohomology are in one-to-one correspondence with the elements of the cohomology group $H^{d+1}(G,U(1))$. Let us now focus on the case when the global symmetry can be written as a direct-product form of two groups:
	\begin{equation}
	G \cong Q \times K.
	\end{equation}
The SPT phases for this symmetry group are still labeled by the elements of the cohomology group
	\begin{equation} \label{eq:hgc}
	H^{d+1}(Q \times K,U(1)).
	\end{equation}
We can now use the K\"{u}nneth formula for group cohomology~\cite{Chen2014} to expand Eq.~(\ref{eq:hgc}) as
	\begin{eqnarray}
	H^{d+1}(Q \times K,U(1)) &\cong& H^{d+1}(Q , H^0(K,U(1))) \nonumber \\ &\times& H^{d}(Q , H^1(K,U(1))) \nonumber \\ &\vdots& \nonumber \\
	&\times&  H^{1}(Q , H^d(K,U(1))) \nonumber \\ &\times&  H^{0}(Q , H^{d+1}(K,U(1)))
	\end{eqnarray}
In other words, the SPT phases labeled by the elements of the group Eq.~(\ref{eq:hgc}) and hence SPT phases protected by $G$ can be written in terms of a collection of $d+2$ group elements of the  smaller groups
	\begin{eqnarray}
	\alpha &\in& H^{d+1}(Q \times K,U(1)) \nonumber \\
	&\equiv& \{\alpha_0, \alpha_1,\ldots,\alpha_{d+1} \} \\
	\alpha_k &\in& H^{d+1-k}(Q , H^k(K,U(1))).
	\end{eqnarray}
	
	\begin{figure}[!htb]
		\includegraphics[width=45mm]{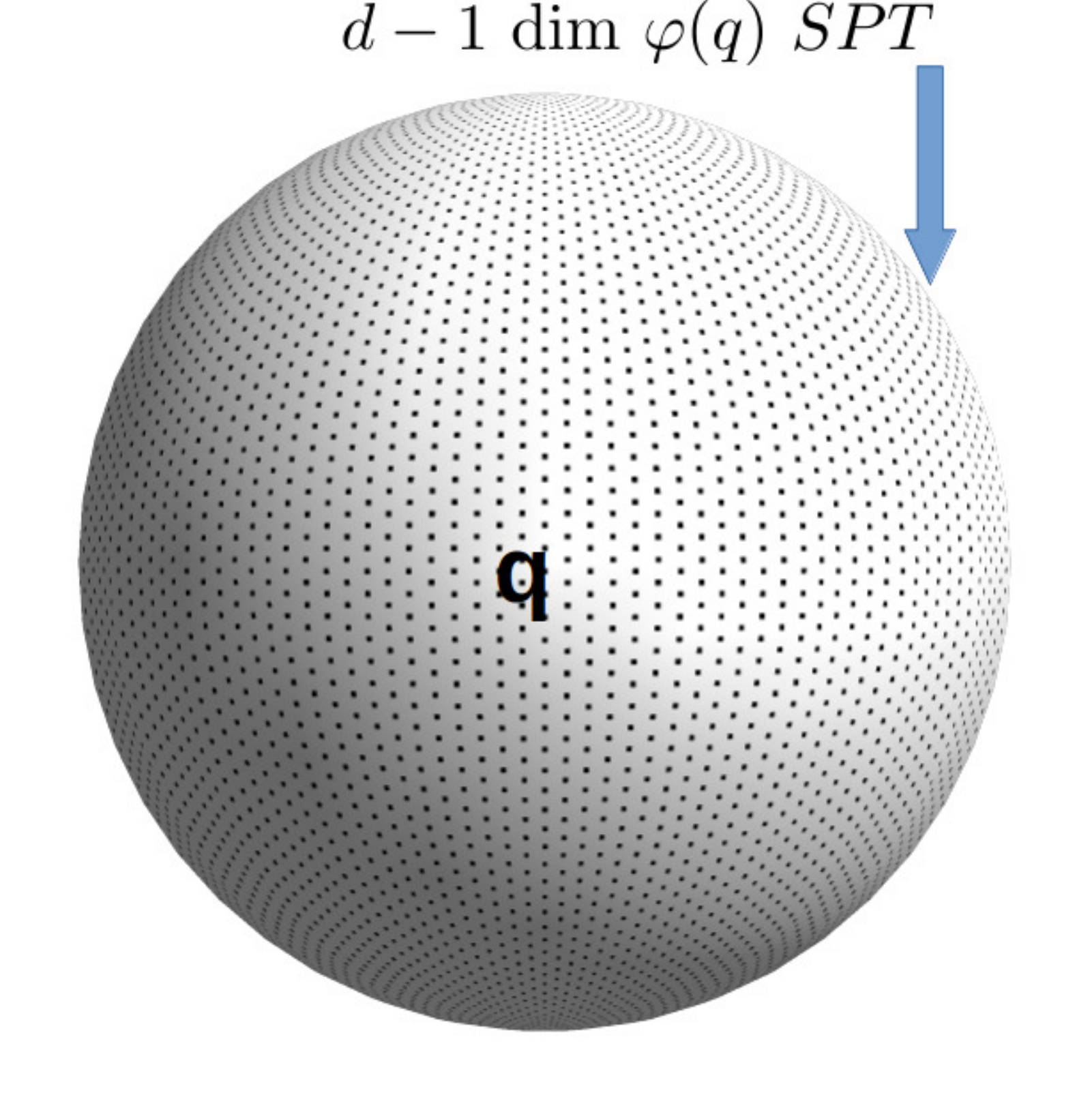}
		\caption{Single domain. \label{fig:single_domain_ddim}}
	\end{figure}

	\begin{figure}[!htb]
		\includegraphics[width=65mm]{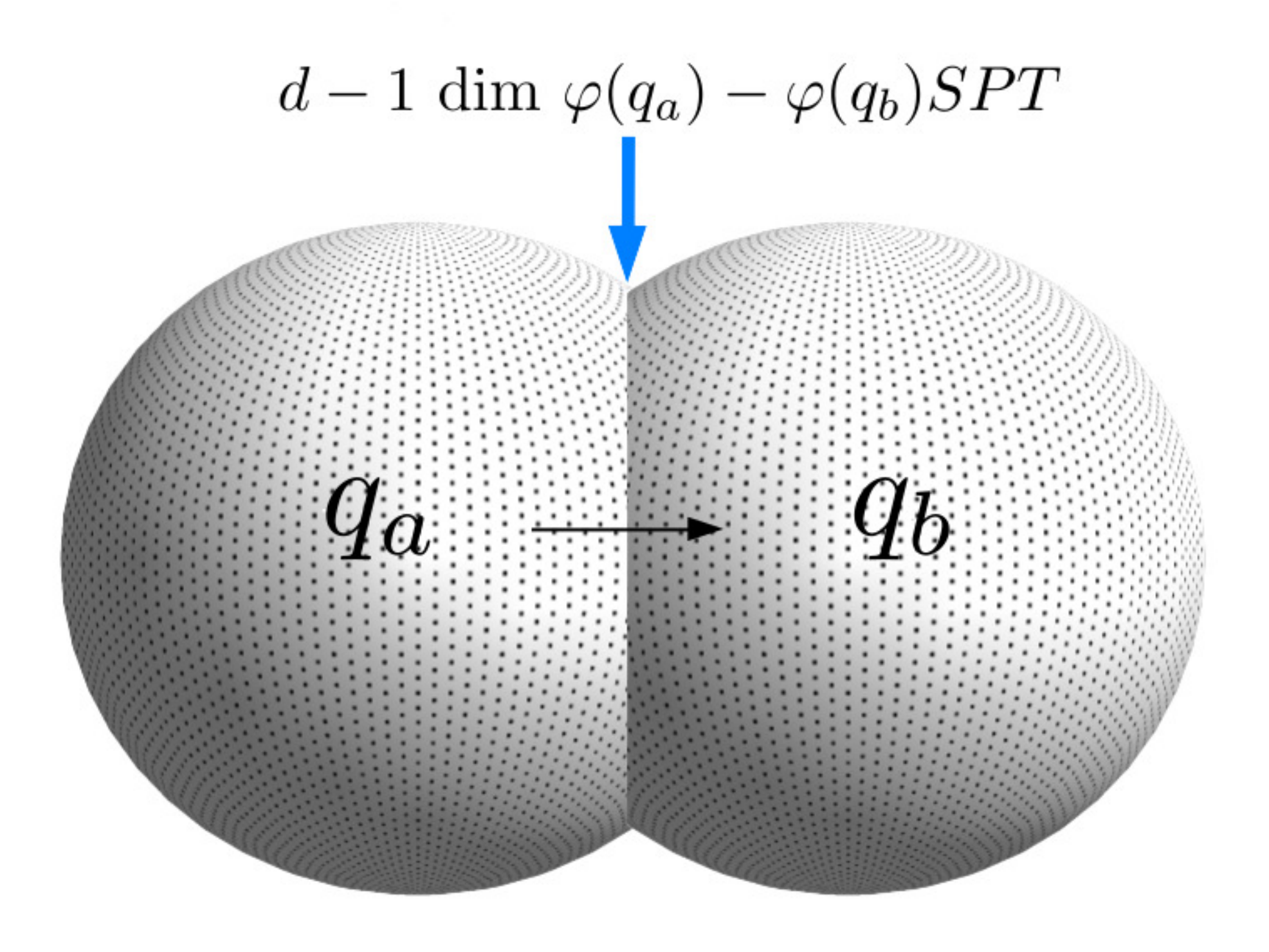}
		\caption{Two intersecting domains. \label{fig:two_domains_ddim}}
	\end{figure}

	\begin{figure}[!htb]
		\includegraphics[width=65mm]{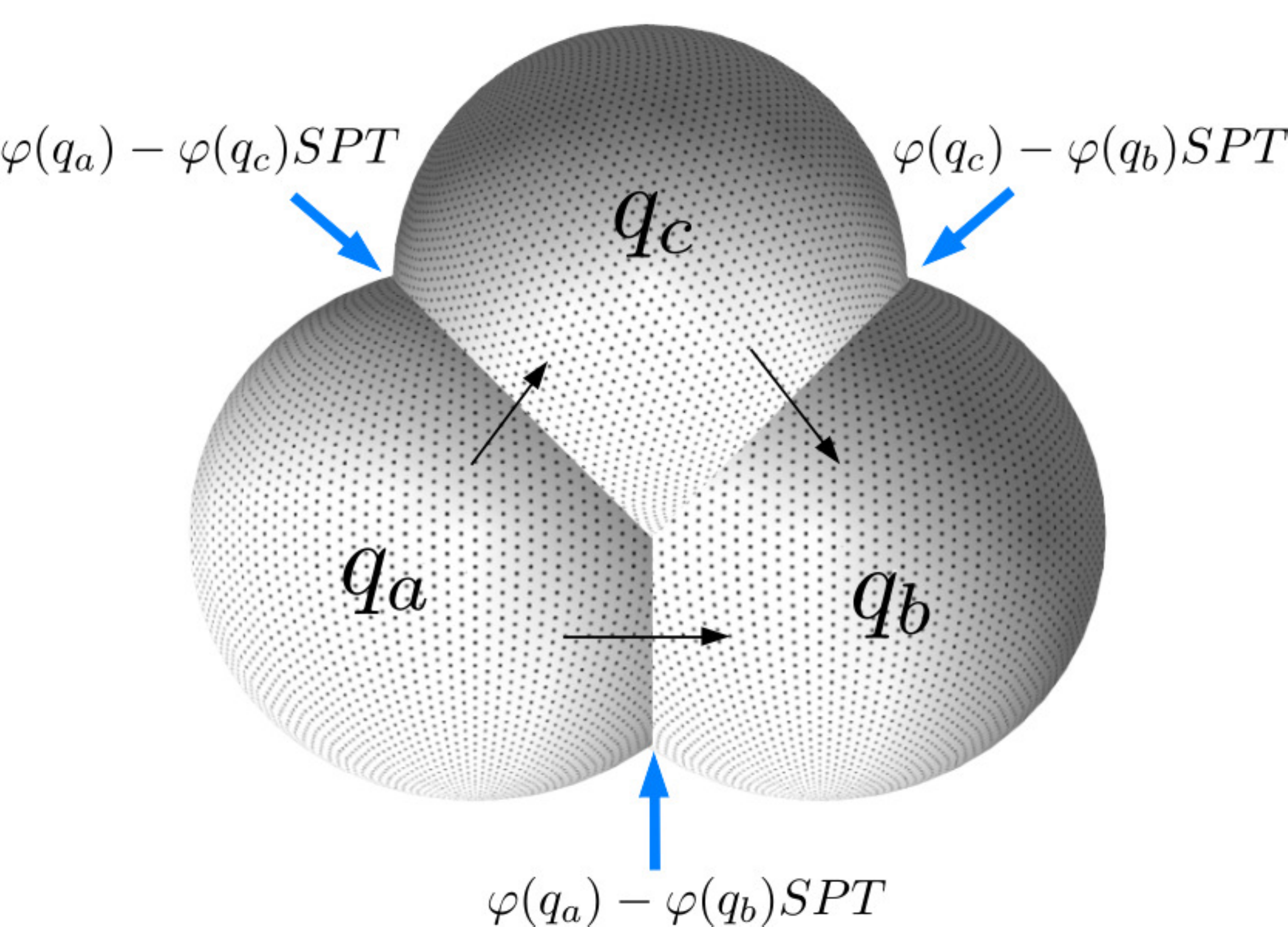}
		\caption{Three intersecting domains. \label{fig:three_domains_ddim}}
	\end{figure}	
	
Let us consider some special cases
\begin{itemize}
	\item $\alpha = \{\alpha_0,1,1, \ldots,1 \}$  : This class of SPT phases are labeled by 
	\begin{equation}
	\alpha_0 \in H^{d+1}(Q , H^0(K,U(1))) \cong H^{d+1}(Q,U(1))
	\end{equation}
	which are simply SPT phases protected by $Q$.
	\item $\alpha = \{1,1,\ldots,1,\alpha_{d+1} \}$  : This class of SPT phases are labeled by 
	\begin{equation}
	\alpha_{d+1} \in H^0(Q , H^{d+1}(K,U(1))) \cong H^{d+1}(K,U(1))
	\end{equation}
	which are simply SPT phases protected by $K$.
	\item $\alpha = \{\alpha_0,1,\ldots,1,\alpha_{d+1} \}$  : This class of SPT phases are obtained by layering an SPT phase protected by $Q$ with an SPT phase protected by $K$.
\end{itemize}
The simplest decorated domain wall SPT phase corresponds to the case of $\alpha = \{1,1,1, \ldots,\varphi,1 \}$. This class of SPT phases are labeled by 
\begin{equation}
\varphi \in H^{1}(Q , H^d(K,U(1))).
\end{equation}
We will henceforth restrict our attention to these phases. The authors of Ref.~\cite{Chen2014} provide a nice physical interpretation of these phases in terms of domains of spins that transform under a representation of $Q$ decorated by domain walls of $d$-1 dimensional SPT phases protected $K$. 

To understand this better, let us first look at the meaning of $H^{1}(Q , H^d(K,U(1)))$. For the cases we are interested in \thatis when $Q$ does not act on $K$, this is simply the group of group homomorphisms,
\begin{equation}\label{eq:homomorphism_QK}
\varphi: Q \rightarrow H^d(K,U(1)).
\end{equation}
This means, that each $\varphi \in H^{1}(Q , H^d(K,U(1)))$ corresponds to a unique group homomorphism Eq.~(\ref{eq:homomorphism_QK}). Note that the elements of the abelian group, $H^d(K,U(1))$ label the different $d$-1 dimensional SPT phases protected by $K$. This is precisely the information the authors of~\cite{Chen2014} use to construct ground state SPT wavefunction amplitudes. For each basis state $\ket{q}$ associated to a domain that transforms as a regular representation of $Q$, they decorate the boundary using a lower dimensional $K$ SPT phase corresponding to the image of $q$ under $\varphi$. In other words, the boundary of a domain $q$ is decorated by the SPT phase $\varphi(q)$ (see Fig.~\ref{fig:single_domain_ddim}). 

When two domains share a common boundary, the SPT phase associated to the domain-wall can be unambiguously fixed by specifying  the orientation to the domain wall. In other words, the domain-wall between domains corresponding to $q_a$ and $q_b$ is decorated by the SPT $\varphi(q_a q_b^{-1}) =\varphi(q_a)- \varphi(q_b)$ (see Fig.~\ref{fig:two_domains_ddim}). Furthermore, a triple intersection of domain-walls reflects the group structure of the lower-dimensional SPT phases, labeled by $H^d(K,U(1))$, as shown in Fig.~\ref{fig:three_domains_ddim}.  The DDW wavefunction can be viewed as a superposition of networks (specified by the domain spin configuration) of lower-dimensional $K$-SPT phases. By measuring domain spins in the canonical basis of the regular representation, we leave behind intersecting domain-walls of lower-dimensional SPT phases, removing the superposition. We remark that this is the essence of the reduction from SPT states to graph-like states, considered in Sec.~\ref{sec:1st_step}. In the next section, we focus on the specific case of our interest in 2 dimensions. We refer the reader to Ref.~\cite{Chen2014} for more details on the general construction.

\subsection{Constructing $\mathbb{Z}_d^3$ DDW SPT ground states}
We now focus exclusively on a 2D DDW SPT phase protected by $G = \mathbb{Z}_d^3$ with $Q = \mathbb{Z}_d$ and $K = \mathbb{Z}_d^2$, where $d$ is a prime number, and construct fixed-point wavefunctions. The data that classifies DDW phases for this group is $H^1(\mathbb{Z}_d,H^2(\mathbb{Z}_d^2,U(1)))$. We will find that  the non-trivial fixed-point wave functions we obtain using DDW picture in this section are precisely the ones we have been studying for universal quantum computation, thereby proving that the latter ones are indeed  non-trivial SPT states. 

\subsubsection{$\zd^2$ SPT phases in 1D.}

First, let us look at the lower dimensional SPT phases labeled by $H^2(\mathbb{Z}_d^2,U(1))$ that are used to decorate the walls of the 2d domains. Once again, using the K\"{u}nneth formula, we find that 
\begin{equation}
H^2(\mathbb{Z}_d^2,U(1)) \cong H^1(\mathbb{Z}_d,H^1(\mathbb{Z}_d,U(1)))
\end{equation}
and all other groups in the expansion are trivial. It is a remarkable fact that all 1D SPT phases protected by $\zd^2$ can themselves be represented as DDWs. Let us construct the wave functions for these SPT phases which we will later use to decorate 2D domains with. $H^1(\mathbb{Z}_d,U(1))$ is the group of one dimensional irreps of $\mathbb{Z}_d$, labeled as $\chi^\alpha$. $\mathbb{Z}_d$ is a group of order $d$, isomorphic to  integers under addition modulo $d$: 
\begin{eqnarray}
\mathbb{Z}_d &=& \{0,1,\ldots,{d-1}\}.
\end{eqnarray}
 The irreps of $\zd$ are generated using the $d^{th}$ roots of unity:
\begin{eqnarray}
\chi^\alpha(k) &=& e^{2 \pi i \alpha k/d},~~\alpha,k~=1\ldots d.
\end{eqnarray}
The $d$ one-dimensional irreps of $\zd$ themselves also have a $\zd$ group structure:
\begin{equation}
\chi^\alpha(g) \chi^\beta(g) \cong \chi^{\alpha+\beta}(g).
\end{equation}
In other words, $H^1(\mathbb{Z}_d,U(1)) \cong \zd$. Furthermore, $H^1(\mathbb{Z}_d,H^1(\mathbb{Z}_d,U(1)))$ are the homomorphisms from $\zd$ to the irreps of $\zd$,
\begin{eqnarray}
\xi: \zd \rightarrow H^1(\mathbb{Z}_d,U(1)),
\end{eqnarray}
which labels distinct ways of associating a $\zd$ irrep to each $\zd$ element. There are $d$ ways of doing this which we will label by $\xi_0,\ldots,\xi_{d-1}$
\begin{eqnarray}
\xi_s: m \rightarrow \chi^{ms}(k) = e^{2 \pi i m s k/d}, \nonumber\\
s,m,k=0\ldots{d-1}
\end{eqnarray}

This means that there are $d$ SPT phases ($d-1$ non-trivial ones) in 1D protected by $\zd^2$. These SPT phases themselves have a $\zd$ structure and we shall label them by $s=0,1,\ldots,d-1$ and are specified by how we associate $\zd$ irreps to $\zd$ domain walls. 

\begin{figure}[!htb]
	\includegraphics[width=65mm]{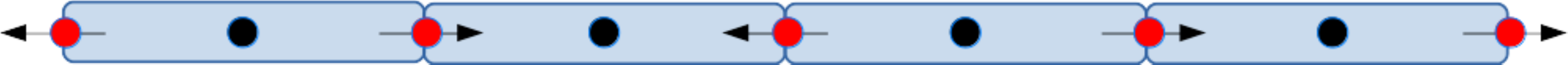}
	\caption{1D DDW state. \label{fig:1d_DDW}}
\end{figure}

\begin{figure}[!htb]
	\includegraphics[width=65mm]{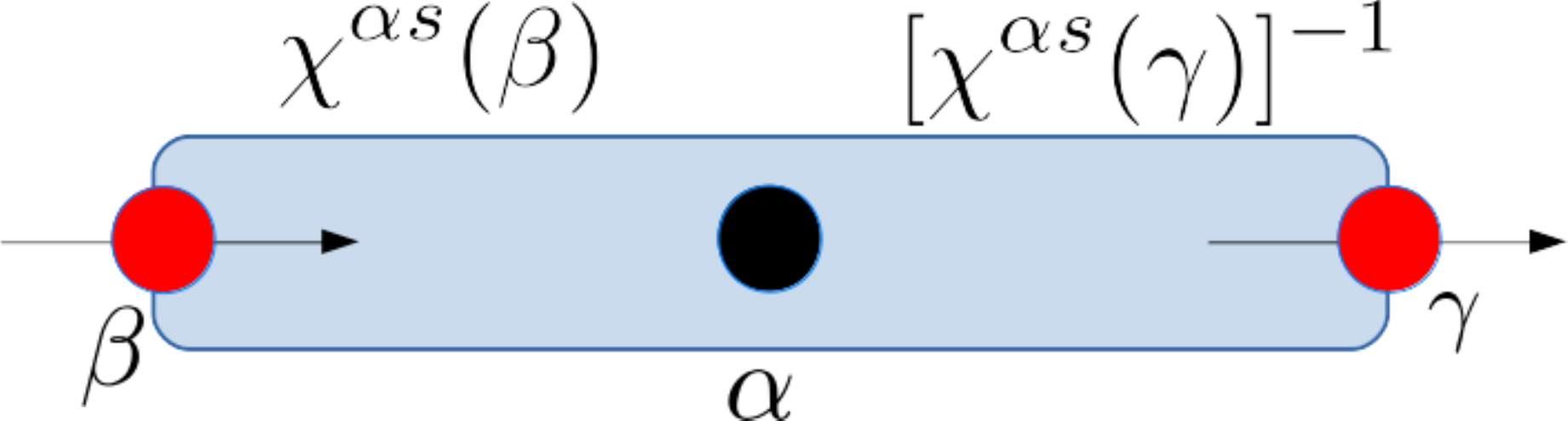}
	\caption{A single domain of 1D DDW state. \label{fig:1d_single}}
\end{figure}

To write fixed-point wave functions, let us consider a 1D lattice with domains (edges) and domain walls (vertices) that both transform as $d$ dimensional regular representations of $\zd$ as shown in Fig.~\ref{fig:1d_DDW}, with orientations of the domain wall specified by the arrows in the figure. 
A non-trivial SPT wavefunction, $\ket{\psi_s}$ is obtained by operating a trivial SPT wavefunction $\ket{\psi_0}$ with a diagonal operator $\mathcal{W}_s$ which cannot be written as a \emph{symmetric} finite depth unitary quantum circuit. 
\begin{eqnarray}
\ket{\psi_s} &=& \mathcal{W}^s \ket{\psi_0}, \\
\ket{\psi_0} &=& \prod_{e=1}^\text{edges} \frac{1}{\sqrt{d}} \sum_{\alpha=0}^{d-1} \ket{\alpha}_e \prod_{v=1}^{\text{vertices}} \frac{1}{\sqrt{d}} \sum_{\beta=0}^{d-1} \ket{\beta}_v ,
\end{eqnarray}
\begin{multline}
\mathcal{W}^s = \prod_{e=1(l,r \in \partial e)}^{\text{domains}} \sum_{\alpha,\beta,\gamma=1}^d [\chi^{\alpha s}(\beta)]^{\theta(l)} [\chi^{\alpha s}(\gamma)]^{\theta(r)}\\
\outerproduct{\alpha}{\alpha}_e \outerproduct{\beta}{\beta}_l \outerproduct{\gamma}{\gamma}_r.
\end{multline}
Note that for each domain, we assign orientation to the boundary of the domain as $\theta = \pm 1$ if the arrow is pointing towards or away from the center of the domain. An example of the wave function amplitude is shown in Fig.~\ref{fig:1d_single}. We have colored domain qudits with black and domain wall qudits with red.
Note that we can rewrite $\mathcal{W}_s$ using a controlled two-qudit operator
\begin{eqnarray}
\mathcal{W}^s &=& \prod_{e=1(l,r \in \partial e)}^{\text{edges}} [CZ^{s}]^{\theta(l)}_{el} [CZ^{s}]^{\theta(r)}_{er}, \nonumber \\
 &=& \prod_{e=1(l,r \in \partial e)}^{\text{edges}} \left[ [CZ]^{\theta(l)}_{el} [CZ]^{\theta(r)}_{er} \right]^s, \\
CZ_{ab} &=& \sum_{\alpha=0}^{d-1} \outerproduct{\alpha}{\alpha}_a \sum_{\beta=0}^{d-1} e^{2 \pi i  \alpha \beta /d} 
 \outerproduct{\beta}{\beta}_b, \nonumber \\
&=&  \sum_{\alpha=0}^{d-1} \outerproduct{\alpha}{\alpha}_a Z^{\alpha }_b, \\
Z &=& \sum_{\beta=0}^{d-1} e^{2 \pi i  \beta /d} 
\outerproduct{\beta}{\beta}.
\end{eqnarray}
Note that, for $\mathbb{Z}_{d=2}$, we get the well-known one-dimensional cluster state with this construction~\cite{Chen2014, Yoshida2017}.

\subsubsection{$\zd^3$ DDW in 2D.}

\begin{figure}[!htb]
	\includegraphics[width=65mm]{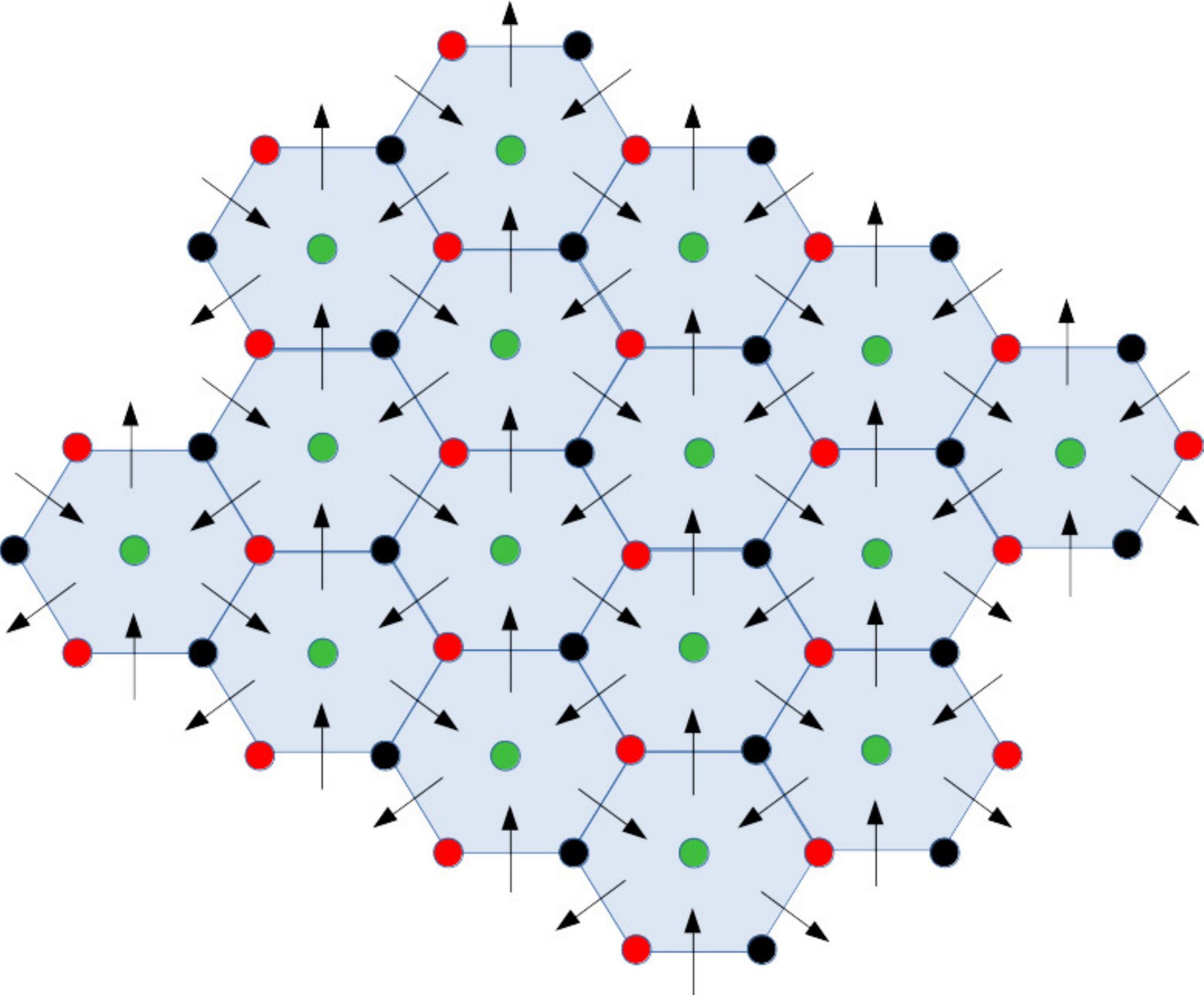}
	\caption{2D DDW state. \label{fig:2d_DDW}}
\end{figure}

Let us now finally construct the $\zd^3$ DDW SPT states. These are labeled by elements of $H^1(\mathbb{Z}_d,H^2(\mathbb{Z}_d^2,U(1)))$. These are the homomorphisms of $\zd$ to the group of $d$ 1D $\zd^2$ SPT phases,
	\begin{equation}
	\varphi: \zd \rightarrow H^2(\mathbb{Z}_d^2,U(1)).
	\end{equation}
This is again formally the group of homomorphisms $\zd \rightarrow \zd$ and hence, there are again $d$ ways of doing this,
	\begin{equation}
	\varphi_k: s \rightarrow ks,~~k,s = 0,\ldots, {d-1}.
	\end{equation}
This means that there are $d$ DDW SPT phases (d-1 non-trivial ones) labeled $k=0,\ldots,d-1$. Let us build fixed-point wave functions for them. We work with a 2D trivalent lattice and place $\zd$ regular representations on all faces (domains) and vertices. We can also fix orientations of all domain boundaries arbitrarily. Let us choose the honeycomb lattice and the orientation shown in Fig.~\ref{fig:2d_DDW} to eventually match up with the state considered. 

\begin{figure}[!htb]
	\begin{tabular}{c}
		\includegraphics[width=65mm]{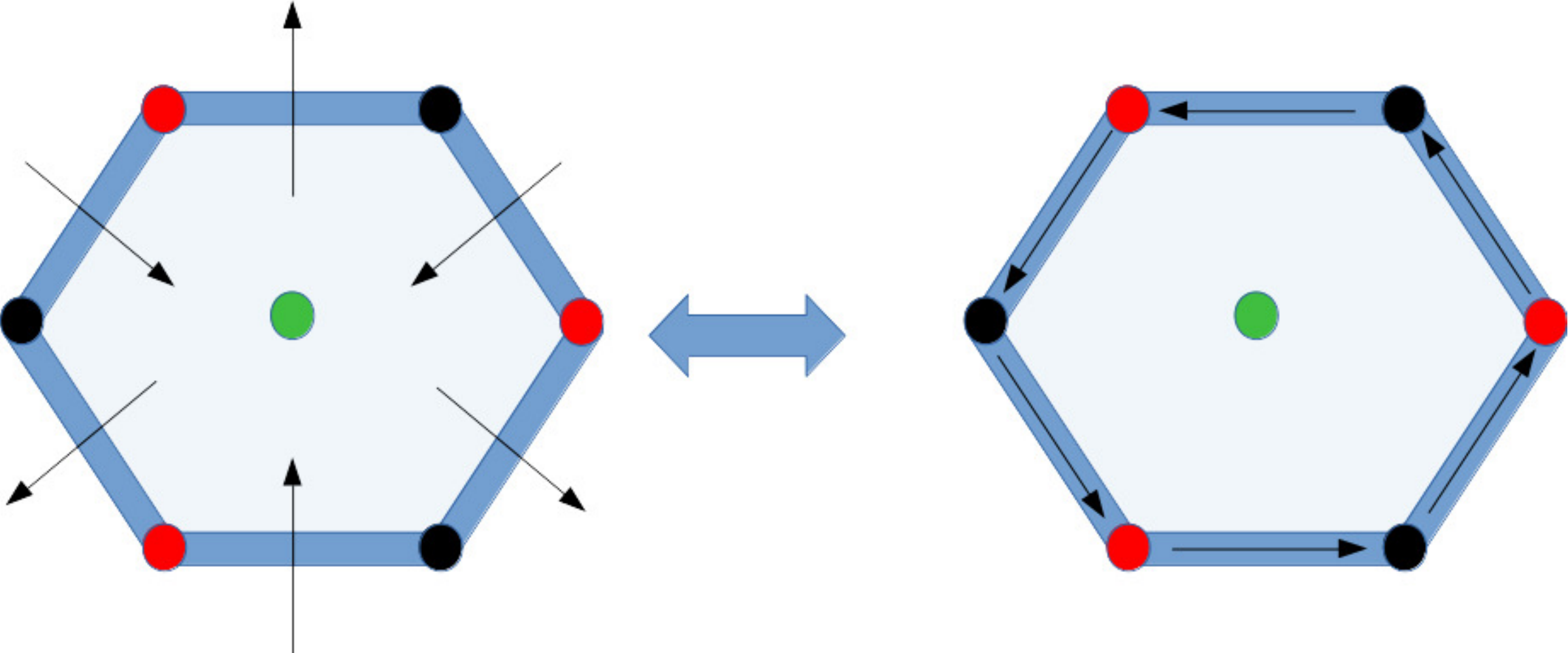}			
	\end{tabular}
	\caption{Orientation rules for 1D domain walls. \label{fig:2d1dorientation}}
\end{figure}

The SPT wave function labeled by $k = 0 \ldots d-1$ is obtained again by operating the trivial state with a diagonal operator $\mathcal{U}_k$ that cannot be written as a finite depth $\zd^3$ symmetric circuit, where
	\begin{eqnarray}
	\ket{\psi_k} &=& \mathcal{U}^k \ket{\psi_0},\\
	\ket{\psi_0} &=& \prod_{f=1}^\text{faces} \frac{1}{\sqrt{d}} \sum_{\alpha=0}^{d-1} \ket{\alpha}_f \prod_{v=1}^{\text{vertices}} \frac{1}{\sqrt{d}} \sum_{\beta=0}^{d-1} \ket{\beta}_v,
	\end{eqnarray}
	
	\begin{eqnarray}
	\mathcal{U}^k &=& \prod_{f = 1}^{\text{faces}} \sum_{\alpha=0}^{d-1} \outerproduct{\alpha}{\alpha}_f \mathcal{W}_{\partial f}^{\varphi_k(\alpha)} \nonumber \\
	&=& \prod_{d = 1}^{\text{faces}} \sum_{\alpha=0}^{d-1} \outerproduct{\alpha}{\alpha}_f \mathcal{W}_{\partial f}^{k\alpha} \nonumber \\
	&=& \prod_{d = 1}^{\text{faces}} \left[ \sum_{\alpha=0}^{d-1} \outerproduct{\alpha}{\alpha}_f \mathcal{W}_{\partial f}^{\alpha} \right]^k, 
	\end{eqnarray}
and $\mathcal{W}_s$ is as defined before. Note that the orientations, $\theta$ in the 1D DDW that is required to specify $\mathcal{W}^s$ is inherited from that of the orientation of the 2D domain walls by assigning a consistent rule. We use a simple ``towards green $\implies$ towards black" rule demonstrated in Fig.~\ref{fig:2d1dorientation}

Using the orientation in Fig.~\ref{fig:2d_DDW}, and the rule specified in Fig.~\ref{fig:2d1dorientation}, we can write $\mathcal{U}^k$ as follows using the three qudit controlled operators:
	\begin{eqnarray}
	\mathcal{U}^k &=& \prod_{\triangle} CCZ^{k} \prod_{\triangledown} CCZ^{-k}, \\
	CCZ &=& \sum_{i=0}^{d-1} \outerproduct{i}{i} CZ^i,
	\end{eqnarray}
It is now clear that the states we constructed in Sec.~\ref{sec:triangularSPT} are  $\ket{\psi_k} = \mathcal{U}^k \ket{\psi_0}$. The upper and lower triangles are obtained  by connecting `boundary' vertices (black and red dots) of hexagons to the centers.

\begin{figure*}[!htb]
	\begin{subfigure}{0.3\textwidth}
		\centering
		\includegraphics[height=50mm,width=50mm]{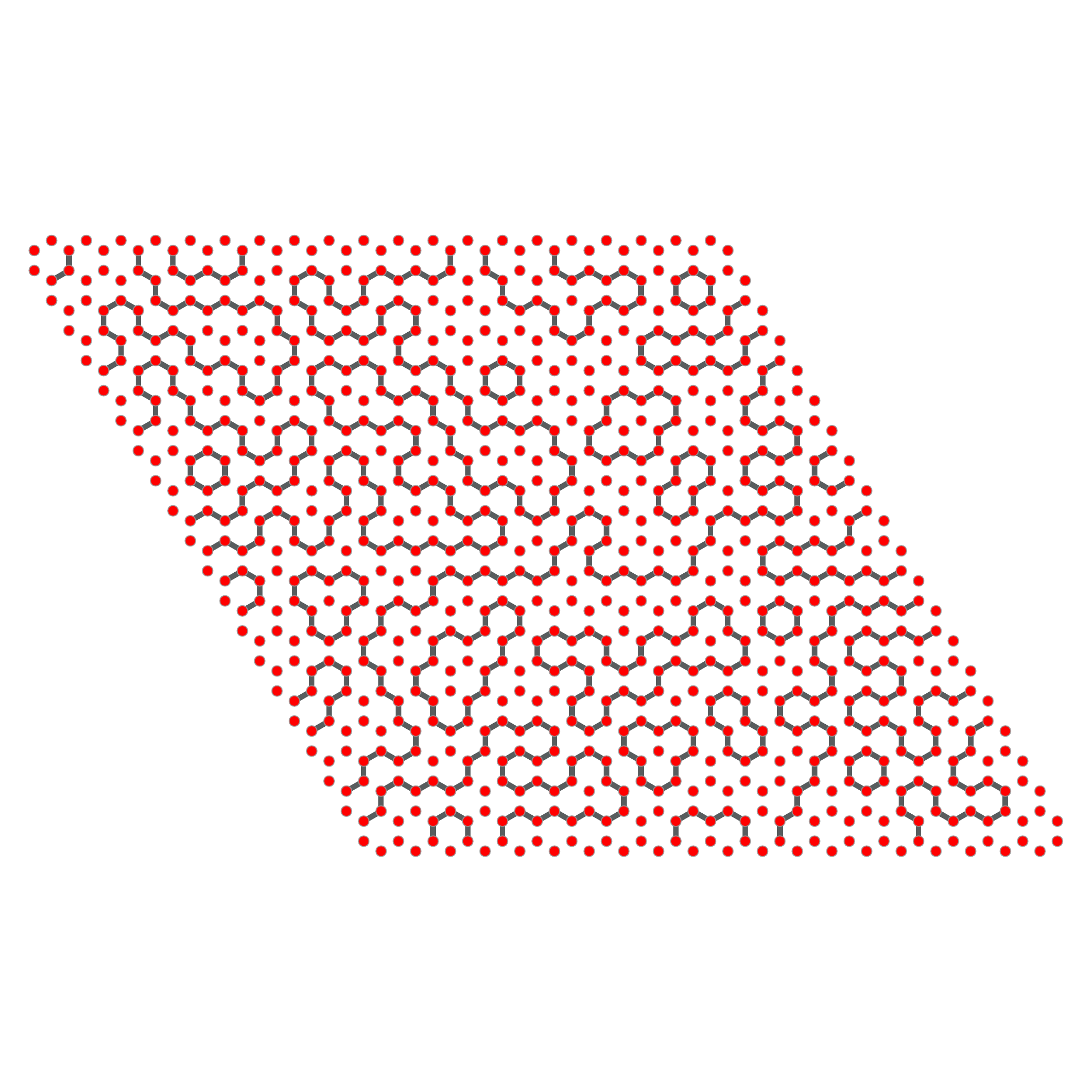}
		\caption{}
	\end{subfigure}
	\hspace{0.5mm}
	\begin{subfigure}{0.3\textwidth}
		\centering
		\includegraphics[height=50mm,width=50mm]{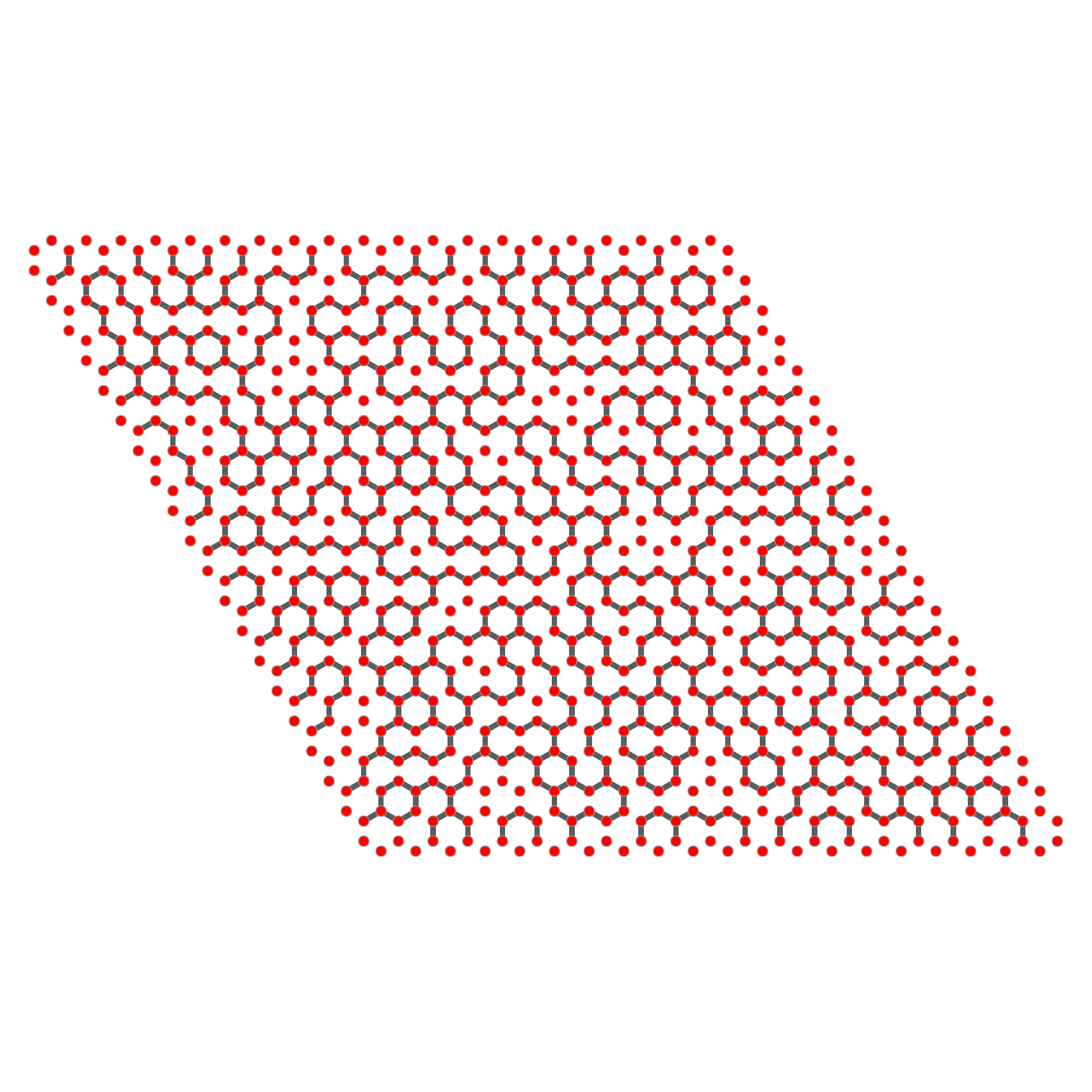}
		\caption{}
	\end{subfigure}
	\hspace{0.5mm}
	\begin{subfigure}{0.3\textwidth}
		\centering
		\includegraphics[height=50mm,width=50mm]{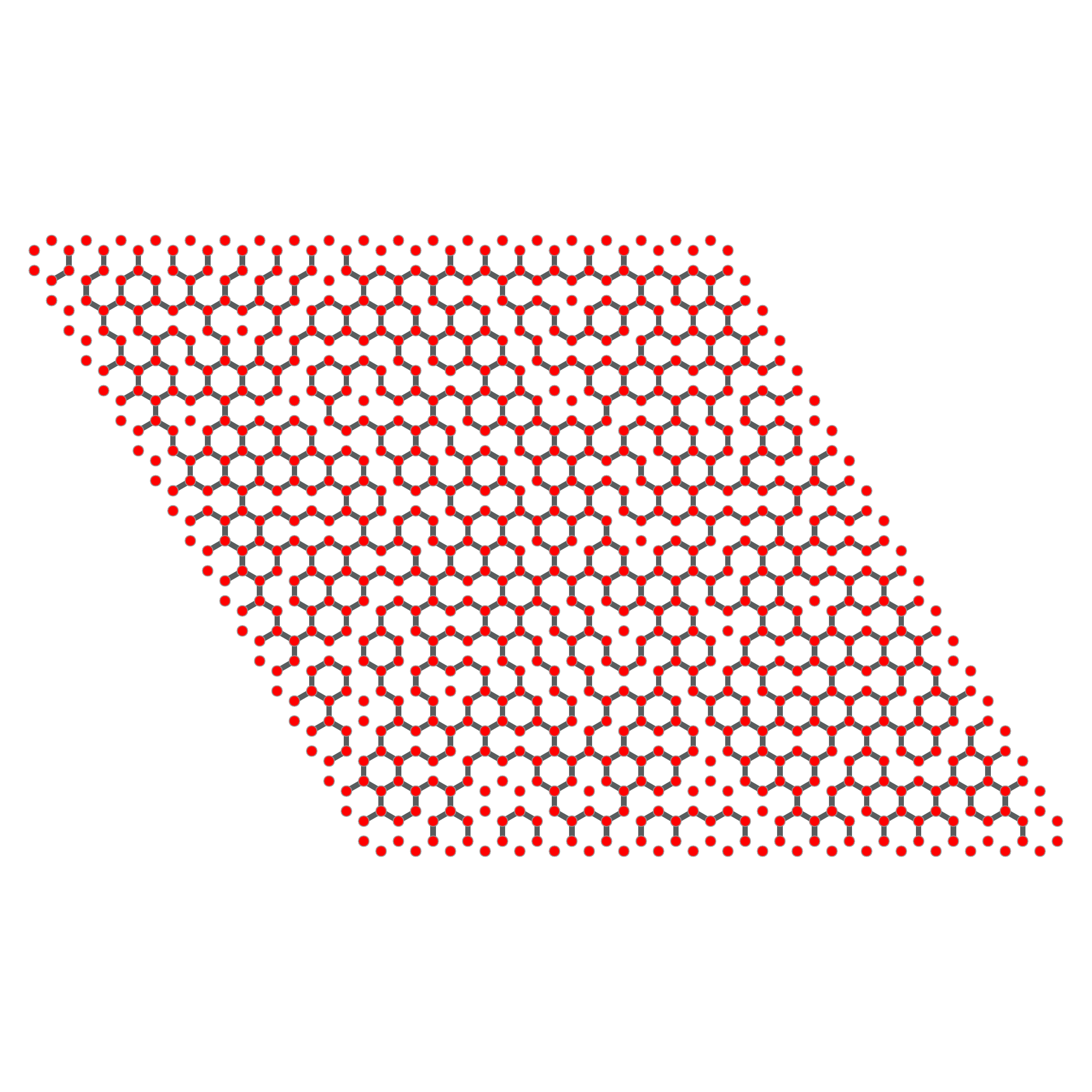}
		\caption{}
	\end{subfigure} \\
	\vspace{1mm}
	\begin{subfigure}{0.3\textwidth}
		\centering
		\includegraphics[height=40mm,width=40mm]{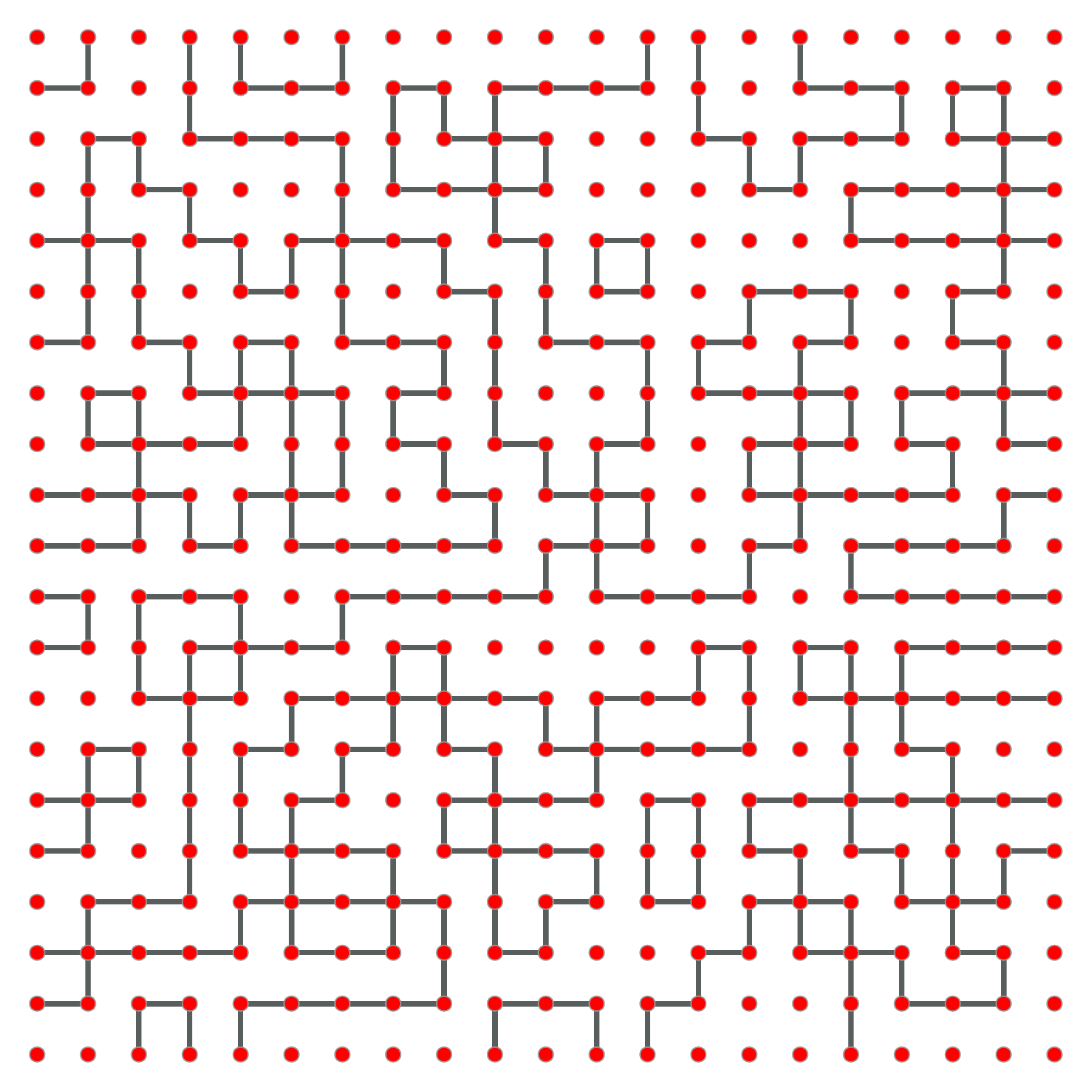}
		\caption{}
	\end{subfigure}
	\hspace{0.5mm}
	\begin{subfigure}{0.3\textwidth}
		\centering
		\includegraphics[height=40mm,width=40mm]{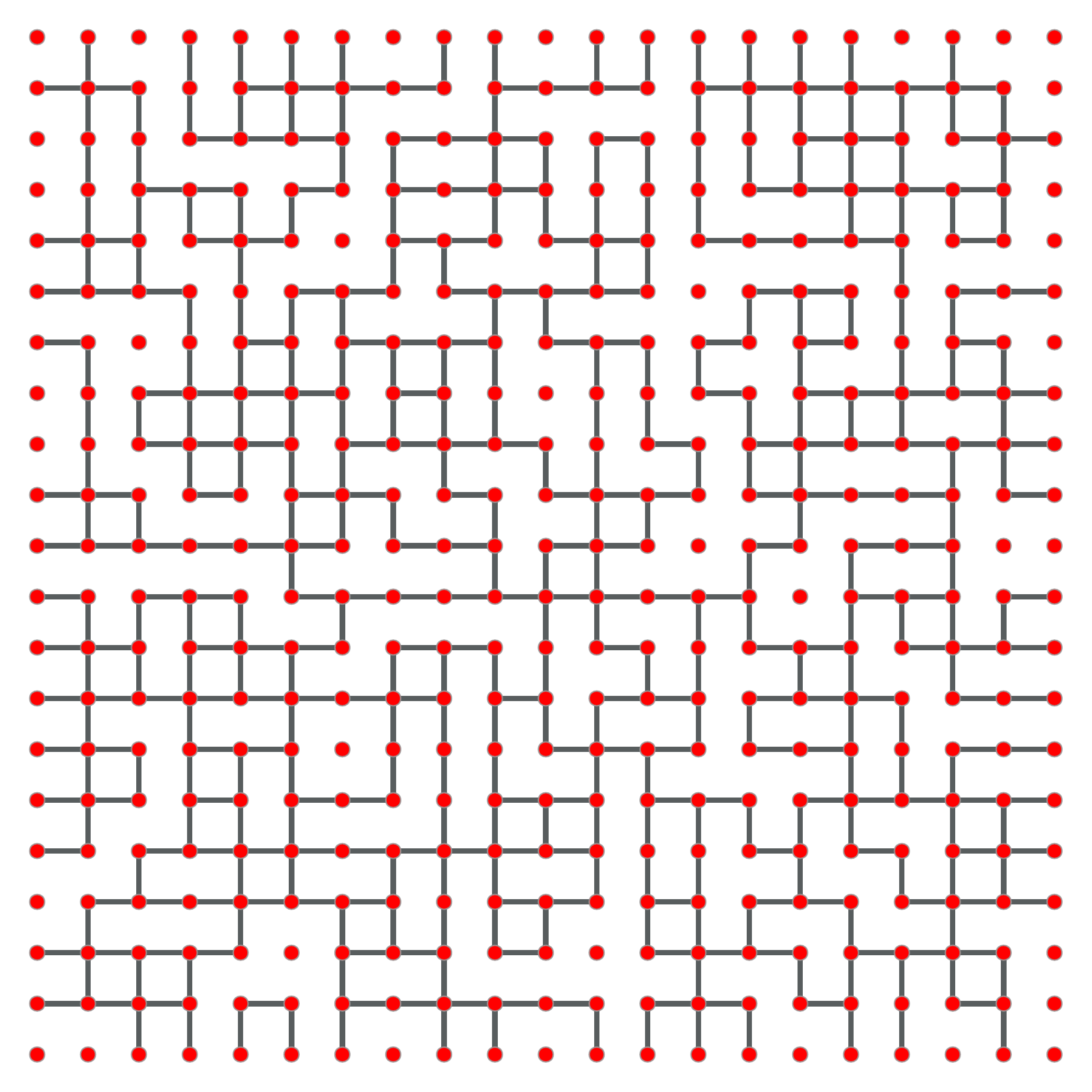}
		\caption{}
	\end{subfigure}
	\hspace{0.5mm}
	\begin{subfigure}{0.3\textwidth}
		\centering
		\includegraphics[height=40mm,width=40mm]{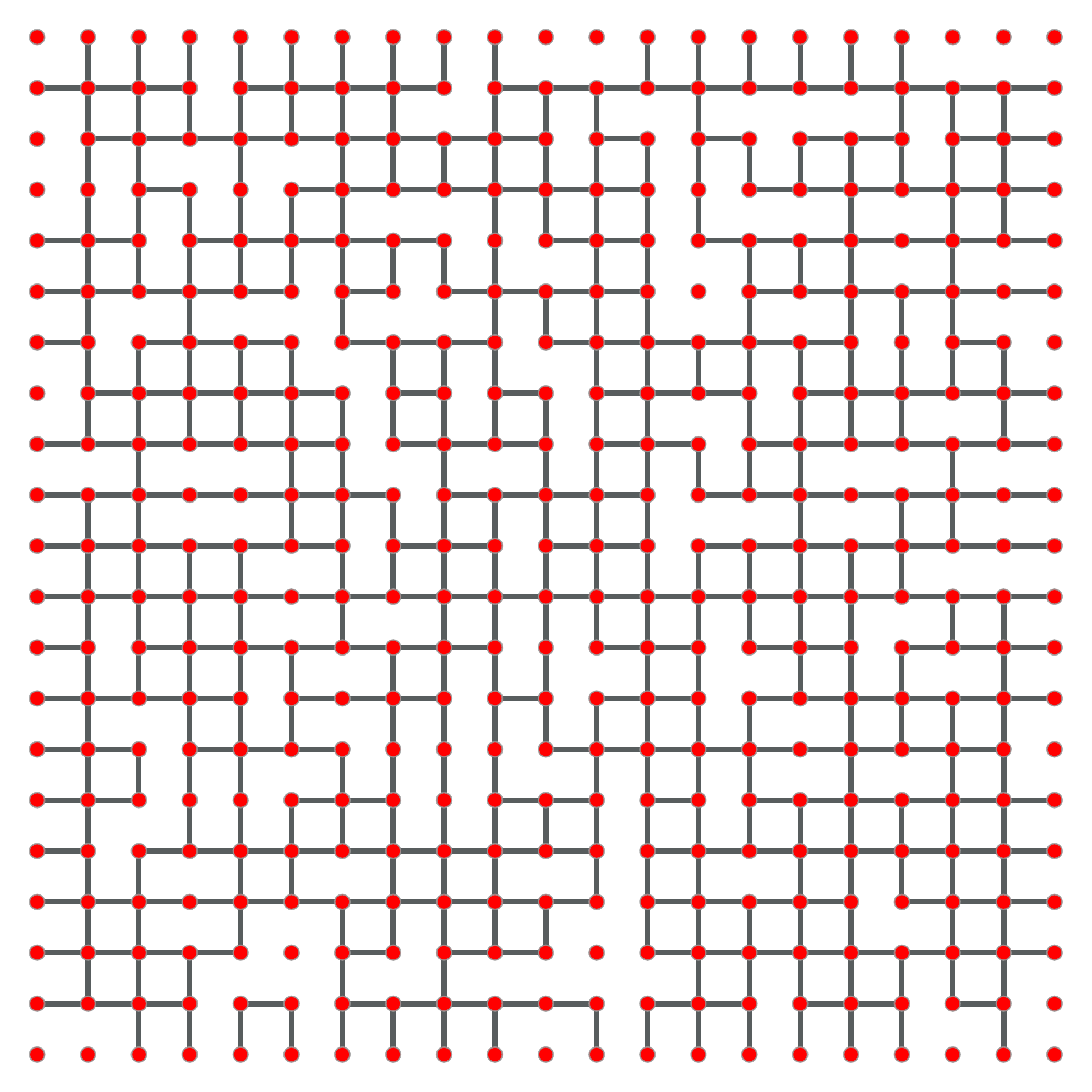}
		\caption{}
	\end{subfigure}
	\caption{Examples of connected structure for the random graph-like state generated after measurements. The top figures are embedded in honeycomb after measuring triangular SPT state, for $d=2, \, 3, \, 5$ respectively. The bottom figures are embedded in square after measuring Union-Jack state. \label{fig:after_measurement}}
	\end{figure*}

\section{Reduction of random graph} \label{app:convert_procedure}

After measuring qudits marked red on the triangular SPT state, as described in Sec.~\ref{sec:1st_step}, we obtain different random graph-like states depending on the measurement outcomes. Their percolation properties were studied in Sec.~\ref{sec:simulation}. Here we include several examples in Fig.~\ref{fig:after_measurement}, which demonstrate that the network connection indeed improves for larger values of $d$. Note that in the $d=2$ case no junction of three edges is present and hence paths connecting top and bottom and paths connecting left and right cannot both exist.

\begin{figure}[h]
		\begin{subfigure}[t]{0.22\textwidth}
			\centering
			\includegraphics[height=20mm,width=40mm]{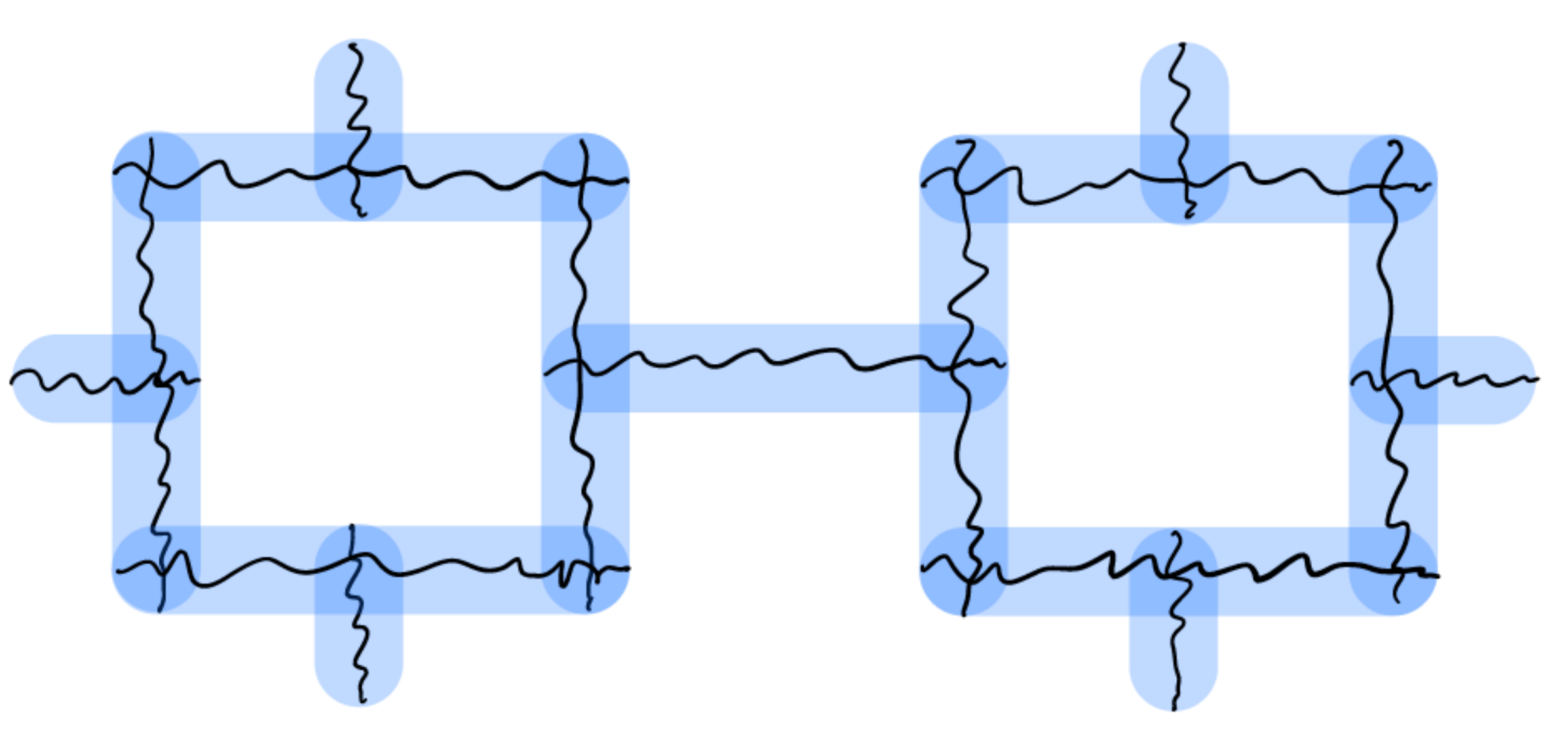}
			\caption{The network structure obtained by finding a path through in sufficiently large regions.}
			\label{Wei_1}
		\end{subfigure}
		\hspace{1mm}
		\begin{subfigure}[t]{0.22\textwidth}
			\centering
			\includegraphics[height=22mm,width=44mm]{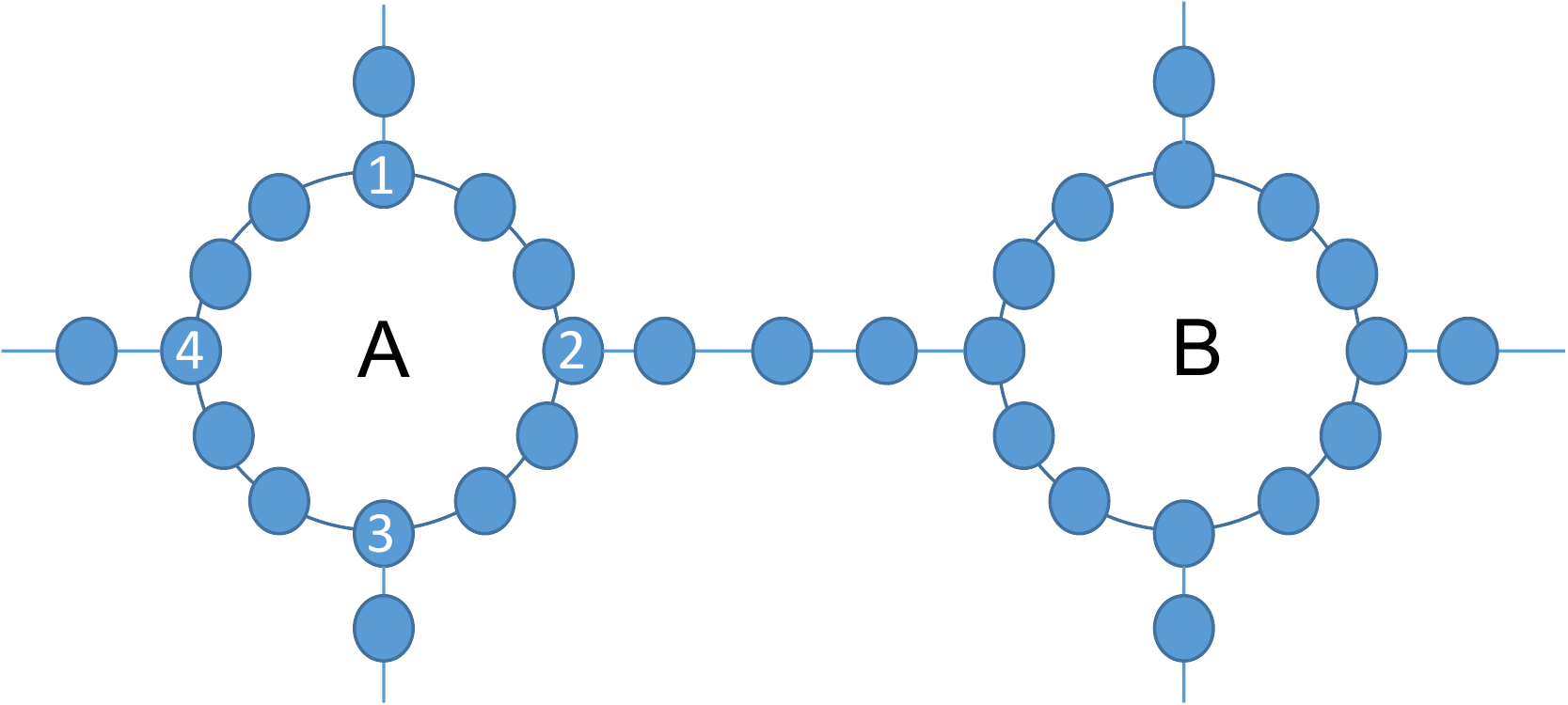}
			\caption{The structure after removing all excessive edges.}
			\label{Wei_2}
		\end{subfigure} \\
		\vspace{1mm}
		\begin{subfigure}[t]{0.22\textwidth}
			\centering
			\includegraphics[height=30mm,width=40mm]{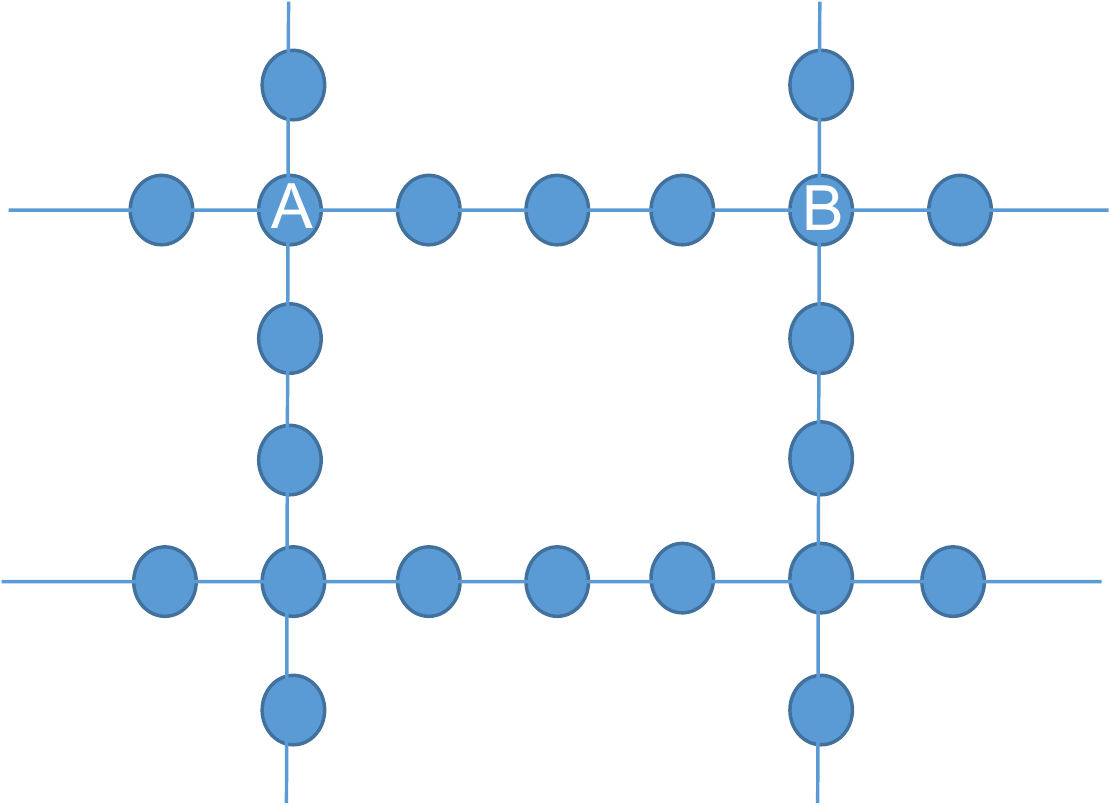}
			\caption{The structure after merging the junctions on each ring.}
			\label{Wei_3}
		\end{subfigure}
		\hspace{1mm}
		\begin{subfigure}[t]{0.22\textwidth}
			\centering
			\includegraphics[height=25mm,width=35mm]{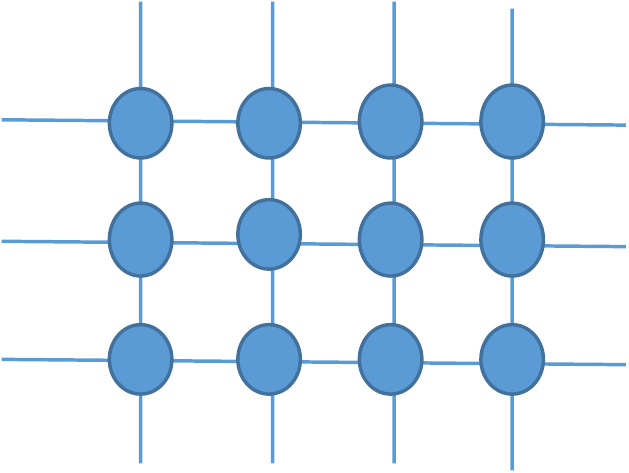}
			\caption{The square structure we need.}
			\label{Wei_4}
		\end{subfigure}
		\caption{The steps of converting a random graph state to a cluster-like state on a square.}
		\label{Wei}
	\end{figure}

Based on the rules we describe in Sec.~\ref{sec:rules}, one is able to convert a random planar graph state to a qudit cluster-like state on a square lattice. The detailed converting procedure is given in Ref.~\cite{Wei2012}, and here we summarize it for completeness. For a graph in the percolated phase, we can always find a path connecting from top to bottom and another connecting from left to right in a $l \times l$ square with sufficiently large $l$. The number of such paths is macroscopic \cite{Browne2008}. Therefore, the network structure in Fig.~\ref{Wei_1} is present in our random graph state, which we can obtain by measuring the other unwanted qudits in $Z$ basis. The ideal structure would be single wires joint together with T-shaped junctions. However, measuring the unwanted qudits cannot remove the excessive entangling edges between the remaining qudits. Now the following three step are taken to convert this net to the qudit cluster-like state on a square lattice, illustrated in Fig.~\ref{Wei}.

\begin{figure}[h]
		\begin{subfigure}[t]{0.22\textwidth}
			\centering
			\includegraphics[height=10mm,width=40mm]{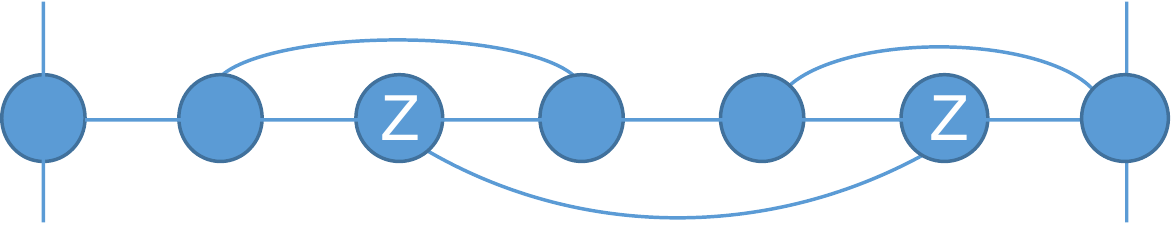}
			\caption{Diagram of a wire. The qudits that are to be measured in $Z$ basis are marked.}
		\end{subfigure}
		\hspace{1mm}
		\begin{subfigure}[t]{0.22\textwidth}
			\centering
			\includegraphics[height=20mm,width=35mm]{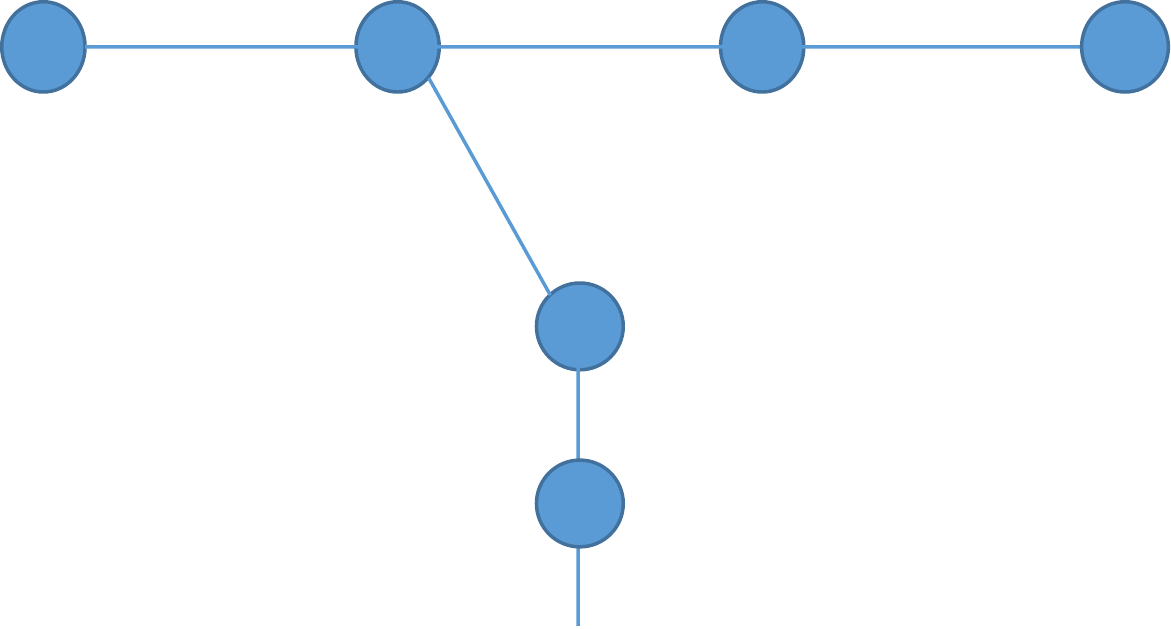}
			\caption{$W_c$ is attached to only one qudit on $W_l \cup W_r$: this junction is already T-shaped.}
			\label{junction_1}
		\end{subfigure} \\
		\vspace{1mm}
		\begin{subfigure}[t]{0.22\textwidth}
			\centering
			\includegraphics[height=20mm,width=35mm]{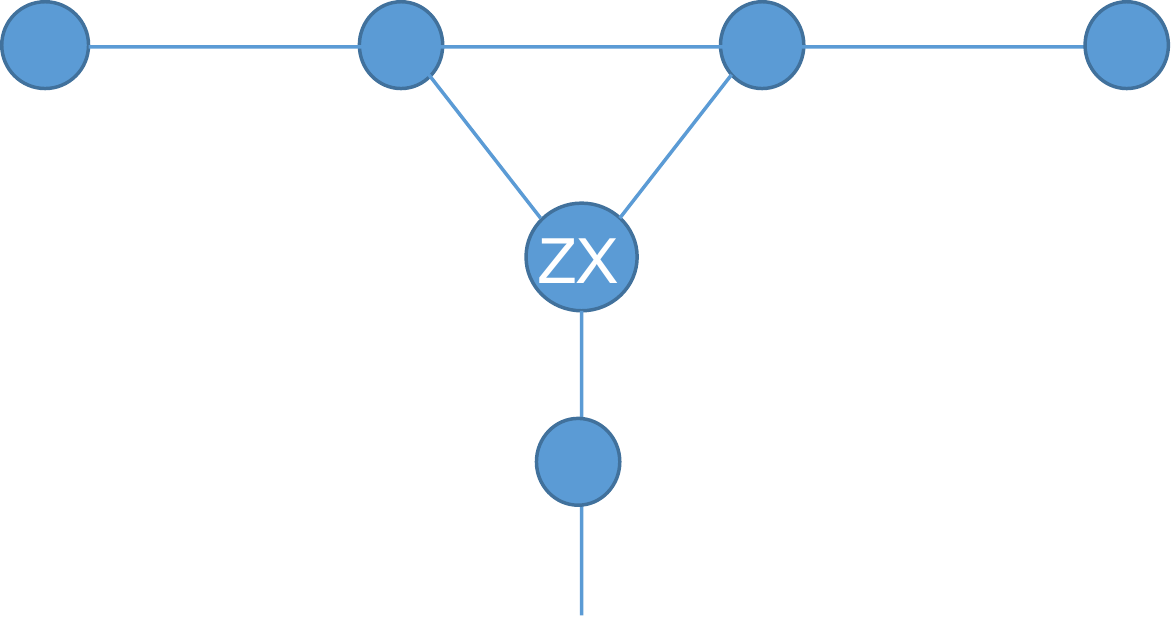}
			\caption{$W_c$ is attached to two adjacent qudits on $W_l \cup W_r$. The one marked is to be measured in $ZX^k$ basis with chosen $k$, explained in Sec.~\ref{sec:rules}. }
			\label{junction_2}
		\end{subfigure}
		\hspace{1mm}
		\begin{subfigure}[t]{0.22\textwidth}
			\centering
			\includegraphics[height=20mm,width=35mm]{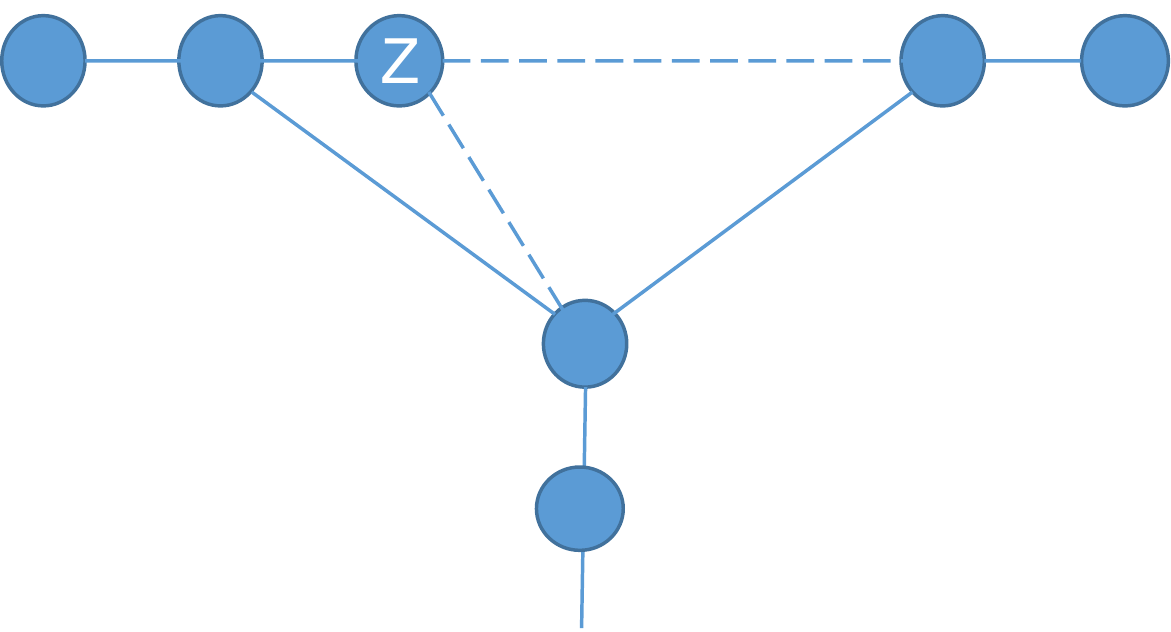}
			\caption{$W_c$ is attached to at least two qudits on $W_l \cup W_r$, with one or more qudit between the two that may be attached to $W_c$: measure in $Z$ basis all the qudits between the two outmost connections with $W_c$.}
			\label{junction_3}
		\end{subfigure}
		\caption{Rules for removing excessive edges, where (b), (c), (d) are the three types of junction with excessive edges between $W_l$ and $W_r$ omitted.}
		\label{remove}
	\end{figure}

\begin{enumerate}
	\item \textit{Clean the unwanted edges.} \\
		This step converts structure in Fig.~\ref{Wei_1} to the graph in Fig.~\ref{Wei_2}. The excessive edges are divided into two types: at a junction and within a wire. Junction is of some general T-shape but can have extra edges attached to it. Wire is the chain of qudits between two junctions. For each wire, we can start from left junction and move to the rightmost neighbour one by one, each time measuring all qudits between them in $Z$ basis. For each junction, to which three wires $W_l$ (left), $W_r$ (right) and $W_c$ (center) are attached, we label the excessive edges between any two wires by $E_{lr}$, $E_{lc}$ and $E_{rc}$. We then find the qudit in, for example $W_c$, that is the furthest from junction and connected to an excessive edge; and we measure the qudits between that one and the junction in $Z$ basis. This leads to three types, each of which can be dealt with, illustrated in Fig.~\ref{remove}.
	\item \textit{Reduce each ring-shaped four-leg junction to one-qudit crossing.} \\
		This takes Fig.~\ref{Wei_2} to Fig.~\ref{Wei_3} by merging one of the four T-shaped junctions with the others one by one. First we take junctions 1 and 4, which are not nearest neighbours, and measure the qudits between them in $Z$ basis. Then we measure the qudits between junction 1 and 2 in $Y$ (in our case $ZX^k$) basis until there is one left unmeasured. From this the chain is shortened but the connection between 1 and 2 is maintained. Then we measure the qudit in between them and qudit 2 in $X$ basis, merging 1 and 2. Repeating this we can merge 1 and 3, 4. The ring structure now becomes a four-leg crossing.                                                                               
	\item \textit{Shorten the distance between adjacent crossings.} \\
		By this we arrive at Fig.~\ref{Wei_4}. Similar to the last step, the qudits between any two crossings can be measured either in $Y$ (in our case $ZX^k$) basis or $X$ basis (where two adjacent ones are measured together), until all qudits at crossing are nearest neighbours to another four qudits at crossing. This gives the square structure.
\end{enumerate}

Note that in the whole process each edge could be any one from the $d-1$ possible entangling gates, and the final product is a cluster-like state instead of the cluster state. We remark that for our random graph embedded in a honeycomb lattice, the first step can be omitted. We can always select out the desired network structure without excessive edges. Also we should emphasize that even though measurements in $ZX^k$ basis and in $Y$ basis are slightly different, the two rules needed are still valid, demonstrated in Fig.~\ref{convert}.

\section{Qudit cluster-like states}\label{app:cluster-like}
	
	\subsection{Proof of the theorem} \label{proof}

	Following the proof in Ref.~\cite{Raussendorf2003, Zhou2003}, we first look at the case where the input state is an eigenstate of $Z_1, Z_2, ..., Z_n$ labelled by $Z$-measurement outcomes $\{t\}$, which can be written as:
		\be
		|\psi(\mathrm{in})\rangle_{C_I} = (\sqrt{d})^n P_{M_{C_I}^\prime}^{\{t\}} |+\rangle_{C_I},
		\ee
	where the basis for preparation $M_{C_I}^\prime = \{ Z_i, i \in C_I \}$ differs from measurement basis $M_{C_I} = \{ X_i, i \in C_I \}$. Then we have:
		\be
		\left(\prod_i |s_i\rangle\right)_{C_I \cup C_M} |\psi(\mathrm{out})\rangle_{C_O} \propto P_{M_{C_I}}^{\{s\}} P_{M_{C_I}^\prime}^{\{t\}} |\psi\rangle_{C(g)},
		\ee
	where $\ket{(\mathrm{out})}$ is a normalized state. With $P_{M_{C_I}}^{\{s\}} P_{M_{C_I}^\prime}^{\{t\}}$ acting on both sides of Eqs.~(\ref{thm1}, \ref{thm2}), we get:
		\begin{align}
		(UX_i^{q_i}U^\dagger)_{C_O} |\tilde{\psi}(\mathrm{out})\rangle_{C_O} = \varpi^{-p_is_i-\lambda_{x, i}} |\psi(\mathrm{out})\rangle_{C_O}
		\label{key1} \\
		(UZ_i^{p_ir_i}U^\dagger)_{C_O} |\psi(\mathrm{out})\rangle_{C_O} = \varpi^{q_ir_it_i-\lambda_{z, i}} |\psi(\mathrm{out})\rangle_{C_O},
		\label{key2}
		\end{align}
	where $|\tilde{\psi}(\mathrm{out})\rangle$ is the output resulting from a different input state:
		\be
		|\tilde{\psi}(\mathrm{in})\rangle = X_i^{\dagger p_i} |\{t\}\rangle,
		\ee
	\thatis $\ket{\tilde{\psi}(\mathrm{in})}$ is the input state specified by $\{\tilde{t}\}=\{\dots, t_i+p_i, \dots\}$, all other elements the same as those in $\{t\}$. $P_{M_{C_I}}^{\{s\}} P_{M_{C_I}^\prime}^{\{t\}} \ket{\psi}_{C(g)}$ is not zero as a consequence of $X$ and $Z$ bases being mutually unbiased. We then need the commutation relations between $S_c$ and $X, Z$ \cite{Hall2006}:
		\be
		S_cX=X^cS_c, \, S_cZ=Z^{c^{-1}}S_c.
		\ee
	We see that Eq.~(\ref{key2}) is an eigenvalue equation of operator $(UZ_i^{p_ir_i}U^\dagger)$, with eigenvalue $\varpi^{q_ir_it_i-\lambda_{z, i}}$. There are $n$ of these equations labelled by $i$. The solution is therefore:
		\be
		|\psi(\mathrm{out})\rangle = e^{i\delta(t)} U \left( \underset{i}{\bigotimes} X_i^{\lambda_{z, i}p_i^{-1}r_i^{-1}}S_{q_ip_i^{-1}, i} \right) \ket{\psi(\mathrm{in})},
		\label{result1}
		\ee
	with $e^{i\delta(t)}$ being an arbitrary phase factor which may depend on $\{t\}$. Since $\{t\}$ is not specified, this equation applies to the input state defined by $\{\tilde{t}\}$ as well:
		\be
		|\tilde{\psi}(\mathrm{out})\rangle = e^{i\delta(\tilde{t})} U \left( \underset{i}{\bigotimes} X_i^{\lambda_{z, i}p_i^{-1}r_i^{-1}}S_{q_ip_i^{-1}, i} \right) \ket{\tilde{\psi}(\mathrm{in})}.
		\label{result2}
		\ee
	Substituting Eqs.~\ref{result1} and \ref{result2} into Eq.~\ref{key1}, we obtain:
		\be
		e^{i\delta(\tilde{t})} = e^{i\delta(t)} \varpi^{-s_ip_i-\lambda_{x, i}},
		\ee
	or equivalently,
		\be
		\delta(\tilde{t}) - \delta(t) = \frac{2\pi}{d} (-s_ip_i-\lambda_{x, i}),
		\label{eq:recursion}
		\ee
	for $\{\tilde{t}\}=\{\dots, t_i+p_i, \dots\}$. We could get this phase difference by adding $Z_i$ operator into the solution:
		\begin{align}
		|\psi(\mathrm{out})\rangle & = e^{i\delta^{\prime}(t)} U Z_i^{-(s_ip_i+\lambda_{x, i})q_i^{-1}} \times \nonumber \\
			& \left( \underset{k}{\bigotimes} X_k^{\lambda_{z, k}p_k^{-1}r_k^{-1}}S_{q_kp_k^{-1}, k} \right) \ket{\psi(\mathrm{in})},
		\end{align}
	where $\delta^{\prime}(t)$ now has no $t_i$ dependence. For each $i$ from $1$ to $n$, Eq.~\ref{eq:recursion} holds so we can put in $Z_i$ operator for each $i$. In the end we get:
		\begin{align}
		& |\psi(\mathrm{out})\rangle = e^{i\eta} U \times \nonumber \\
			& \left( \underset{i}{\bigotimes} Z_i^{-(s_ip_i+\lambda_{x, i})q_i^{-1}} X_i^{\lambda_{z, i}p_i^{-1}r_i^{-1}}S_{q_ip_i^{-1}, i} \right) \ket{\psi(\mathrm{in})},
		\label{eq:result}
		\end {align}
	where $\eta$ does not depend on any of the $t_i$'s. Therefore we can define the unimportant global phase to be 1, and prove the theorem. Alternatively, one can employ the trick of considering another input state $\overset{n}{\underset{i=1}{\bigotimes}} \ket+_i$ and use linearity to relate its phase of output state $e^{i\chi}$ to the phase $e^{i\eta(t)}$ of the output resulting from input labelled by $\{t\}$. To satisfy this one obtain $e^{i\chi} = \frac{1}{d^n}  \sum_{t} e^{i\eta(t)}$~\cite{Raussendorf2003, Zhou2003}.	
	In the next section we are going to use this theorem and realize various types of gates.

	\subsection{Realization of gates} \label{gates_realization}
	
	\begin{figure}
		\centering
		\begin{subfigure}[t]{0.45\textwidth}
			\centering
			\includegraphics[height=10mm,width=20mm]{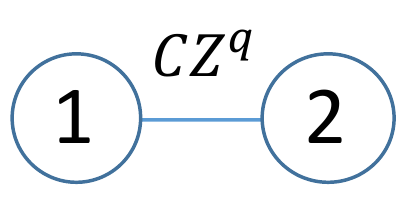}
			\caption{Can realize teleportation. \label{teleportation1}}
		\end{subfigure}
		\hspace{2mm}
		\begin{subfigure}[t]{0.45\textwidth}
			\centering
			\includegraphics[height=15mm,width=35mm]{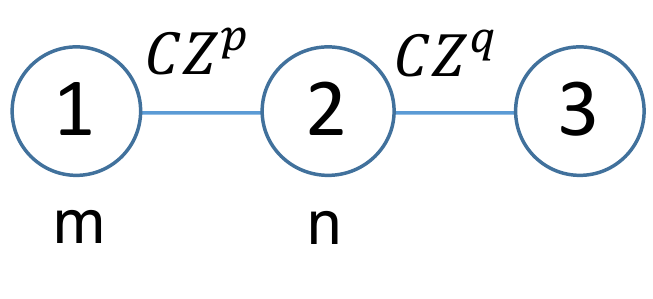}
			\caption{Can realize identity. \label{identity1}}
		\end{subfigure} \\
		\vspace{10mm}
		\begin{subfigure}[t]{0.45\textwidth}
			\centering
			\includegraphics[height=10mm,width=60mm]{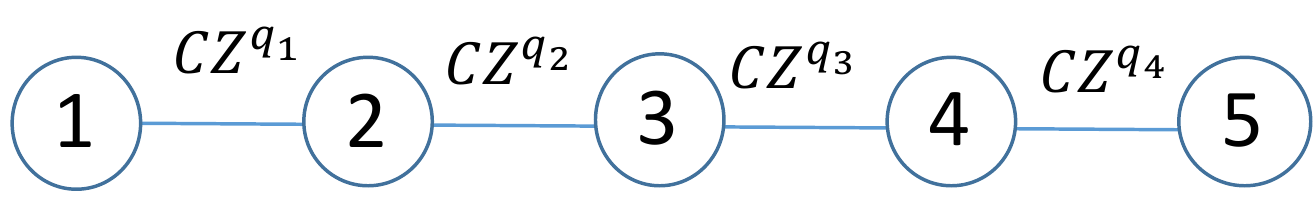}
			\caption{Can realize $X^\alpha(m)$, and two types of Clifford group elements. \label{5sites1}}
		\end{subfigure}
		\hspace{2mm}
		\begin{subfigure}[t]{0.45\textwidth}
			\centering
			\includegraphics[height=10mm,width=70mm]{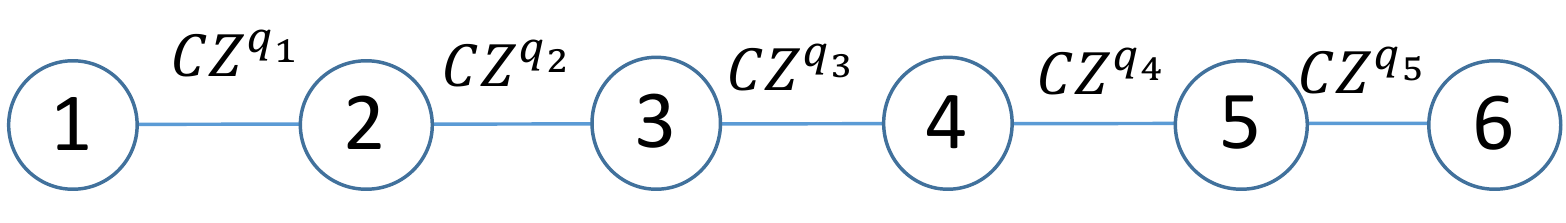}
			\caption{Can realize one type of Clifford group elements. \label{6sites1}}
		\end{subfigure} \\
		\vspace{10mm}
		\begin{subfigure}[t]{0.45\textwidth}
			\centering
			\includegraphics[height=20mm,width=35mm]{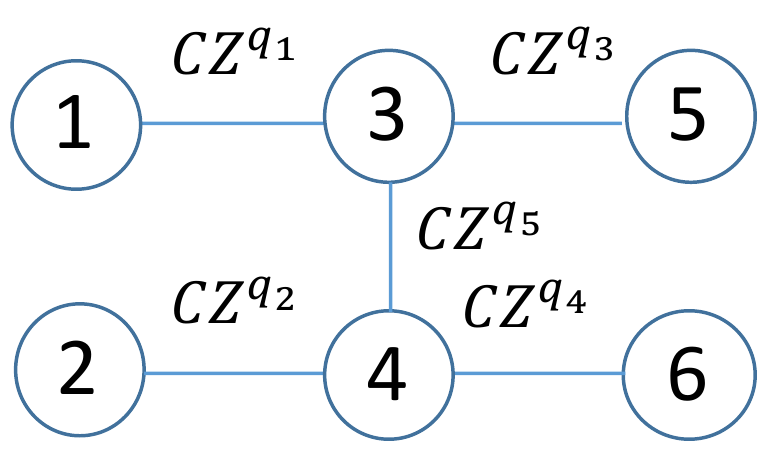}
			\caption{Can realize an imprimitive 2-qudit gate. \label{2gate1}}
		\end{subfigure}
		\caption{Qudit cluster-like states that can realize various gates up to byproduct operators. Each edge is labelled with the gate that entangles the neighboring qudits.}
	\end{figure}
	
	Having established the theorem, we will see how it can be utilized to prove realization of some gates. The goal is to realize the complete collection of one-qudit gates, and one imprimitive two-qudit gate. We then closely follow the procedure in Ref.~\cite{Zhou2003} and see how their measurement patterns still hold in our case. In short, to simulate all one-qudit gates we need to realize gates of the form:
		\begin{align}
		Z^\alpha(m) = \sum_n |n\rangle \varpi^{\alpha(n+m_nd)} \langle n|, \, \alpha \in \mathds{R} \\
		X^\alpha(m) = \sum_n |+_n\rangle \varpi^{\alpha(n+m_nd)} \langle +_n|, \, \alpha \in \mathds{R},
		\end{align}
	where $m$ is a collection of $d$ integers; and four types of Clifford group elements. We merge their first type and third type to one, and review the definitions below. The first type is:
		\begin{align}
		U^{(1n)}ZU^{(1n)\dagger} = \varpi^{-n(d-1)/2} ZX^n \\
		U^{(1n)}XU^{(1n)\dagger} = X.
		\end{align}
	With $n=1$ it recovers the third type in Ref.~\cite{Zhou2003}. The second type is:
		\begin{align}
		U^{(n1)}ZU^{(n1)\dagger} = \varpi^{-n(d-1)/2} Z^nX \\
		U^{(n1)}XU^{(n1)\dagger} = Z^\dagger.
		\end{align}
	The last one is:
		\begin{align}
		WZW^\dagger = Z \\
		WXW^\dagger = \varpi^{-(d-1)/2} ZX.
		\end{align}
	We will see that on a linear cluster-like state of five qudits or six qudits these gates can be realized. We then give an example of an imprimitive two-qudit gate realized on a six-qudit cluster-like state.
	
\begin{figure}[h]
		\centering
		\includegraphics[height=40mm,width=80mm]{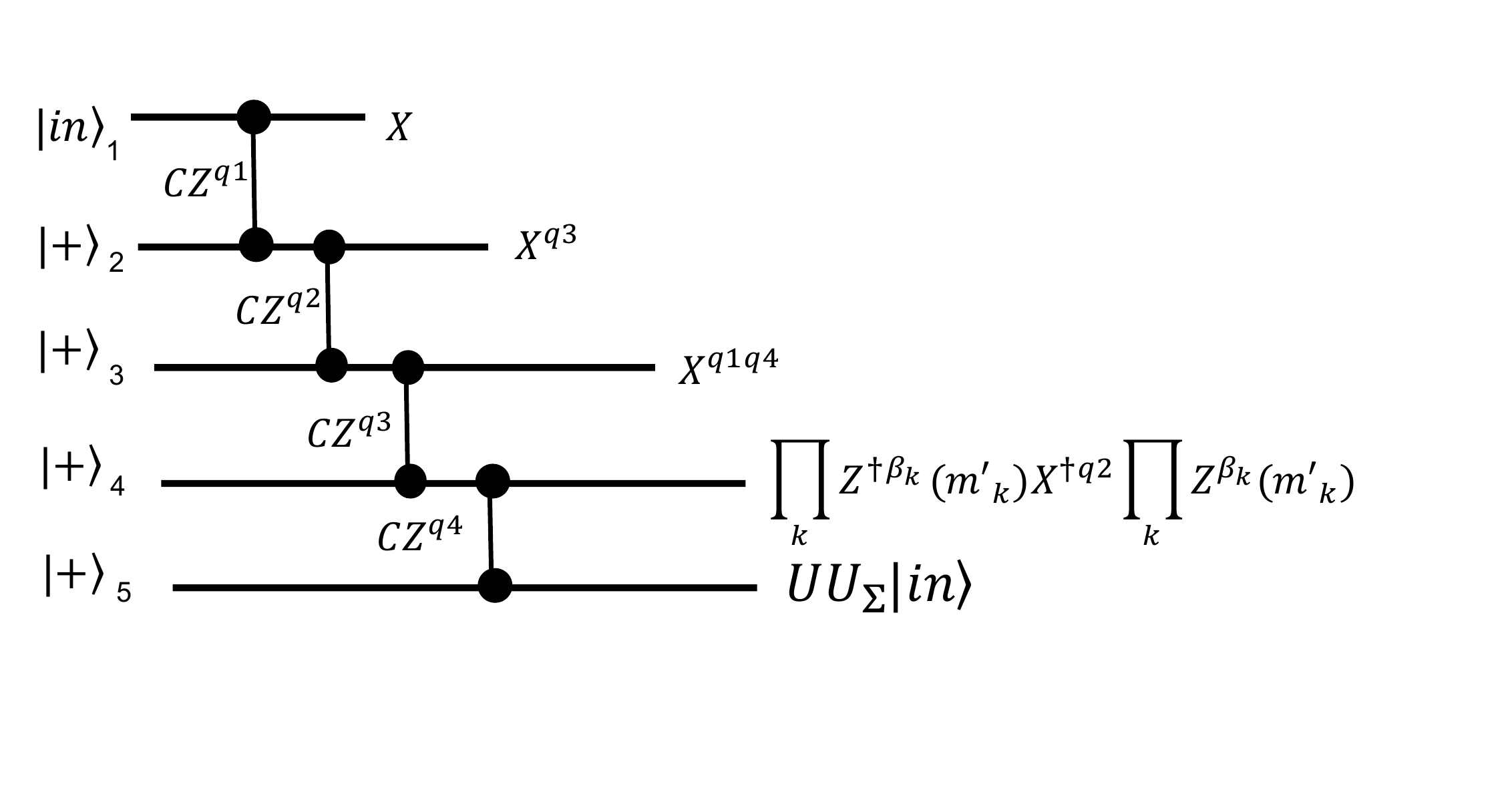}
		\caption{View realization of gate $X_5^\alpha(m)$ from the perspective of teleportation.}
		\label{tele}
	\end{figure}

	Let us first examine a cluster-like state of five qudits, as shown in Fig.~\ref{5sites1}. Eq.~(\ref{stabilizer}) holds for each of the five sites:
		\begin{align}
		X_1^\dagger Z_2^{q_1} |\phi\rangle_C = |\phi\rangle_C \\
		Z_1^{q_1} X_2^\dagger Z_3^{q_2} |\phi\rangle_C = |\phi\rangle_C \\
		Z_2^{q_2} X_3^\dagger Z_4^{q_3} |\phi\rangle_C = |\phi\rangle_C \\
		Z_3^{q_3} X_4^\dagger Z_5^{q_4} |\phi\rangle_C = |\phi\rangle_C \\
		Z_4^{q_4} X_5^\dagger |\phi\rangle_C = |\phi\rangle_C.
		\end{align}
	In order to achieve what appears on the left hand side of Eqs.~(\ref{thm1}, \ref{thm2}), we construct some operators from these stabilizers. As explained earlier, with stabilizers changed to their Hermitian conjugates the eigenvalue equations still hold. We then take Hermitian conjugate of some stabilizers, raise them to certain powers, and combine them with some powers of other stabilizers. As a consequence,
		\begin{align}
		X_1^{q_2q_4} X_3^{\dagger q_1q_4} X_5^{q_1q_3} |\phi\rangle_C = |\phi\rangle_C \label{5site1} \\
		Z_1^{\dagger q_1q_3} X_2^{q_3} X_4^{\dagger q_2} Z_5^{q_2q_4} |\phi\rangle_C = |\phi\rangle_C \label{5site2}.
		\end{align}
	First we show that on this state we can simulate gates $X^\alpha(m), \, Z^\alpha(m)$. We want to prove another eigenvalue equation:
		\be
		Z_4^{\dagger\beta}(m^\prime) X_5^\alpha(m) |\phi\rangle_C = |\phi\rangle_C,
		\label{phase}
		\ee
	where $\beta$ and $m^\prime$ are uniquely determined by $\alpha$ and $m$ for given $q_4$. We define some numbers before giving these relations. For $1 \leq q_4 \leq d-1$ there exists an integer $1 \leq q_4^{-1} \leq d-1$ which satisfies $q_4^{-1}q_4=kd+1, \, k \in \mathds{Z}$. For given $q_4$, any $0 \leq n \leq d$, we can find a unique $\bar{n}$ between 0 and $d$ such that
		\be
		q_4n=\bar{n} \, mod \, d \label{bar}.
		\ee
	Now we have:
		\begin{align}
		\beta=\frac{\alpha}{q_4^{-1}} \label{beta} \\
		m^\prime_n = kn + q_4^{-1}(m_{\bar{n}}+\frac{\bar{n}-q_4n}{d}) \label{m_prime}.
		\end{align}
		First we notice that $|\phi\rangle_C = E \overset{5}{\underset{i=1}{\bigotimes}} |+\rangle_i$ where the entangling operation is:
		\be
		E = CZ_{12}^{q_1} CZ_{23}^{q_2} CZ_{34}^{q_3} CZ_{45}^{q_4}.
		\ee
	Therefore Eq.~(\ref{phase}) is equivalent to
		\be
		(E^\dagger Z_4^{\dagger\beta}(m^\prime) X_5^\alpha(m) E) \overset{5}{\underset{i=1}{\bigotimes}} |+\rangle_i = \overset{5}{\underset{i=1}{\bigotimes}} |+\rangle_i,
		\ee
	so we can prove this equation instead. Writing this explicitly and after some manipulation,
		\begin{align}
		& (CZ_{45}^{q_4\dagger} Z_4^{\dagger\beta}(m^\prime) X_5^\alpha(m) CZ_{45}^{q_4}) |+\rangle_4 \otimes |+\rangle_5 \nonumber \\
		& = \frac{1}{\sqrt{d}}\sum_{n} |n\rangle_4 \otimes |+\rangle_5 \times e^{i\frac{2\pi}{d}[\alpha(\bar{n}+m_{\bar{n}}d)-\beta(n+m_nd)]},
		\end{align}
	where $\bar{n}$ is given by Eq.~(\ref{bar}). For the right hand side to equal $|+\rangle_4 \otimes |+\rangle_5$ we need:
		\be
		\alpha(\bar{n}+m_{\bar{n}}d)-\beta(n+m_nd)=0.
		\ee
	We can check that Eqs.~(\ref{beta}, \ref{m_prime}) satisfy this. Eq.~(\ref{phase}) is then proved. Making use of it, combined with Eqs.~(\ref{5site1}, \ref{5site2}), we obtain the following:
		\begin{align}
		& X_1^{q_2q_4} X_3^{\dagger q_1q_4} [X_5^\alpha(m) X_5^{q_1q_3} X_5^{\dagger\alpha}(m)] |\phi\rangle_C = |\phi\rangle_C \\
		& Z_1^{\dagger q_1q_3} X_2^{q_3} [Z_4^{\dagger\beta}(m^\prime) X_4^{\dagger q_2} Z_4^\beta(m^\prime)] \times \nonumber \\
			& [X_5^\alpha(m) Z_5^{q_2q_4} X_5^{\dagger\alpha}(m)] |\phi\rangle_C = |\phi\rangle_C.
		\end{align}
	The theorem then tells us that this realizes the gate $X_5^\alpha(m)U_\Sigma$. To find the byproduct operator, we see that
		\be
		U_\Sigma = Z^{-(s_1q_2q_4+s_3)q_1^{-1}q_3^{-1}} X^{(s_2+s_4)q_2^{-1}q_4^{-1}} S_{q_1q_3q_2^{-1}q_4^{-1}}.
		\ee
	Since $X^\alpha(m)$ commutes with $X$, we only need to study powers of $Z$ and $S_{q_1q_3q_2^{-1}q_4^{-1}}$. Note that:
		\begin{align}
		Z |+_j\rangle = |+_{j+1}\rangle \\
		S_c |+_j\rangle = |+_{jc^{-1}}\rangle.
		\end{align}
	Therefore $Z^zS_c X^\alpha(m) S_c^{-1}Z^{-z}$ is also diagonal in $X$ basis. We then define $m^{(k)}$ to be a collection of $d$ integers where only non-zero element is $m^{(k)}_k=m_k$, for $k=0, \, ... \, d-1$. The following equation can be solved by $d$ real numbers labelled by $\alpha_k$:
		\begin{align}
		\prod_k X^{\alpha_k}(m^{(k)}) & = Z^{-(s_1q_2q_4+s_3)q_1^{-1}q_3^{-1}} S_{q_1q_3q_2^{-1}q_4^{-1}} \times \nonumber \\
			& X^\alpha(m) S_{q_1q_3q_2^{-1}q_4^{-1}}^{-1} Z^{(s_1q_2q_4+s_3)q_1^{-1}q_3^{-1}}.
		\label{basis}
		\end{align}
	The procedure to realize $X^\alpha(m)$ is as follows: measure qudit 1 in basis $X_1$, qudit 2 in basis $X_2^{q_3}$ and qudit 3 in basis $X_3^{\dagger q_1q_4}$ to obtain $s_1$, $s_2$ and $s_3$; solve Eq.~(\ref{basis}); measure qudit 4 in basis $[\prod_kZ_4^{\dagger\beta_k}(m^{\prime(k)}) X_4^{\dagger q_2} \prod_kZ_4^{\beta_k}(m^{\prime(k)})]$, where $\beta_k$ and $m^{\prime(k)}$ are related to $\alpha_k$ and $m^{(k)}$ by Eqs.~(\ref{beta}, \ref{m_prime}). The final result is:
		\begin{align}
		& \left( \prod_k X^{\alpha_k}(m^{(k)}) \right) U_\Sigma = \nonumber \\
			& Z^{-(s_1q_2q_4+s_3)q_1^{-1}q_3^{-1}} X^{(s_2+s_4)q_2^{-1}q_4^{-1}} S_{q_1q_3q_2^{-1}q_4^{-1}} X^\alpha(m).
		\end{align}
	It will be interesting to see how we can view this realization from the perspective of teleportation. In Fig.~\ref{tele} we see that the measurement basis of the last step is determined by previous measurements.
	
	The five-qudit cluster-like state also enables us to realize the first and the last types of Clifford group elements. Starting from Eqs.~(\ref{5site1}, \ref{5site2}), we can form some operators by applying the operator in Eq.~(\ref{5site1}) $q_1^{-1}q_3^{-1}$ times, the operator in Eq.~(\ref{5site2}) $q_2^{-1}q_4^{-1}$ times, and the stabilizer $X_5^\dagger Z_4^{q_4}$ $n$ times.
		\begin{align}
		X_1^{q_2q_4q_1^{-1}q_3^{-1}} X_3^{\dagger q_4q_3^{-1}} X_5 |\phi\rangle_C = |\phi\rangle_C \\
		Z_1^{\dagger q_1q_3q_2^{-1}q_4^{-1}} X_2^{q_3q_2^{-1}q_4^{-1}} (Z_4^{q_4n}X_4^{q_4^{-1}})^\dagger (Z_5 X_5^n) |\phi\rangle_C = |\phi\rangle_C.
		\end{align}
	From the theorem we can realize the first type by measuring in the bases \{$X_1$, $X_2^{q_3q_2^{-1}q_4^{-1}}$, $X_3^{\dagger q_4q_3^{-1}}$, $(\varpi^{-n(d-1)/2}Z_4^{q_4n}X_4^{q_4^{-1}})^\dagger$\}. Denoting the outcomes by $\{s_1, s_2, s_3, s_4\}$, we have:
		\be
		U_\Sigma = Z^{-s_1q_2q_4q_1^{-1}q_3^{-1}-s_3} X^{s_2+s_4} S_{q_1q_3q_2^{-1}q_4^{-1}}.
		\ee
	Commuting $U^{(1n)}$ through powers of $Z$ and $X$, we see that the only effects are change in exponents of $Z$ and $X$ and a global phase. The non-trivial consequence comes from commuting through $S_{q_1q_3q_2^{-1}q_4^{-1}}$. We examine $\tilde{U}^{(1n)} = S_cU^{(1n)}S_c^{-1}$:
		\begin{align}
		\tilde{U}^{(1n)} Z \tilde{U}^{(1n)\dagger} = \varpi^{nc(c-d)/2} ZX^{nc^2} \\
		\tilde{U}^{(1n)} X \tilde{U}^{(1n)\dagger} = X,
		\end{align}
	where for simplicity we define:
		\be
		c=q_1q_3q_2^{-1}q_4^{-1} \label{c_defn}.
		\ee
	By measuring in the bases \{$X_1$, $X_2^{q_3q_2^{-1}q_4^{-1}}$, $X_3^{\dagger q_4q_3^{-1}}$, $(\varpi^{nc(c-d)/2}Z_4^{q_4nc^2}X_4^{q_4^{-1}})^\dagger$\} instead, we obtain the desired form:
		\be
		\tilde{U}^{(1n)} U_\Sigma = \varpi^{ncz(cz-d)/2} Z^{z} X^{s_2+s_4+nc^2z} S_c U^{(1n)},
		\ee
	where $z=-s_1q_2q_4q_1^{-1}q_3^{-1}-s_3$. Similarly, we can also form the following two equations from Eqs.~(\ref{5site1}, \ref{5site2}):
		\begin{align}
		& X_1^{q_2q_4q_1^{-1}q_3^{-1}} (Z_3^{q_3q_4^{-1}c^{-2}} X_3^{\dagger q_4q_3^{-1}}) \times \nonumber \\
			& X_4^{\dagger q_4^{-1}c^{-2}} (Z_5^{c^{-2}} X_5) |\phi\rangle_C = |\phi\rangle_C \\
		& Z_1^{\dagger q_1q_3q_2^{-1}q_4^{-1}} X_2^{q_3q_2^{-1}q_4^{-1}} X_4^{\dagger q_4^{-1}} Z_5 |\phi\rangle_C = |\phi\rangle_C,
		\end{align}
	where $c$ is defined in the same way as Eq.~(\ref{c_defn}). Measuring in the bases \{$X_1$, $X_2^{q_3q_2^{-1}q_4^{-1}}$, $(\varpi^{c^{-1}(d-c^{-1})/2} Z_3^{q_3q_4^{-1}c^{-2}} X_3^{\dagger q_4q_3^{-1}})$, $X_4^{\dagger q_4^{-1}}$\} then allows us to realize $\tilde{W}U_\Sigma$, where $\tilde{W} = S_cWS_c^{-1}$. From this we get byproduct operator:
		\begin{align}
		\tilde{W} U_\Sigma & = \varpi^{c^{-2}(s_2+s_4)(s_2+s_4-cd)/2} \times \nonumber \\
		& Z^{-s_1c^{-1}-s_3+s_2c^{-2}} X^{s_2+s_4} S_c W.
		\end{align}
	
	For the second type, we need a linear cluster-like state of six qudits, like in Fig.~\ref{6sites1}. The stabilizer eigenvalue equations are:
		\begin{align}
		X_1^\dagger Z_2^{q_1} |\phi\rangle_C = |\phi\rangle_C \\
		Z_1^{q_1} X_2^\dagger Z_3^{q_2} |\phi\rangle_C = |\phi\rangle_C \\
		Z_2^{q_2} X_3^\dagger Z_4^{q_3} |\phi\rangle_C = |\phi\rangle_C \\
		Z_3^{q_3} X_4^\dagger Z_5^{q_4} |\phi\rangle_C = |\phi\rangle_C \\
		Z_4^{q_4} X_5^\dagger Z_6^{q_5} |\phi\rangle_C = |\phi\rangle_C \\
		Z_5^{q_5} X_6^\dagger |\phi\rangle_C = |\phi\rangle_C.
		\end{align}
	From them we get the following two equations:
		\begin{align}
		& X_1^{q_2q_4q_1^{-1}q_3^{-1}q_5^{-1}} X_3^{\dagger q_4q_3^{-1}q_5^{-1}} X_5^{q_5^{-1}} Z_6^\dagger |\phi\rangle_C = |\phi\rangle_C \\
		& Z_1^{\dagger q_1q_3q_5q_2^{-1}q_4^{-1}} X_2^{q_3q_5q_2^{-1}q_4^{-1}} (Z_4^{q_4q_5^{-1}ne^2}X_4^{\dagger q_5q_4^{-1}}) \times \nonumber \\
			& X_5^{\dagger q_5^{-1}ne^2} (Z_6^{ne^2} X_6) |\phi\rangle_C = |\phi\rangle_C,
		\end{align}
	where $e=q_1q_3q_5q_2^{-1}q_4^{-1}$. This realizes $\tilde{U}^{(n1)} U_\Sigma$ where $\tilde{U}^{(n1)}$ is defined by:
		\begin{align}
		\tilde{U}^{(n1)} Z \tilde{U}^{(n1)\dagger} = \varpi^{ne(e-d)/2} Z^{ne^2} X \\
		\tilde{U}^{(n1)} X \tilde{U}^{(n1)\dagger} = Z^\dagger.
		\end{align}
	If we measure in bases \{$X_1$, $X_2^{q_3q_5q_2^{-1}q_4^{-1}}$, $X_3^{\dagger q_4q_3^{-1}q_5^{-1}}$, $(\varpi^{ne(d-e)/2}Z_4^{q_4q_5^{-1}ne^2}X_4^{\dagger q_5q_4^{-1}})$, $X_5^{q_5^{-1}}$\}. And we have $\tilde{U}^{(n1)}S_e=S_{e^{-1}} U^{(n1)}$. Therefore, defining $x=-s_1e^{-1}-s_3-s_5$ for simplicity,
		\begin{align}
		\tilde{U}^{(n1)} U_\Sigma & = \varpi^{x(\frac{1}{2}ne^2x-s_2-s_4)} \times \nonumber \\
			& Z^{ne^2x} X^x S_{e^{-1}} U^{(n1)}.
		\end{align}
	
	Now Let us make use of the theorem again to show realization of an imprimitive two-qudit gate. We look at the six-qudit cluster-like state in Fig.~\ref{2gate1}. Again, we give the stabilizer eigenvalue equations first:
		\begin{align}
		X_1^\dagger Z_3^{q_1} |\phi\rangle_C = |\phi\rangle_C \\
		X_2^\dagger Z_4^{q_2} |\phi\rangle_C = |\phi\rangle_C \\
		Z_1^{q_1} X_3^\dagger Z_5^{q_3} Z_4^{q_5} |\phi\rangle_C = |\phi\rangle_C \\
		Z_2^{q_2} X_4^\dagger Z_6^{q_4} Z_3^{q_5} |\phi\rangle_C = |\phi\rangle_C \\
		X_5^\dagger Z_3^{q_3} |\phi\rangle_C = |\phi\rangle_C \\
		X_6^\dagger Z_4^{q_4} |\phi\rangle_C = |\phi\rangle_C.
		\end{align}
	After some manipulation that involves finding inverse in $\zd$ we reach:
		\begin{align}
		X_1^{-q_1^{-1}q_3} X_5 |\phi\rangle_C = |\phi\rangle_C \\
		Z_1^{\dagger -q_1q_3^{-1}} X_3^{\dagger q_3^{-1}} Z_5 X_6^q |\phi\rangle_C = |\phi\rangle_C \\
		X_2^{-q_2^{-1}q_4} X_6 |\phi\rangle_C = |\phi\rangle_C \\
		Z_2^{\dagger -q_2q_4^{-1}} X_4^{\dagger q_4^{-1}} Z_6 X_5^q |\phi\rangle_C = |\phi\rangle_C,
		\end{align}
	where $q=q_3^{-1}q_4^{-1}q_5$. Obtaining measurement outcomes $s_3$ and $s_4$ on qudit 3 in the $X_3^{q_3^{-1}}$ basis and qudit 4 in the $X_4^{q_4^{-1}}$ basis respectively, we find that the effect of gate $U$ is:
		\begin{align}
		UX_5U^\dagger = X_5 \label{eq:imprimitive1} \\
		UX_6U^\dagger = X_6 \label{eq:imprimitive2} \\
		UZ_5U^\dagger = Z_5 X_6^q \label{eq:imprimitive3} \\
		UZ_6U^\dagger = Z_6 X_5^q. \label{eq:imprimitive4}
		\end{align}
	It can then be verified that $U$ is the equivalent of $CZ^q$ gate in the $X$ basis: $\tilde{U}(q)=\sum_{j, k}\varpi^{qjk} |+_j\rangle_5\langle+_j| \otimes |+_k\rangle_6\langle+_k|$. By commuting $U_\Sigma$ through $\tilde{U}(q)$ we find that:
		\begin{align}
		& \tilde{U}(q) Z_5^{s_1q_1^{-1}q_3}X_5^{s_3}S_{-q_1q_3^{-1}, 5} Z_6^{s_2q_2^{-1}q_4}X_6^{s_4}S_{-q_2q_4^{-1}, 6} \nonumber \\
		& = Z_5^{z_5}X_5^{x_5}S_{-q_1q_3^{-1}, 5} Z_6^{z_6}X_6^{x_6}S_{-q_2q_4^{-1}, 6} \tilde{U}(q_1^{-1}q_2^{-1}q_5),
		\end{align}
	where $z_5, \, x_5, \, z_6, \, x_6$ are integers that can be found according to Eqs.~\ref{eq:imprimitive1}-\ref{eq:imprimitive4}. The actual gate realized is $\tilde{U}(q_1^{-1}q_2^{-1}q_5)$, which is indeed imprimitive. Combining all the one-qudit gates above, we can simulate any arbitrary one-qudit gate. Together with the imprimitive two-qudit gate, we are able to simulate any gate.
	
	The entanglement structure to implement the universal gates can be cut from that on the square lattice, by measuring certain sites in the $Z$ basis. Therefore the qudit cluster-like state on the square lattice or any other regular lattices is universal.

\end{document}